\RequirePackage[T1]{fontenc}
\documentclass[11pt, urlcolor=blue, linkcolor=blue]{article} 
\usepackage{cite}
\usepackage{float}
\usepackage[utf8]{inputenc}

\usepackage{multicol}
\usepackage{multirow}
\usepackage{tablefootnote}
\usepackage{amsmath, amsthm, amssymb,slashed,mathtools,tabu}
\usepackage{pifont}
\newcommand{\cmark}{\ding{51}}
\newcommand{\xmark}{\ding{55}}

\usepackage{ifpdf}
\ifpdf
  \usepackage[pdftex]{graphicx}
  \usepackage{epstopdf}
\else
  \usepackage[dvips]{graphicx}
\fi
\parskip=\baselineskip

\usepackage[usenames, dvipsnames]{color}
\usepackage[svgnames]{xcolor}
\usepackage{colortbl}
\usepackage[colorlinks,citecolor=RoyalBlue, urlcolor=RoyalBlue, linkcolor=RoyalBlue ]{hyperref} %BlueViolet  NavyBlue RoyalBlue MidnightBlue

\allowdisplaybreaks[1]

\numberwithin{equation}{section}

\usepackage{booktabs}

\theoremstyle{definition}

\newcommand{\cblue}[1]{\textcolor{blue}{#1}}

\newcommand{\ccblue}[1]{\textcolor{blue}{#1}}

\definecolor{amber}{rgb}{1.0, 0.49, 0.0}

\newcommand{\ccorange}[1]{\textcolor{amber}{#1}}

\definecolor{mygray}{gray}{0.6}

\usepackage{upgreek}
\usepackage{bbm}

\newcommand\finline[3][]{\begin{myfont}[#1]{#2}#3\end{myfont}}%
\newenvironment{myfont}[2][]{\csname#2\endcsname[#1]}{}

\usepackage{slashed}
\usepackage[makeroom]{cancel}
\usepackage[normalem]{ulem}
\usepackage{soul}
\newcommand{\stkout}[1]{\ifmmode\text{\sout{\ensuremath{#1}}}\else\sout{#1}\fi}

\usepackage{sseq}
\usepackage[all,cmtip]{xy}
\usepackage{tikz-cd}
\usepackage{tikz}
\usetikzlibrary{matrix}

\newcommand{\bea}{\begin{eqnarray}}
\newcommand{\eea}{\end{eqnarray}}
\def\be{\begin{equation}}
\def\ee{\end{equation}}

\newcommand{\e}{\hspace{1pt}\mathrm{e}}
\newcommand{\ii}{\hspace{1pt}\mathrm{i}\hspace{1pt}}

\def\CP{{\mathbb{CP}}}

\newcommand{\nn}{\nonumber}

\definecolor{red}{rgb}{1,0,0}
\definecolor{blue}{rgb}{0,0,1}
\definecolor{dblue}{rgb}{0,0,0.4}
\definecolor{green}{rgb}{0,1,0}
\definecolor{black}{rgb}{0,0,0}
\definecolor{white}{rgb}{1,1,1}

\definecolor{brn}{rgb}{.8,.4,.0}
\definecolor{redo}{rgb}{1,.5,.0}
\definecolor{ddgrn}{rgb}{0,0.4,0}
\definecolor{dgrn}{rgb}{0,0.55,0}
\definecolor{dbl}{rgb}{0,0,0.5}

\definecolor{lightgray}{gray}{0.9}

\usepackage[bbgreekl]{mathbbol}
\usepackage{amscd}

\newcommand{\Z}{\mathbb{Z}}
\newcommand{\C}{\mathbb{C}}
\newcommand{\R}{\mathbb{R}}

\newcommand{\dd}{\hspace{1pt}\mathrm{d}}
\newcommand{\<}{\langle} 
\renewcommand{\>}{\rangle} 
\newcommand{\ket}[1]{{|#1\rangle}}
\newcommand{\bra}[1]{{\langle #1|}}
\newcommand{\braket}[2]{{\langle #1 |#2\rangle}}

\newcommand{\Refe}[1]{Ref.~\cite{#1}}
 
\newcommand{\eq}[1]{(\ref{#1})} 
 
\newcommand{\Eqn}[1]{Eqn.~(\ref{#1})} 

\newcommand{\Tr}{{\rm Tr}} 
 
\renewcommand{\Im}{{\rm Im}} 
\renewcommand{\Re}{{\rm Re}}

\newcommand{\prt}{\partial}

\newcommand{\bpm}{\begin{pmatrix}}
\newcommand{\epm}{\end{pmatrix}}
\newcommand{\bmm}{\begin{matrix}}
\newcommand{\emm}{\end{matrix}}

\newcommand{\cB}{ {\cal B} }
 
\newcommand{\cD}{ {\cal D} }

\def\CA{{\cal A}}
\def\CB{{\cal B}}
\def\CC{{\cal C}}

\def\CO{{\cal O}}

\def\Z{{\mathbb{Z}}}

\def\R{{\mathbb{R}}}
\def\C{{\mathbb{C}}}

\def\Tr{{\mathrm{Tr}}}

\def \Hom{\operatorname{Hom}}

\def \Im{\operatorname{Im}}
\def \H{\operatorname{H}}
\def \id{\operatorname{id}}

\def \Z{\mathbb{Z}}
\def \Pin{\mathrm{Pin}}

\def \CP{\mathbb{CP}}

\newcommand{\Sec}[1]{Sec.~\ref{#1}}

\usetikzlibrary{decorations.markings}
\usetikzlibrary{positioning}
\usetikzlibrary{shadings}

\newcommand{\SO}{{\rm SO}}
\newcommand{\Spin}{{\rm Spin}}
\newcommand{\U}{{\rm U}}
\newcommand{\SU}{{\rm SU}}
\newcommand{\PSU}{{\rm PSU}}

\renewcommand{\O}{{\rm O}}

\newcommand{\Cl}{{\rm Cl}}

\newcommand{\rE}{{\rm E}}
\newcommand{\rF}{{\rm F}}
\newcommand{\rN}{{\rm N}}

\def\Sq{\mathrm{Sq}}

\def\B{\mathrm{B}}

\def\TP{\mathrm{TP}}

\usepackage{datetime}
\usepackage{enumitem} 
\usepackage{moreenum}

\newcommand{\Wfootnote}[1]{%
\let\oldthefootnote=\thefootnote%
\stepcounter{mpfootnote}%
\addtocounter{footnote}{-1}%
\renewcommand{\thefootnote}{{W}} %{{W$^+$}} %{{W$^+\sharp$}}
\footnote{#1}%
\let\thefootnote=\oldthefootnote%
}

\newcommand{\naturalfootnote}[1]{%
\let\oldthefootnote=\thefootnote%
\stepcounter{mpfootnote}%
\addtocounter{footnote}{-1}%
\renewcommand{\thefootnote}{{W$^-\natural$}}
\footnote{#1}%
\let\thefootnote=\oldthefootnote%
}

\newcommand{\flatfootnote}[1]{%
\let\oldthefootnote=\thefootnote%
\stepcounter{mpfootnote}%
\addtocounter{footnote}{-1}%
\renewcommand{\thefootnote}{{W$^-\flat$}}
\footnote{#1}%
\let\thefootnote=\oldthefootnote%
}

\DeclareRobustCommand\sWang
{\cblue{\includegraphics[height=3.2ex]{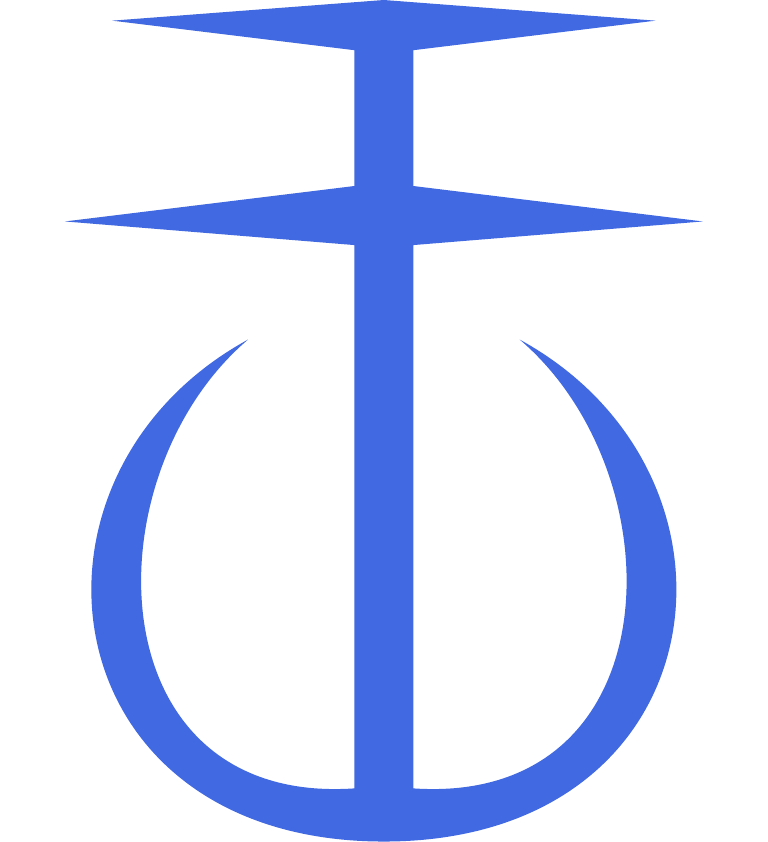}}}

\DeclareRobustCommand\sYou
{\cblue{\includegraphics[height=3.3ex]{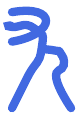}}}

\newcommand{\Wangfootnote}[1]{%
\let\oldthefootnote=\thefootnote%
\stepcounter{mpfootnote}%
\addtocounter{footnote}{-1}%
\renewcommand{\thefootnote}{\sWang}
\footnote{#1}%
\let\thefootnote=\oldthefootnote%
}

\newcommand{\Youfootnote}[1]{%
\let\oldthefootnote=\thefootnote%
\stepcounter{mpfootnote}%
\addtocounter{footnote}{-1}%
\renewcommand{\thefootnote}{\sYou}
\footnote{#1}%
\let\thefootnote=\oldthefootnote%
}

\usepackage{comment}
\def\bZ{{\mathbf{Z}}}

\usepackage{upgreek}
\newcommand{\Fig}[1]{Fig.~\ref{#1}}

\newcommand{\SM}{{\rm SM}}

\newcommand{\GUT}{{\rm GUT}}

\def \DSpin{\mathrm{DSpin}}

\newcommand{\PS}{\text{PS}}
\newcommand{\LR}{\text{LR}}
\newcommand{\GG}{\text{GG}}
\newcommand{\rL}{\text{L}}
\newcommand{\rR}{\text{R}}

\newcommand{\diag}{{\rm diag}}

\usepackage{mathrsfs}
\usepackage{esint} %\oiint

\newcommand{\Table}[1]{Table \ref{#1}}

\newcommand\ointint{\begingroup \displaystyle \unitlength 1pt
\int\mkern-7mu\begin{picture}(0,3)\put(0,3){\oval(10,8)}\end{picture}
\mkern-8mu\int\endgroup}
\usepackage{pdflscape}

\newcommand{\eqs}[1]{\begin{equation}\begin{split}#1\end{split}\end{equation}}

\newcommand{\g}{\mathfrak{g}}
\newcommand{\z}{\mathfrak{z}}

\newcommand{\smat}[1]{\left(\begin{smallmatrix}#1\end{smallmatrix}\right)}

\begin{document}

\begin{titlepage}
%\begin{flushright}
%/yymm.nnnn
%\end{flushright}
%\vskip1.25in

\vspace*{0cm}
\begin{center}

%{\bf\LARGE{ 
%Hidden Topological Force: 
%\\[8mm]
%Anomaly Constraints 
%Beyond Standard Model
% \\[10mm]
%}}

{\bf\LARGE{ 
Gauge Enhanced Quantum Criticality 
\\[8mm]
%and Morphogenesis 
%Beyond 
Between Grand Unifications:
%Beyond Grand Unifications
}\\[10mm]
\Large{Categorical Higher Symmetry Retraction}
}

\vskip0.5cm 
\Large{\quad Juven Wang$^{1}$\Wangfootnote{%e-mail: 
{\tt jw@cmsa.fas.harvard.edu} \quad\quad\quad\quad\; 
 \includegraphics[width=2.2in]{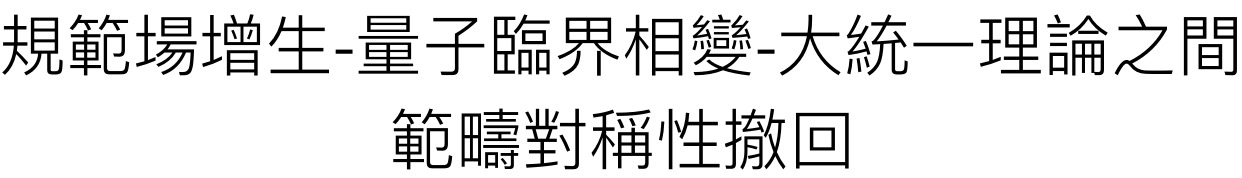}
%\\[2mm]
} 
}
 \Large{\quad\quad Yi-Zhuang You$^{2}$\Youfootnote{
 {\tt  yzyou@ucsd.edu}\\[2mm]
 \strut \hfill Dedicate to Subir Sachdev (60), Xiao-Gang Wen (60), %\\[2mm]
  %\strut \hfill 
  Edward Witten (70), and Shing-Tung Yau (72),\\[2mm]
 \strut \hfill \quad \quad \quad 
%Dedicate to 
anniversaries of various researchers mentioned in: 
\href{https://www.youtube.com/results?search_query=quantum+crticiality+standard+model+Juven+Wang+Yizhuang+You}{Related presentation 
videos available online}
%\href{https://www.youtube.com/results?search_query=quantum+crticiality+standard+model+ultra+unification}
 }
 } 
\\[2.75mm]  
\vskip.5cm
{ {\small{\textit{$^{1}${Center of Mathematical Sciences and Applications, Harvard University,  Cambridge, MA 02138, USA}}}}
}\\
{ {\small{\textit{$^{2}${Department of Physics, University of California, San Diego, CA 92093, USA}}}}
}

\end{center}

\vskip 0.5cm
\baselineskip 16pt
\begin{abstract}

Prior work %[arXiv:2106.16248] 
\cite{Wang2106.16248} 
shows that the Standard Model (SM) naturally arises near a gapless
quantum critical region 
between Georgi-Glashow (GG) $su(5)$ and Pati-Salam (PS) $su(4) \times su(2) \times su(2)$ models of quantum vacua (in a phase diagram or moduli space), 
by implementing a modified $so(10)$ Grand Unification (GUT)
with a Spin(10) gauge group plus a new discrete Wess-Zumino Witten term matching 
a 4d nonperturbative global mixed gauge-gravity $w_2 w_3$ anomaly. 
%(captured by a mod 2 $w_2 w_3$ Stiefel-Whitney class of 5d invertible topological quantum field theory). %(iTQFT).
In this work, we include Barr's flipped $u(5)$ model into the quantum landscape, showing these four GUT-like models arise near the quantum criticality near SM.
The SM and GG models can have either 15 or 16 Weyl fermions per generation, 
while the PS, flipped $u(5)$, and the modified $so(10)$ have 16n Weyl fermions.
Highlights include: 
First, we find the precise GG or 
flipped $u(5)$ gauge group requires to 
redefine a
Lie group U(5)$_{\hat q} \equiv (\SU(5) \times_{\hat q} \U(1))/\Z_{\hat q}$ with $\hat q=2$ or 3 
(instead of non-isomorphic analog $\hat q=1$ or 4), and different $\hat q$ are related by multiple covering.
Second, for 16n Weyl fermions,
we show that the GG and flipped $u(5)$
are two different symmetry-breaking vacua of the same order parameter
separated by a first-order Landau-Ginzburg transition.
We also show that 
analogous 3+1d 
deconfined quantum criticalities,
both between GG and PS,
and between the flipped $u(5)$ and PS,
are beyond Landau-Ginzburg paradigm.
Third,
topological quantum criticality occurs by tuning between the 15n vs 16n scenarios.
{Fourth, we explore the {generalized higher global symmetries in the SM and GUTs}. Gauging the $\Z_2$ flip symmetry between GG and flipped $u(5)$ models, 
leads to a potential categorical higher symmetry that is a non-invertible global symmetry: 
within a gauge group $\big[ (\U(1)_{X_1} \times_{\Z_{4,X}} \U(1)_{X_2}) \rtimes \Z_2^{\rm flip} \big]$,
the fusion rule of 2d topological surface operator splits. 
Even if the mixed anomaly between $\Z_2$ flip symmetry and two %1-form 
U(1) magnetic 1-symmetries at IR is absent, 
the un-Higgs  
Spin(10) at UV 
%we find %obstructions 
%the retraction on %eliminating 
retracts 
this categorical symmetry.}\\

\flushright

\end{abstract}

\end{titlepage}

\pagenumbering{arabic}
\setcounter{page}{2}
    
\tableofcontents

\newpage

\section{Introduction and Summary}
\label{sec:IntroductionandSummary}

\subsection{Unity of Gauge Forces vs Many Dualities}

\begin{figure}[!ht] %[htbp]
\centering
(a)\includegraphics[width=0.4\textwidth]{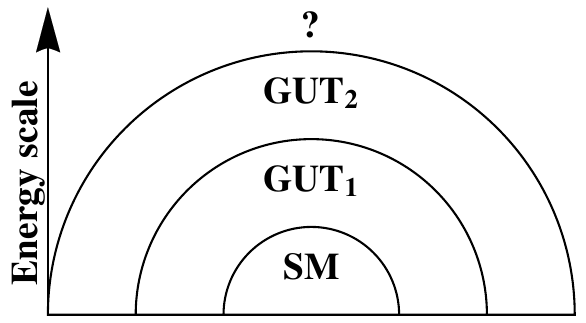}
\quad
(b)\includegraphics[width=0.42\textwidth]{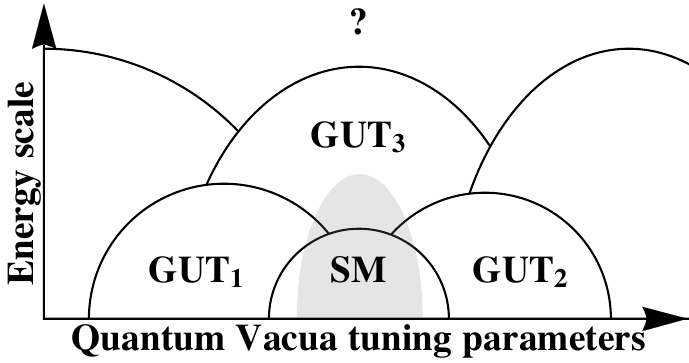}
\caption{(a) ``Unity of Gauge Forces'' perspective seeks for a single unified dynamically gauged internal symmetry at high energy, 
which is a more \emph{kinematic} or \emph{static} issue of gauge theories towards higher energy. %\newline
(b) Our ``quantum competing criticality'' perspective \cite{Wang2106.16248, YouWang2202.13498}
suggests that the SM is 
a low energy quantum vacuum arising from the quantum competition of various neighbor GUT-like vacua (which can also appear at higher energy).
Especially when there are constraints from topological terms or nonperturbative global anomalies,
the SM arises near a gapless quantum critical region (schematically shown as the shaded gray area). 
The gapless quantum critical region induces new beyond-SM gapless modes, Dark Gauge forces, or excitations.
This perspective is more on the \emph{dynamics, criticality,} or \emph{phase transition} issue of gauge theories.
}
\label{fig:two-views}
\end{figure}

One of the open unsolved problems in fundamental physics and high-energy physics (HEP) is:
\bea
\text{``If the strong, electromagnetic, and weak forces of the Standard Model are unified at high energies,}\cr
\text{by which gauge group (of the gauged internal symmetry) is this unification governed?''}\nn
\eea
We may quote this perspective as ``{\bf Unity of Gauge Forces}.''
It is conventional to %research attitude  %ritual practice
regard our quantum vacuum in the 4-dimensional spacetime (denoted as 4d or 3+1d) governed by one of the candidate $su(3)\times su(2)\times u(1)$ Standard Models (SMs) \cite{Glashow1961trPartialSymmetriesofWeakInteractions, Salam1964ryElectromagneticWeakInteractions, Salam1968, Weinberg1967tqSMAModelofLeptons},
while lifting towards one of some Grand Unification-like (GUT-like) structure
\cite{Georgi1974syUnityofAllElementaryParticleForces, Pati1974yyPatiSalamLeptonNumberastheFourthColor,
Fritzsch1974nnMinkowskiUnifiedInteractionsofLeptonsandHadrons, SenjanovicMohapatra1975, 
Barr1982flippedSU5, Derendinger1984flippedSU5, WilczekZee1981iz1982Spinors} or String Theory at higher energy scales.\footnote{{Throughout our article,
we denote $n$d for $n$-dimensional spacetime, or $n'+1$d as an $n'$-dimensional space and 1-dimensional time. 
We also denote the Lie algebra in the lower case such as the $so(10)$ Lie algebra in the $so(10)$ GUT 
\cite{Fritzsch1974nnMinkowskiUnifiedInteractionsofLeptonsandHadrons} , but denote their Lie group in the capital case such as Spin(10).}} 
This perspective may be schematically summarized as  \Fig{fig:two-views} (a).

However, gauge symmetry is not a physical symmetry (unlike the global symmetry) but only a gauge redundancy to describe interactions between matters;
thus the gauge group is not physical nor universal. 
Furthermore, it is widely known that there are many different dual descriptions of the same physical theories via different gauge groups.
We may quote this perspective as ``{\bf Many Dualities}.'' 

This raises the conflict between the above two perspectives: How could we ask for the 
governing gauge group for the {\bf Unity of Gauge Forces}, if gauge groups are not universal, and if there are {\bf Many Dualities} of possible different gauge theory realizations of the same unification?
Partly motivated to provide a resolution of this conceptual conflict, 
our prior work \cite{Wang2106.16248} initiates an alternative viewpoint:
we propose that the SM vacuum may be a low energy quantum vacuum arising from the quantum competition of various neighbor higher-energy GUT or other unified theories' vacua in an immense quantum phase diagram,
schematically summarized as  \Fig{fig:two-views} (b).
Here we highlight and summarize some of the viewpoints of Ref.~\cite{Wang2106.16248}: 
\begin{enumerate}[leftmargin=.mm]
\item When we treat the internal symmetry as a global symmetry (or in the weakly gauged or ungauged limit), it is physically sensible to ask ``what is the 
{\bf Unity of the Governing Internal Symmetry Group} ${{G}_{\text{internal}} }$?''
In this global symmetry limit, the immense quantum phase diagram in fact not only contains many different GUTs in the same Hilbert space with same 't Hooft anomalies \cite{tHooft1979ratanomaly}
of their internal global symmetries,
but also give rise to the SM near the quantum criticality\footnote{Let us clarify the terminology on
{\bf criticalities} vs {\bf phase transitions}. \\
$\bullet$ The {\bf criticality} means the system with gapless excitations (gapless thus critical, sometimes conformal) and with an infinite correlation length, 
it can be either (i) a {\bf continuous phase transition} as an unstable critical point/line/etc. as an unstable renormalization group (RG) fixed point which has at least one relevant perturbation in the phase diagram, 
or (ii) a {\bf critical phase} as a stable critical region controlled by a stable RG fixed point which does not have any relevant perturbation
in the phase diagram.\\
$\bullet$ The {\bf phase transition} \cite{subirsachdev2011book} means the phase interface between two (or more) bulk phases in the phase diagram.
The phase transition can be a {\bf continuous phase transition} (second order or higher order, with gapless modes)
or a {\bf discontinuous phase transition} (first-order, without gapless modes, and with a finite correlation length).\\
In the phase diagram, the spacetime dimensionality of the phase interface is the same as that of the bulk phase.
This is in contrast with the one-lower spacetime dimensional physical interface or physical boundary of the bulk phase.
\label{ft:criticality-phase-transition}
} 
between different GUTs.
The quantum field theories (QFTs), sharing the same 't Hooft anomalies especially with the same global symmetries,
are believed to live in the same phase diagram with the ``same'' Hilbert space, possibly by adding new degrees of freedom at the short-distance or the higher energy. 
These QFTs are deformable to each other via symmetry-preserving interactions --- known as the \emph{deformation class of QFTs}, particularly advocated by 
Seiberg \cite{NSeiberg-Strings-2019-talk}. The \emph{deformation class of quantum gravity theory} is also proposed, by McNamara-Vafa \cite{McNamara2019rupVafa1909.10355}.
The \emph{deformation class of the standard model} is studied in \cite{WangWanYou2112.14765, WangWanYou2204.08393}.

\item Since the SM arises near the quantum criticality between different GUTs, 
it makes sense to study the emergent (gauged or global) symmetries and dualities of QFTs at this quantum criticality.
We can further gauge the internal symmetry to be a gauge theory, thus we can study possible {\bf Many Dualities} of these gauge theories.

\item The above two different viewpoints highlight the validity of {\bf Unity of Internal Symmetry} and {\bf Many Dualities} respectively, but now in the same quantum phase diagram and same framework (\Fig{fig:two-views} (b)).
Moreover, various GUTs may encounter \emph{stable gapless quantum critical regions} (the shaded gray area in \Fig{fig:two-views} (b)) to enter other neighbor GUTs, if we apply the idea that the
quantum criticality is protected by the 't Hooft anomalies of some spacetime-internal symmetries:\footnote{The notation $G_1 \ltimes_{N_{\text{shared}}} G_2 \equiv \frac{G_1 \ltimes G_2}{{N_{\text{shared}}}}$ means modding out their common normal subgroup 
${N_{\text{shared}}}$.
The $\ltimes$ is a semidirect product (as a generalization of direct product) to specify a particular group extension.}
\bea
\bar{G} \equiv {{G_{\text{spacetime} }} \ltimes_{{N_{\text{shared}}}}  {{G}_{\text{internal}} } }
\equiv  ({\frac{{G_{\text{spacetime} }} \ltimes  {{G}_{\text{internal}} } }{{N_{\text{shared}}}}}).
\eea
The spacetime ${G_{\text{spacetime} }}$ and internal ${{G}_{\text{internal}}}$ symmetries may share a common normal subgroup ${N_{\text{shared}}}$.
If there is a 't Hooft anomaly in $\bar{G}$ (here 4d, denoted as a one-higher dimensional 
invertible topological quantum field theory [iTQFT] defined on some $\bar{G}$-structure manifold in 5d with a partition function ${\bZ}$),
this iTQFT may be written schematically as
\bea \label{eq:schematic-Z}
{\bZ}[\cB_{\bar{G}}] =
{\bZ}[\cB_{\bar{G}_{\GUT_1}} \smile  \cB_{\bar{G}_{\GUT_2}}  \smile \dots ]  ,
\eea
where $\cB_{\bar{G}}$ is the background field coupling to 
the spacetime-internal symmetry ${\bar{G}}$.\footnote{The background field $\cB_{\bar{G}}$ may be organized as their dependence on various GUT subgroup's 
 background fields, say $\cB_{\bar{G}_{\GUT_1}}, \cB_{\bar{G}_{\GUT_2}}$, etc.
Here the cup product $\smile$ is for cohomology classes (analogous to the wedge product $\wedge$ for differential forms).
In general, the internal symmetry group of various GUT models have subgroups overlapped, so their background fields $\cB_{\bar{G}}$ also overlapped
--- the expression \eq{eq:schematic-Z} is only schematic.}
When we have the symmetry breaking from 
${\bar{G}}$ to ${\bar{G}_{\GUT_j}}$, etc.,
we are only left with the valid background field $\cB_{\bar{G}_{\GUT_j}}$
while other background fields must be turned off --- thus the 't Hooft anomaly may be canceled to zero 
by this symmetry breaking.
In contrast, if we preserve the full ${\bar{G}}$ (happening especially at the quantum critical region in between GUT phases),
and if we have other possibilities to cancel the  't Hooft anomaly other than by symmetry breaking,
then we must have \emph{no symmetric trivial gapped phase} between neighbor GUT phases,
which likely resulting in \emph{gapless quantum criticalities} in many cases. 
In general, these phases are called  \emph{nontrivial phases}: 
(1) symmetry-preserving gapless, 
(2) symmetry-preserving gapped topologically order with a low energy TQFT,
(3) symmetry-breaking gapless or gapped,
or (4) their mixed combinations. 
In short, some nontrivial state of matter must be {at the phase transition or the quantum criticality} to match the  't Hooft anomaly.

{In fact, the {\bf deconfined quantum criticality} (DQC) in 2+1d in condensed matter physics \cite{SenthildQCP0311326, Wang2017Deconfined} is the 
incarnation and reminiscence of this above idea of 't Hooft anomaly protected quantum criticality (See  \Refe{Wang2017Deconfined} and Appendix C of \Refe{Wang2106.16248} for 
a contemporary QFT overview of DQC). Recently the DQC phenomena in 3+1d are explored in
\cite{BiSenthil1808.07465, Wan2018djlW2.1812.11955, BiLakeSenthil1910.12856, WangYouZheng1910.14664},
where Ref.~\cite{Wang2106.16248} provides the first 3+1d DQC analog in the SM and beyond-the Standard Model (BSM) physics.
}

\item In particular, when the internal symmetry is treated as a global symmetry,
Ref.~\cite{Wang2106.16248} %develops this Remark 3's idea  and 
shows that between two GUT models, 
Georgi-Glashow (GG) $su(5)$ \cite{Georgi1974syUnityofAllElementaryParticleForces} 
and Pati-Salam (PS) $su(4) \times su(2) \times su(2)$ \cite{Pati1974yyPatiSalamLeptonNumberastheFourthColor},
we can utilize a mod 2 class of nonperturbative global mixed gauge-gravity 't Hooft anomaly in 4d captured by
the 5d iTQFT written in Stiefel-Whitney (SW) characteristic classes:\footnote{The 
$w_j$ is the $j$-th Stiefel-Whitney (SW) characteristic class.
The $w_j(TM)$ is the SW class of spacetime tangent bundle $TM$ of manifold $M$. 
The $w_j(V_{G})$ is the SW class of the principal $G$ bundle.
{The 5-manifold that detects $w_2(TM)w_3(TM)$
is a Dold manifold $\CP^2 \rtimes S^1$ or
a Wu manifold $\SU(3)/\SO(3)$ which yields a path integral ${\bf Z}(M^5)=-1$ \cite{WangWenWitten2018qoy1810.00844, WanWang2018bns1812.11967, Wan2019oyr1904.00994}}.
\label{footnote:w2w3anomaly}
}
\bea \label{eq:w2w3-2}
{\bf Z}(M^5)=
{\exp(\ii \pi \int_{M^5} w_2w_3)}=
{\exp(\ii \pi \int_{M^5} w_2(TM)w_3(TM))=}\exp(\ii \pi \int_{M^5} w_2(V_{\SO(10)})w_3(V_{\SO(10)}) ).
\;\;
\eea
This 4d 't Hooft anomaly requires the spacetime-internal global symmetry 
$(\Spin \times_{\Z_2^F}\Spin(10))$ on a 4-manifold $M^4$, 
captured by this 5d bulk invertible TQFT \cite{WangWen2018cai1809.11171, WangWenWitten2018qoy1810.00844} living on a 5-manifold $M^5$ with the anomaly-inflow bulk-boundary correspondence $\prt M^5= M^4$.

This mixed gauge-gravitational anomaly is tightly related to {the new SU(2) anomaly} \cite{WangWenWitten2018qoy1810.00844} due to the bundle constraint
$w_2w_3(TM) = w_2w_3(V_{G})$ with $G$ can be substituted by $\SO(3) \subset \SO(10)$ related to the embedding $\SU(2) = \Spin(3) \subset \Spin(10)$.
This mod 2 class $w_2w_3$ global anomaly has been checked to be absent in the conventional $so(10)$ GUT by \Refe{WangWen2018cai1809.11171, WangWenWitten2018qoy1810.00844};
thus the conventional $so(10)$ GUT is free from 4d anomaly classified by the $d=5$-th cobordism group 
defined in Freed-Hopkins \cite{Freed2016},
\bea
\Omega^{d}_{\bar{G} } \equiv \TP_d(\bar{G} ),
\eea
here with $d=5$, $\bar{G}=(\Spin \times_{\Z_2^F}\Spin(10))$, and 
$\Omega^{5}_{(\Spin \times_{\Z_2^F}\Spin(10))}=\Z_2$ generated by ${\exp(\ii \pi \int_{M^5} w_2w_3)}$.
However, Ref.~\cite{Wang2106.16248} modifies the $so(10)$ GUT by appending a new 4d Wess-Zumino-Witten (WZW) term
with this $w_2w_3$ global anomaly in order to realize the SM vacuum as the quantum criticality phenomenon between the neighbor GG SU(5) GUT and PS vacua:\\
$\bullet$ On either GG SU(5) or PS sides of the quantum phases, the $w_2w_3$ anomaly is matched by breaking the internal Spin(10) symmetry down to their GUT subgroups.\\
$\bullet$ But at the critical (gapless) region between GG and PS quantum phases, the full Spin(10) symmetry can be preserved, while the $w_2w_3$ anomaly is matched by the BSM sector
from the new 4d WZW topological term (constructed out of the GUT-Higgs fields or their fractionalized partons) living on a 4d boundary of a 5d bulk. \\
$\bullet$ {Since the mod 2 class $w_2w_3$ anomaly is matched by the sector of GUT-Higgs fields (or their fractionalized partons) and their 4d WZW term alone,
we just need to ensure the anomaly index from GUT-Higgs WZW sector contributes 1 mod 2. If each generation of 16 SM Weyl fermions associates with its own GUT-Higgs field,
then the generation number n times of 16 SM Weyl fermions with n GUT-Higgs field requires a constraint 
${\rm{n}} = 1 \mod 2$ to match the $w_2 w_3$ anomaly, where ${\rm{n}} = 3$ generation indeed works. 
However, in general, we can just introduce a single or any odd number of GUT-Higgs field sectors (independent from the ${\rm{n}}=3$ of SM)
to match the $1 \mod 2$ class of $w_2 w_3$ anomaly.}

$\bullet$ The dynamics of this modified $so(10)$ GUT with WZW term can be fairly complicated, giving rise to many possible \emph{gapless} phases or \emph{gapped} TQFT phases at low energy, enumerated in \cite{Wang2106.16248}.

\end{enumerate}

In this present work, we continue developing from Ref.~\cite{Wang2106.16248} to include
another GUT model: the flipped $su(5)$ model (originally proposed by Barr \cite{Barr1982flippedSU5} and others \cite{Derendinger1984flippedSU5})
and the left-right (LR) model \cite{SenjanovicMohapatra1975}
into our quantum landscape or quantum phase diagram, exploring further the neighbors of the critical region near SM.
The GG model can have either choice of a gauge group of SU(5) or U(5), but Barr's flipped model must require a U(5).
We should emphasize that the U(5) Lie group of GG or Barr's model
requires a certain refined Lie group that we name $\U(5)_{\hat q=2}$, see \Sec{sec:Refined-U5-group} for details.
In many cases, we do require the additional $u(1)$ gauge sector in addition to the $su(5)$ gauge sector of GG or Barr's model,
we thus call the corresponding models as the GG $u(5)$ model and Barr's flipped $u(5)$ model.

The SM and GG $su(5)$ models can have either choice of 15 or 16 Weyl fermions for each generation. 
In contrast, in order to be consistent with the SM data constraint,
the PS, the GG $u(5)$, the flipped $u(5)$, and the modified $so(10)$ have 16 Weyl fermions per generation.
In this article, we mainly focus on the scenarios all with 16n Weyl fermions with n the number of generations.\footnote{
{Readers can find \Refe{Freed0607134, GarciaEtxebarriaMontero2018ajm1808.00009, 
2019arXiv191011277D, WW2019fxh1910.14668} and \cite{WangWen2018cai1809.11171} 
for the systematical studies on the nonperturbative global anomalies of various SM and GUT via generalized cohomology or cobordism theories.}\newline
In contrast to the scenarios of SM or GUT with 16n Weyl fermions, 
Ref.~\cite{JW2006.16996, JW2008.06499, JW2012.15860} considers the SM or GUT 
with 15n Weyl fermions and with a discrete variant of baryon minus lepton number ${\bf B}-{\bf L}$ symmetry \cite{Wilczek1979hcZee} preserved. 
Ref.~\cite{JW2006.16996, JW2008.06499, JW2012.15860} then suggests the missing 16th Weyl fermions can be substituted 
by additional symmetry-preserving 4d or 5d gapped topological quantum field theories (TQFTs), 
or by the symmetry-preserving 4d gapless interacting conformal field theories (CFTs), or 
other symmetry-breaking sectors (e.g., the right-handed neutrinos), to saturate a certain $\Z_{16}$ global anomaly.
We will comment about the topological phase transitions between the 15n to 16n Weyl fermions at the very end in \Sec{sec:conclude}.
}

\subsection{Outline: The plan of the article}

In \Sec{sec:Refined-U5-group}, we clarify the U(5) Lie group structure of the GG or flipped models.
They should be both refined as $\U(5)_{\hat q}$ gauge theories, with ${\hat q}=2,3$, which is non-isomorphic to ${\hat q}=1,4$.
For applications, there we point out group theoretical facts like $\Spin(10) \supset \U(5)_{\hat q=2,3}$ but $\Spin(10) \not\supset \U(5)_{\hat q=1,4}$.

\noindent
In \Sec{sec:GUT}, we clarify 
{various GUT models as different vacua or different phases} of QFTs, and present their
{representations of SMs and {five} GUT-like models in a unified \Table{table:fermionAll}}.
Additional details of {quantum numbers and representations of SMs and GUTs} are provided in Appendix \ref{sec:SM-GUT-table-app}.

\noindent
In \Sec{sec:Quantum-Landscape},
we organize various SM or GUT models 
in a {quantum landscape or in a quantum phase diagram}.
The parameter space of the quantum phase diagram can be specified for example via 
the GUT-Higgs condensation. Thus the parameter space is also a moduli space.
In \Sec{sec:Embedding-Web},
we provide
the {embedding web by their internal symmetry groups}.
In \Sec{sec:QuantumPhaseDiagram},
we derive a 
{quantum phase diagram based on the mother effective field theory}
of a modified $so(10)$ GUT of \cite{Wang2106.16248}.
As a toy model, we clarify the quantum phase structures
when {the internal symmetries are treated as global symmetries}
in \Sec{sec:QuantumPhaseDiagram-Internal-global}.
Then we clarify the quantum phase structures
when {the internal symmetries are dynamically gauged}
in \Sec{sec:QuantumPhaseDiagram-Internal-gauged}.

\noindent
In \Sec{sec:HigherSymmetries},
as the ordinary internal symmetry is dynamically gauged,
the outcome gauge theory can have the generalized global symmetries \cite{Gaiotto2014kfa1412.5148}.
We systematically explore these generalized global symmetries (the higher symmetries) of SM and GUT. 

\noindent
In \Sec{sec:CategoricalSymmetries},
we study the potential non-invertible global symmetries
\cite{Verlinde1988sn, Moore1988qv, Frohlich0607247, DavydovKong2010rm1004.4725, BhardwajTachikawa1704.02330, ChangLinShaoWangYin1802.04445, NguyenTanizakiUnsal2101.02227, NguyenTanizakiUnsal2104.01824, MonteroRudelius2104.07036, ChoiCordovaHsinLamShao2111.01139, KaidiOhmoriZheng2111.01141}
(also known as {categorical symmetries} \cite{ThorngrenWang1912.02817, GaiottoKulp2008.05960, KongLanWenZhangZheng2005.14178, KomargodskiOhmoriRoumpedakisSeifnashri2008.07567, ThorngrenWang2106.12577}) of SM and GUT. 
We show that part of the gauge structure of the GG $\U(5)$ and the flipped $\U(5)$ gauge theories, 
with their $\Z_2^{\rm flip}$ symmetry gauged,
contains a gauge sector $\big[ (\U(1)_{X_1} \times_{\Z_{4,X}} \U(1)_{X_2}) \rtimes \Z_2^{\rm flip} \big]$.
There is a non-invertible global symmetry exhibited by the topological 2-surface operators
as the gauge-invariant symmetry generators of their magnetic 1-symmetries. But we show that 
this categorical symmetry is retracted thus disappears, when we embed the theory in the modified $so(10)$ GUT of Spin(10).

\noindent
In \Sec{sec:conclude}, we conclude and comment on future research directions.

\noindent
In  Appendix \ref{app:Embedding}, we provide various matrix representations of Lie algebras and Lie groups of 
SU(5), $\U(5)_{\hat q}$, SO(10), and Spin(10). Then we describe how they could embed each other properly.
In  Appendix \ref{app:Flipping},
we show the flipping isomorphism between two $\U(5)_{\hat q=2}$ (the GG's and the flipped model's)
while both $\U(5)_{\hat q=2}$ can be embedded inside the Spin(10).
Then we show that the intersection of two $\U(5)_{\hat q=2}$ contains the SM Lie groups,
while the minimal Lie group union of two $\U(5)_{\hat q=2}$ is exactly the Spin(10).

%\newpage
\section{Refined $\U(5)_{\hat q}$ gauge theory}
%\subsection{On the refined $\U(5)_{\hat q}$ gauge theory}
\label{sec:Refined-U5-group}
%\label{sec:refined-UN-gauge}

Here we point out there are in fact different non-isomorphic versions of U(5) Lie groups (and their corresponding gauge theories) that we should refine and redefine them
as several $\U(5)_{\hat q}$ with ${\hat q} \in \Z$:
\be \label{eq:U5q}
\U(5)_{\hat q} \equiv \frac{\SU(5) \times_{\hat q} \U(1) }{\Z_5} \equiv {\SU(5) \times_{\Z_5, \hat q} \U(1) } \equiv
\{ (g, \e^{\ii \theta}) \in \SU(5) \times \U(1) \big\vert  ( \e^{\ii \frac{2 \pi n}{5}} \mathbb{I}, 1) \sim ( \mathbb{I}, \e^{\ii \frac{2 \pi n {\hat q}}{5}} ), n\in \Z_5 \}.
\ee
We use two data $(g, \e^{\ii \theta})$ to label the $\SU(5) \times \U(1)$ group elements respectively,
while we identify $( \e^{\ii \frac{2 \pi n}{5}} \mathbb{I}, 1) \sim ( \mathbb{I}, \e^{\ii \frac{2 \pi n {\hat q}}{5}} )$ for $n\in \Z_5$,
with a rank-5 identity matrix $\mathbb{I}$.
Different identifications of the generator of the center $Z(\SU(5))=\Z_5$ with the $\U(1)$ charge ${\hat q}$ in principle give rise to different Lie groups.
All different $\U(5)_{\hat q}$ obey the group extension as
 the short exact sequence
$1 {\to} {\SU(5)} {\to} {\U(5)_{\hat q}} \overset{\det}{\to} {\U(1)'} {\to} 1$ where
${\U(1)'} \equiv \frac{\U(1)}{\Z_{5 {\hat q} }} $
is related to modding out ${\Z_{5 {\hat q} }}$ of the $\U(1)$ defined in ${\U(5)_{\hat q}}$ in \eq{eq:U5q},  
while different ${\hat q}$ identifications specify different $\U(1)$ actions on the $\SU(5)$.
But there are in fact the following group isomorphisms
\bea \label{eq:U5-isomorphism}
\U(5)_{\hat q}  \cong \U(5)_{-\hat q}  \cong \U(5)_{5 m \pm \hat q} 
\eea
for any $m\in \Z$.\footnote{We can prove the group isomorphism by the following. For 
$\U(5)_{\hat q}$ and $\U(5)_{-\hat q}$ defined in \eq{eq:U5q}, we can map 
$h_j = (g_j, \e^{\ii \theta_j}) \in \U(5)_{\hat q}$ and define the group homomorphism map
$f(h_j)=(g_j, \e^{-\ii \theta_j})$. Then we check that the group homomorphism
$f(h_1) \cdot f(h_2) = f(h_1 \cdot h_2)$ is true:
On the left-hand side, $f(h_1) \cdot f(h_2)=(g_1 g_2, \e^{-\ii (\theta_1+\theta_2)})$.
On the right-hand side, $h_1 \cdot h_2=(g_1 g_2, \e^{\ii (\theta_1+\theta_2)})$ thus
$f(h_1 \cdot h_2)=(g_1 g_2, \e^{-\ii (\theta_1+\theta_2)})$.
In addition, it is injective (one-to-one) and surjective (onto) thus a bijective 
group homomorphism. It would only be bijective if the map 
$f(h_j)=(g_j, \e^{\pm \ii \theta_j})$, either a trivial identity map or the outer automorphism on the U(1) part;
thus $\U(5)_{\hat q}  \cong \U(5)_{-\hat q}$.
The only exception of other isomorphism is when we shift $\hat q$ to ${5 m + \hat q}$ (which does not modify the identification in \eq{eq:U5q}), 
thus we prove 
$\U(5)_{\hat q}  \cong \U(5)_{-\hat q}  \cong \U(5)_{5 m \pm \hat q}$.
}
Thus among general ${\hat q} \in \Z$, we have \emph{three distinct} non-isomorphic types of $\U(5)_{\hat q}$ group for any $m\in \Z$:
\bea
(1) &&\U(5)_{1} \cong \U(5)_{4} \cong \U(5)_{5 m + 1} \cong \U(5)_{5 m - 1}.\cr 
(2) &&\U(5)_{2} \cong  \U(5)_{3} \cong \U(5)_{5 m+ 2} \cong \U(5)_{5 m - 2}.\cr
(3) &&\U(5)_{0} \cong \U(5)_{5 m} \cong {\SU(5) \times \U(1) }.  
\eea
We emphasize the distinctions of these three different non-isomorphic $\U(5)_{\hat q}$ Lie groups (and their gauge theories)
somehow seem not yet been carefully examined in the previous high-energy particle physics literature.
Here we make attempts to address their Lie group differences in the context of gauge theories and GUTs. 
Several comments are in order:\footnote{We provide the detailed mathematical proofs of many statement listed here in a separate work (jointly with Zheyan Wan et al) in \cite{ZWtoappear}.}
\begin{enumerate} 
\item $\U(5)_{\hat q}$ as a $k$-sheeted covering space of $\U(5)_{k \hat q}$: If we compare the definition of
$\U(5)_{\hat q}$ and $\U(5)_{k \hat q}$, we find that from \eq{eq:U5q},
the $\U(5)_{\hat q}$ identifies
$( \e^{\ii \frac{2 \pi }{5}} \mathbb{I}, 1) \sim ( \mathbb{I}, \e^{\ii \frac{2 \pi {\hat q}}{5}} )$
while
the
$\U(5)_{k \hat q}$ identifies
$( \e^{\ii \frac{2 \pi }{5}} \mathbb{I}, 1) \sim ( \mathbb{I}, \e^{\ii \frac{2 \pi k {\hat q}}{5}} )$.
If the $\U(5)_{\hat q =1}$ has the periodicity of U(1) as $\theta \in [0, 2 \pi)$,
then the $\U(5)_{\hat q =k}$ has the periodicity of U(1) as $\theta \in [0, \frac{2 \pi}{k})$.
Similarly, 
if the $\U(5)_{\hat q }$ has the periodicity of U(1) as $\theta \in [0, 2 \pi)$,
then the $\U(5)_{k \hat q }$ has the periodicity of U(1) as $\theta \in [0, \frac{2 \pi}{k})$. 
So importantly,\\
$\bullet$ the $\U(5)_{\hat q}$ is a $k$-sheeted covering space of $\U(5)_{k \hat q}$.\\
$\bullet$ the $\U(5)_{\hat q=1}$ is a double covering space of $\U(5)_{\hat q=2}$.\\
$\bullet$ the $\U(5)_{\hat q=2}$ is a double covering space of $\U(5)_{\hat q=4}$.\\ 
$\bullet$ the $\U(5)_{\hat q=1}$ is a quadruple covering space of $\U(5)_{\hat q=4}$, \\
but which goes back to itself because of the isomorphism
$\U(5)_{\hat q= 1} \cong \U(5)_{\hat q= 4}$. \\ 
$\bullet$ the $\U(5)_{\hat q=2}$ is a quadruple covering space of $\U(5)_{\hat q=3}$, \\
but which goes back to itself because of the isomorphism
$\U(5)_{\hat q= 2} \cong \U(5)_{\hat q= 3}$. \\ 
$\bullet$ the $\U(5)_{\hat q}$ is a quadruple covering space of $\U(5)_{4 \hat q}  \cong \U(5)_{4 \hat q \mod 5}$, \\
but which goes back to itself because of the isomorphism
$\U(5)_{\hat q} \cong \U(5)_{4 \hat q}  \cong \U(5)_{4 \hat q \mod 5}$. \\ 
\item It can be shown that
$$
\U(5)_{\hat q=1,4} \subset \SO(10) \text{ but }
\U(5)_{\hat q=1,4} \not \subset  \Spin(10)$$
because the homotopy group maps between $\pi_1(\U(5)_{\hat q=1,4})=\Z$ to $\pi_1(\SO(10))=\Z_2$
cannot be lifted to the $\SO(10)$'s double-cover since $\Spin(10)$ has $\pi_1(\Spin(10))=0$, which violates the lifting criterion (See Proposition 1.33 of Hatcher \cite{hatcher2002algebraic}).

It can be shown that the $\U(5)_{\hat q=2,3}$ as a double cover version of $\U(5)_{\hat q=1,4}$ satisfies:
$$
\U(5)_{\hat q=2,3} \not \subset \SO(10)
\text{ but }
\U(5)_{\hat q=2,3} \subset  \Spin(10).
$$
Since $Z(\Spin(10))=\Z_4$, 
$Z(\SO(10))=\Z_2$, and $Z( \frac{\SO(10)}{\Z_2} )=0$, 
while $\U(5)_{\hat q=2,3}$ is a quadruple cover of itself;
overall we have:\footnote{Here ``$G_1 \hookrightarrow G_1$'' implies the inclusion thus also implies the group embedding ``$G_1 \subset G_2$.''
The vertical arrow ``$\downarrow$'' implies the upper group $\tilde G$ is a double cover of the lower group $G$ such that
we have a nontrivial extension
$1 \to \Z_2 \to \tilde G \to G \to 1$.}
\bea \label{eq:U5-embed}
\xymatrix{
\U(5)_{\hat q=2,3} \ar@{^{(}->}[r] \ar@{^{}->}[d] & 
\Spin(10) \ar@{^{}->}[d] \\
\displaystyle
\U(5)_{\hat q=1,4}   \ar@{^{(}->}[r]  \ar@{^{}->}[d] &\SO(10)  \ar@{^{}->}[d]\\
\displaystyle
\U(5)_{\hat q=2,3}   \ar@{^{(}->}[r] & \frac{\SO(10)}{\Z_2} 
}.
\eea
\item Follow Atiyah-Bott-Shapiro \cite{AtiyahBottShapiro1964}, it is shown that the group homomorphism and the embedding 
$\SU(5) \hookrightarrow \SO(10)$ can be lifted to $\SU(5) \hookrightarrow \Spin(10)$.
Similarly, 
the group homomorphism and the embedding 
$\U(5)_{\hat q=1} \hookrightarrow \SO(10) \times \frac{\U(1)}{\Z_2} = \frac{\Spin(10)\times \U(1)}{\Z_2 \times \Z_2}$
can be lifted to
$\U(5)_{\hat q=1} \hookrightarrow \Spin^c(10)\equiv \frac{\Spin(10)\times \U(1)}{\Z_2}$.
We can double cover or half-cover of these results to obtain: 
\bea \label{eq:U5-embed-2}
\xymatrix{
\U(5)_{\hat q=2,3} \ar@{^{(}->}[r] \ar@{^{}->}[d] & 
\Spin(10)\times \U(1) \ar@{^{}->}[d] \\
\displaystyle
\U(5)_{\hat q=1,4}   \ar@{^{(}->}[r]  \ar@{^{}->}[d] &\Spin^c(10) \equiv\frac{\Spin(10)\times \U(1)}{\Z_2}  \ar@{^{}->}[d]\\
\displaystyle
\U(5)_{\hat q=2,3}   \ar@{^{(}->}[r] &\frac{\Spin(10)\times \U(1)}{\Z_4} 
},
\eea
which says not only the embedding
$\U(5)_{\hat q=1,4} \subset \Spin^c(10)$,
but also the embedding $\U(5)_{\hat q=2,3} \subset \Spin(10)\times \U(1)$
and $\U(5)_{\hat q=2,3} \subset \frac{\Spin(10)\times \U(1)}{\Z_4}$.
\item In a short summary,
for our application on the SM and GUT physics, we shall particularly focus on
these two results:
\bea \label{eq:U5-embed-3}
\U(5)_{\hat q=2,3} \subset \Spin(10), \quad \quad \U(5)_{\hat q=2,3} \subset \frac{\Spin(10)\times \U(1)}{\Z_4}.
\eea
Here
$\frac{\Spin(10)\times \U(1)}{\Z_4}$ can be interpreted as 
$\frac{\Spin(10)\times \U(1)_{X}}{\Z_{4,X}}$ with 
$X \equiv 5({ \mathbf{B}-  \mathbf{L}})-4Y$, including the baryon minus lepton number ${ \mathbf{B}-  \mathbf{L}}$ and the electroweak hypercharge $Y$,
is a good global symmetry respected by SM and the $su(5)$ GUT \cite{Wilczek1979hcZee}. 

We provide a verification on \eq{eq:U5-embed-3} 
via exponential maps of the Lie algebras into these Lie groups embedding in Appendix \ref{app:Embedding}.

\item In order to study the $\U(5)_{\hat q}$ gauge theory, 
we should understand the allowed Wilson line operators and their 
endpoint particle charge representations (if the 1d line can be broken by the 
particle at open ends).
In particular, when the matter field $\psi$ is in the fundamental rep ${\bf 5}$ of SU(5), we can ask which 
U(1) charge representation $Q$ of $\psi$ is allowed.
Because of the identification  $( \e^{\ii \frac{2 \pi n}{5}} \mathbb{I}, 1) \sim ( \mathbb{I}, \e^{\ii \frac{2 \pi n {\hat q}}{5}} )$ (for $n\in \Z_5$)
must act on the $\psi$ in 
$({\bf 5},Q)$  in the same way, regardless of
whether we consider\\
$\bullet$ $( \e^{\ii \frac{2 \pi n }{5}} \mathbb{I}, 1)$ 
on $\psi$  of $({\bf 5},Q)$ 
which sends $\psi$
to $\e^{\ii \frac{2 \pi n  }{5}} \psi$.\\
%%%
 or consider \\
$\bullet$  $( \mathbb{I}, \e^{\ii \frac{2 \pi n {\hat q}}{5}} )$ 
on $\psi$  of 
$({\bf 5},Q)$  
which gives 
$\e^{\ii \frac{2 \pi n  {\hat q} Q}{5}}  \psi$.\\
%%%
The group element identification also means that the
$\e^{\ii \frac{2 \pi n  }{5}}=\e^{\ii \frac{2 \pi n  {\hat q}  Q}{5}}$,
which is true if $\hat q Q = 1 \mod 5$.
Thus we derive the relation between the Lie group $\U(5)_{\hat q}$ and its corresponding matter representation:
\bea\label{eq:U5q-matter-rep}
\begin{tabular}{lcc c}
\hline
& ${\hat q}$ & $Q$ & matter rep $({\bf 5},Q)$, $(\overline{\bf 5},-Q)$\\
\hline
$\U(5)_{{\hat q} = 1}$ & 1 & 1 & $({\bf 5},1)$, $(\overline{\bf 5},-1)$\\
\hline
$\U(5)_{{\hat q} = 2}$ & 2 & 3 & $({\bf 5},3)$, $(\overline{\bf 5},-3)$\\
\hline
$\U(5)_{{\hat q} = 3}$ & 3 & 2  & $({\bf 5},2)$, $(\overline{\bf 5},-2)$\\
\hline
$\U(5)_{{\hat q} = 4}$ & 4 & 4 & $({\bf 5},4)$, $(\overline{\bf 5},-4)$\\
\hline
\end{tabular}.
\eea
\item If we want to choose the appropriate
$\U(5)_{{\hat q}}$ Lie group for the GG $su(5)$ or the flipped $su(5)$ models,
we should consider the group contains
$(\overline{\bf{5}}, -3)$ of $su(5) \times u(1)$.
This \eq{eq:U5q-matter-rep} means that 
the $\U(5)_{{\hat q} = 2}$ is the correct choice.
This matter representation $(\overline{\bf 5},-3)$ is naturally included in
$\U(5)_{{\hat q} = 2}$.
\item We can generalize the above discussions to $\U(N)_{{\hat q}}$ cases to find non-isomorphisms for some of $(N,{\hat q})$. 
\end{enumerate} 

\newpage
\section{Various Grand Unification (GUT) Models as Vacuum Phases}
\label{sec:GUT}

\begin{figure}[h!] %[htbp]
%\begin{center}
  \centering
  \hspace{-.9cm}
 \includegraphics[width=4.in]{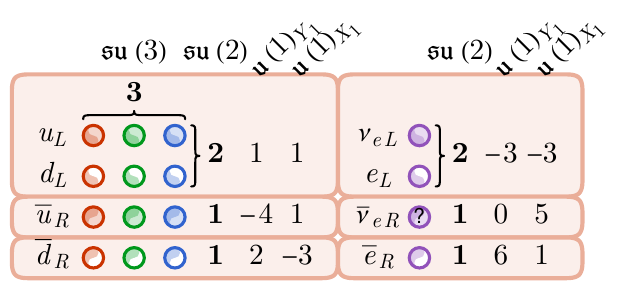}
 \caption{Standard Model with 16n Weyl fermions and their ${{su(3)}_c \times {su(2)}_{\rm L} \times u(1)_{\tilde{Y}}}$ representation (rep):
 $\bar{d}_R \oplus {l}_L  \oplus q_L  \oplus \bar{u}_R \oplus   \bar{e}_R \oplus  \bar{\nu}_R
=
(\overline{\bf 3},{\bf 1})_{2,L} \oplus ({\bf 1},{\bf 2})_{-3,L}  
\oplus
 ({\bf 3},{\bf 2})_{1,L} \oplus (\overline{\bf 3},{\bf 1})_{-4,L} \oplus ({\bf 1},{\bf 1})_{6,L} \oplus {({\bf 1},{\bf 1})_{0,L}}$.}
  \label{fig:SM}
%\end{center}
\end{figure}

In this article,
we require the SM gauge group $G_{\SM_q} \equiv \frac{{\SU(3)}_c \times {\SU(2)}_{\rm L} \times \U(1)_{\tilde{Y}}}{\Z_q}$ with $q=6$,
where the 16 Weyl fermions are in the following representation (see \Fig{fig:SM}):
\bea \label{eq:SM-rep}
\bar{d}_R \oplus {l}_L  \oplus q_L  \oplus \bar{u}_R \oplus   \bar{e}_R \oplus  \bar{\nu}_R
=
(\overline{\bf 3},{\bf 1})_{2,L} \oplus ({\bf 1},{\bf 2})_{-3,L}  
\oplus
 ({\bf 3},{\bf 2})_{1,L} \oplus (\overline{\bf 3},{\bf 1})_{-4,L} \oplus ({\bf 1},{\bf 1})_{6,L} \oplus {({\bf 1},{\bf 1})_{0,L}},
\eea 
written all in the left-handed ($L$) Weyl basis.
Here we use the $\U(1)_{\tilde{Y}}$ hypercharge instead of HEP phenomenology $\U(1)_{{Y}}$ hypercharge which is 1/6 of $\U(1)_{\tilde{Y}}$'s.\footnote{Namely,  
$q_{\U(1)_{{Y}}}=\frac{1}{6} q_{\U(1)_{\tilde{Y}}}$. If we use the hypercharge $\U(1)_{{Y}}$, 
then we have instead: 
$(\overline{\bf 3},{\bf 1})_{\frac{1}{3},L} \oplus ({\bf 1},{\bf 2})_{-\frac{1}{2},L}  
\oplus
 ({\bf 3},{\bf 2})_{\frac{1}{6},L} \oplus (\overline{\bf 3},{\bf 1})_{-\frac{2}{3},L} \oplus ({\bf 1},{\bf 1})_{1,L} \oplus {({\bf 1},{\bf 1})_{0,L}}$.} 
 The $u$ and $d$ are the up and down quarks, 
 the $\nu$ and $e$ are the neutrino and electron.\footnote{This matter content is for the first generation of quarks and leptons.
 We can replace these quarks to the charm $c$ and strange $s$ for the second generation, or to the top $t$ and bottom $b$ for the third generation.
 We can  replace these leptons to the muon $\mu$ and tauon $\tau$ for the second and third generations.}
 The $q_L$ and ${l}_L$ are both of the SU(2)$_{\rm L}$ doublets; the $q_L$ contains the $(u_L,d_L)$
 while the ${l}_L$ contains the $(\mu_L, e_L)$.
 We use the $L$ and $R$ to specify the left/right-handed spacetime spinor of Spin(1,3).
We use the L and R to specify the left or right internal spinor representation, 
such as $su(2)_{\rL}$ of the SM and the $su(2)_{\rL} \times su(2)_{\rR}$ of the Pati-Salam model.
Conventionally, the spacetime $L$ and the internal L are locked in the sense that the spacetime $L$-handed spinor is also the internal $\SU(2)_{\rL}$ doublet,
while the spacetime $R$-handed spinor is the $\SU(2)_{\rL}$ singlet (or $\SU(2)_{\rR}$ doublet in the PS model).
But we can regard the spacetime $R$-handed anti-particle as the $L$-handed particle as written in \eq{eq:SM-rep}.

\begin{figure}[h!] %[htbp]
%\begin{center}
  \centering
  \hspace{-.9cm}
 \includegraphics[width=4.in]{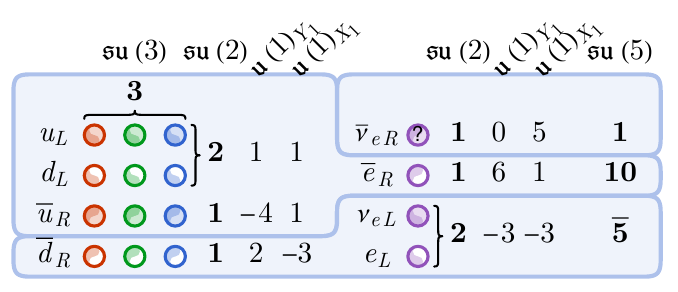}
 \caption{Georgi-Glashow $u(5)$ GUT with 16n Weyl fermions and their
 $u(5)^{\rm 1st}  = su(5)^{\rm 1st} \times u(1)_{X} = su(5)^{\rm 1st} \times u(1)_{X_1}$ rep:
$(\bar{d}_R \oplus {l}_L ) \oplus (q_L  \oplus \bar{u}_R \oplus   \bar{e}_R) \oplus ( \bar{\nu}_R)
=
\overline{\bf 5}_{-3} \oplus {\bf 10}_{1} \oplus {\bf 1}_{5}$.
 }
  \label{fig:GG}
%\end{center}
\end{figure}

\begin{figure}[h!] %[htbp]
%\begin{center}
  \centering
  \hspace{-.9cm}
 \includegraphics[width=4.in]{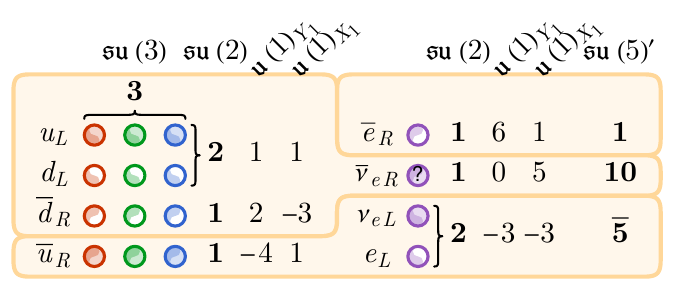}
 \caption{Flipped $u(5)$ GUT with 16n Weyl fermions 
  and their
 $u(5)^{\rm 2nd}  = su(5)^{\rm 2nd} \times u(1)_{\chi} = su(5)^{\rm 1st} \times u(1)_{X_2}$ rep:
$(\bar{u}_R \oplus {l}_L ) \oplus (q_L  \oplus \bar{d}_R \oplus   \bar{\nu}_R) \oplus ( \bar{e}_R)
=
\overline{\bf 5}_{-3} \oplus {\bf 10}_{1} \oplus {\bf 1}_{5}$. Note that the charge of $X_1 = X \neq \chi = X_2$.}
  \label{fig:flip}
%\end{center}
\end{figure}

\begin{figure}[h!] %[htbp]
%\begin{center}
  \centering
  \hspace{-.9cm}
 \includegraphics[width=4.in]{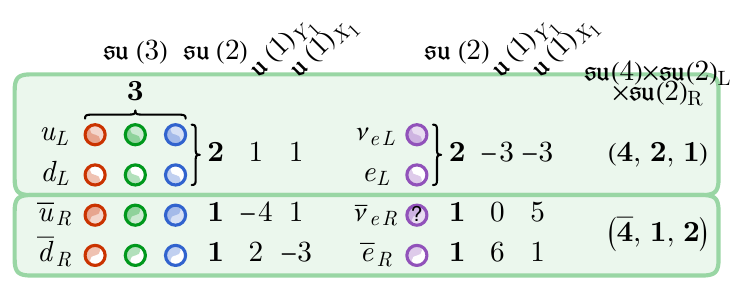}
 \caption{Pati-Salam {$su(4) \times su(2)_{\rm L} \times su(2)_{\rm R}$} model and their rep:
$(q_L \oplus l_L) \oplus (q_R \oplus l_R) 
=
(u_L \oplus d_L \oplus \nu_L \oplus e_L)  \oplus 
(\bar{u}_R \oplus \bar{d}_R \oplus \bar{\nu}_R \oplus \bar{e}_R)
=
({\bf 4}, {\bf 2}, {\bf 1}) \oplus (\overline{\bf 4}, {\bf 1}, {\bf 2})$.}
  \label{fig:PS}
%\end{center}
\end{figure}

\begin{figure}[h!] %[htbp]
%\begin{center}
  \centering
  \hspace{-.9cm}
 \includegraphics[width=4.6in]{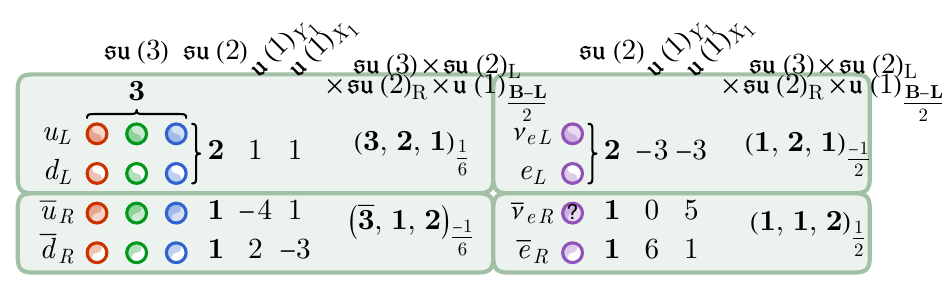}
 \caption{Left-Right {$su(3) \times su(2)_{\rm L} \times su(2)_{\rm R} \times { u(1)_{\frac{ \mathbf{B}-  \mathbf{L}}{2}} }$} model and their rep:
$q_L \oplus l_L \oplus q_R \oplus l_R  
=(u_L \oplus d_L) \oplus (\nu_L \oplus e_L)  \oplus 
(\bar{u}_R \oplus \bar{d}_R) \oplus (\bar{\nu}_R \oplus \bar{e}_R)
=
({\bf 3}, {\bf 2}, {\bf 1})_{\frac{1}{6}} \oplus ({\bf 1}, {\bf 2}, {\bf 1})_{\frac{-1}{2}} \oplus  (\overline{\bf 3}, {\bf 1}, {\bf 2})_{\frac{-1}{6}} \oplus  ({\bf 1}, {\bf 1}, {\bf 2})_{\frac{1}{2}}$.}
  \label{fig:LR}
%\end{center}
\end{figure}

\begin{figure}[h!] %[htbp]
%\begin{center}
  \centering
  \hspace{-.9cm}
 \includegraphics[width=4.in]{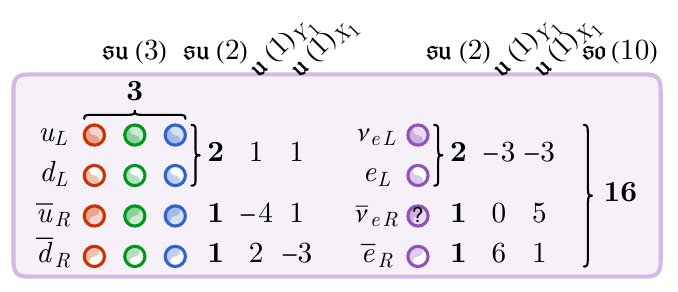}
 \caption{The $so(10)$ GUT with 16 Weyl fermions in the rep
 {${\bf 16}^+$ of $G_{so(10)} \equiv \Spin(10)$ gauge group}.}
  \label{fig:so10}
%\end{center}
\end{figure}

\newpage
\subsection{Georgi-Glashow $u(5)$ vs Flipped $u(5)$ models}

\begin{enumerate}[leftmargin=.mm]
\item \emph{The $su(5)$ or $u(5)$ Grand Unification} ({$su(5)$ or $u(5)$ GUT}):
Georgi-Glashow (GG)
\cite{Georgi1974syUnityofAllElementaryParticleForces}
hypothesized that 
the three SM gauge interactions 
merged into a single electronuclear force at higher energy
under a simple Lie algebra $su(5)$, or precisely a Lie group 
$
G_{\GG}\equiv  \SU(5)
$ 
gauge theory. 
The {su(5) GUT} works for 15n Weyl fermions, also for 16n Weyl fermions (i.e., 15 or 16 Weyl fermions per generation).
The Weyl fermions are in the representation of $u(5)^{\rm 1st}  = su(5)^{\rm 1st} \times u(1)_{X} = su(5)^{\rm 1st} \times u(1)_{X_1}$ 
as (see \Fig{fig:GG}):
\bea \label{eq:5-10-1-1st}
(\bar{d}_R \oplus {l}_L ) \oplus (q_L  \oplus \bar{u}_R \oplus   \bar{e}_R) \oplus ( \bar{\nu}_R)
=
\overline{\bf 5}_{-3} \oplus {\bf 10}_{1} \oplus {\bf 1}_{5},
\eea
More precisely, they are in the representation of 
a refined $\U(5)_{\hat q=2}$ group 
that we carefully define in \Sec{sec:Refined-U5-group}:
\bea \label{eq:U5-1st}
\U(5)_{\hat q=2}^{\text{1st}} \equiv \frac{\SU(5)^{\text{1st}} \times_{\hat q=2} \U(1)_{X} }{\Z_5} = \frac{\SU(5)^{\text{1st}} \times_{\hat q=2} \U(1)_{X_1} }{\Z_5}.
\eea

The 16th Weyl fermion is an extra  neutrino, sterile to the $su(5)$ but not sterile to the $u(1)$ gauge force.
%, also called the right-handed ``sterile'' neutrino. 
The corresponding U(1) is typically called the $X$ as
$
\U(1)_X \equiv \U(1)_{X_1} \equiv  \U(1)_{5 {({\bf B}-{\bf L})} -4 Y}
\equiv \U(1)_{5 {({\bf B}-{\bf L})} -\frac{2}{3} \tilde{Y}}
$ 
which we also call $X_1$ because this corresponds to the U(1) subgroup of the first type of $u(5)$ GUT.

\item \emph{The Barr's  flipped $su(5)$ or $u(5)$ GUT} \cite{Barr1982flippedSU5}:

The Weyl fermions are also in the representation of $u(5)^{\rm 2nd}  = su(5)^{\rm 2nd} \times u(1)_{\chi} = su(5)^{\rm 2nd} \times u(1)_{X_2}$ 
(also precisely a refined $\U(5)_{\hat q=2}$ group defined in \Sec{sec:Refined-U5-group})
but arrange in a different manner (see \Fig{fig:flip}):
\bea  \label{eq:5-10-1-2nd}
(\bar{u}_R \oplus {l}_L ) \oplus (q_L  \oplus \bar{d}_R \oplus   \bar{\nu}_R) \oplus ( \bar{e}_R)
=
\overline{\bf 5}_{-3} \oplus {\bf 10}_{1} \oplus {\bf 1}_{5}.
\eea
More precisely, they are in the representation of 
a refined $\U(5)_{\hat q=2}$ group 
 defined in \Sec{sec:Refined-U5-group}:
\bea \label{eq:U5-2nd}
\U(5)_{\hat q=2}^{\text{2nd}} \equiv \frac{\SU(5)^{\text{2nd}} \times_{\hat q=2} \U(1)_{\chi} }{\Z_5} = \frac{\SU(5)^{\text{2nd}} \times_{\hat q=2} \U(1)_{X_2} }{\Z_5}.
\eea

The corresponding U(1) is typically called the $\chi$, as
$
\U(1)_{\chi} \equiv \U(1)_{X_2}
$ 
which we also call $X_2$ because this corresponds to the U(1) subgroup of the second type of the $u(5)$ GUT (namely the flipped model \cite{Barr1982flippedSU5}).

\end{enumerate}

\subsection{There are only two types of $u(5)$ GUTs}
\label{sec:two-u(5)}

%\emph{There are only two types of $u(5)$ GUTs}:
Given the SM data and fermion representations,
we can prove that there are only two types of $u(5)$ GUTs (both embeddable inside a Spin(10) gauge group)
inside the largest possible internal U(16) group.

We sketch the proof below. {The normalizer of this $\SU(5)$ in $\U(16)$ is in fact 
$$\rN_{\U(16)}(\SU(5)) = \U(5) \times \U(1) \times \U(1)$$ (precisely we need $\U(5)_{\hat q=2}  \times \U(1) \times \U(1)$).}
There are in fact the important four U(1) subgroups in $\U(5)_{\hat q=2}  \times \U(1) \times \U(1)$ listed below.
\begin{enumerate}[leftmargin=.mm]
\item %$\bullet$ 
The $\U(1)_{Y}$ which in our convention is also generated by the 24th generator of the Lie algebra of $u(5)$ 
(precisely there are $\U(1)_{Y_1} \equiv \U(1)_{\tilde Y}$ or $\U(1)_{Y_2}$ depending on which $u(5)$ GUT models we choose).
This $\U(1)_{Y}$ is inside the SU(5). More precisely, the $\U(1)_{Y_1} \subset \SU(5)^{\text{1st}}$ of the GG model,
and $\U(1)_{Y_2} \subset \SU(5)^{\text{2nd}}$ of the Barr's flipped model.

\item %$\bullet$ 
The $\U(1)_{X}$ which in our convention is also generated by the 25th generator of the Lie algebra of $u(5)$ 
(precisely there are $\U(1)_{X_1}$ or $\U(1)_{X_2}$ depending on the which $u(5)$ GUT models we choose).
This $\U(1)_{X}$ is not inside the SU(5), but inside the U(5) (precisely the $\U(5)_{\hat q=2}$ in \Sec{sec:Refined-U5-group}).

\item
{For two additional $\U(1)^2$ in the normalizer $\rN_{\U(16)}(\SU(5)) = \U(5) \times \U(1) \times \U(1)$, 
we can choose one U(1) to act on $ \overline{\bf 5}$ alone, another U(1) to act on ${\bf 1}$ alone.}

Naively, other than the two SM Weyl fermion representation combinations of the 
$\overline{\bf 5} \oplus {\bf 10} \oplus {\bf 1}$ of SU(5) given in 
\eq{eq:5-10-1-1st} and \eq{eq:5-10-1-2nd}, there shall be two more kinds of interpretations (thus totally four kinds):
\bea  \label{eq:5-10-1-3rd} 
(\bar{u}_R \oplus {l}_L )  \oplus (q_L  \oplus \bar{d}_R \oplus   \bar{e}_R) \oplus ( \bar{\nu}_R)
&=&
\overline{\bf 5}_{} \oplus {\bf 10}_{} \oplus {\bf 1}_{},\\
 \label{eq:5-10-1-4th} 
(\bar{d}_R \oplus {l}_L ) \oplus (q_L  \oplus \bar{u}_R \oplus   \bar{\nu}_R) \oplus ( \bar{e}_R)
&=&
\overline{\bf 5}_{} \oplus {\bf 10}_{} \oplus {\bf 1}_{}.
\eea
All these four arrangements can establish the embedding
$\SU(5) \supset {\SU(3)}_c \times {\SU(2)}_{\rm L}$.
However, only the 
\eq{eq:5-10-1-1st} leads to 
$$\SU(5)^{\text{1st}} \supset  G_{\SM_6}$$
and  the \eq{eq:5-10-1-2nd} leads to 
$$\U(5)^{\text{2nd}}_{\hat q=2} \supset  G_{\SM_6}.$$
The \eq{eq:5-10-1-3rd} and  \eq{eq:5-10-1-4th}
both
lead to unsuccessful embeddings:
$$\rN_{\U(16)}(\SU(5)^{\text{3rd}}) = \U(5)^{\text{3rd}}_{\hat q=2} \times \U(1) \times \U(1)
\not \supset G_{\SM_6},
\text{ and } 
\rN_{\U(16)}(\SU(5)^{\text{4th}}) = \U(5)^{\text{4th}}_{\hat q=2} \times \U(1) \times \U(1)
 \not \supset G_{\SM_6},$$
\end{enumerate}
because the linear combinations of their U(1) subgroups cannot give rise to the SM's $\U(1)_{Y}$ or $\U(1)_{\tilde Y}$.
This concludes our proof.

\subsection{Pati-Salam $su(4) \times su(2)_{\rL} \times su(2)_{\rR}$ model}
Pati-Salam (PS) \cite{Pati1974yyPatiSalamLeptonNumberastheFourthColor}
hypothesized that the lepton carries the fourth color, extending SU(3) to SU(4). 
The PS also puts the left $\SU(2)_{\rL}$ and a hypothetical right $\SU(2)_{\rR}$ on the equal footing.
The PS gauge Lie algebra is
$su(4) \times su(2)_{\rL} \times su(2)_{\rR}$, 
and the PS gauge Lie group is 
$$G_{\PS_{q'}}\equiv\frac{\SU(4)_c\times(\SU(2)_\rL\times \SU(2)_\rR)}{\mathbb{Z}_{q'}}
= \frac{\Spin(6) \times \Spin(4)}{\mathbb{Z}_{q'}}$$ 
with the mod $q'=1,2$ depending on the global structure of Lie group.
We require $q'=2$ in order to embed $G_{\PS_{q'}}$ into the Spin(10) group. 
The particle excitations of this PS with 16n Weyl fermions are constrained by the representation of 
$G_{\PS_{q'}}$as (see \Fig{fig:PS}):
\bea
(q_L \oplus l_L) \oplus (\bar{q}_R \oplus \bar{l}_R) 
=
(u_L \oplus d_L \oplus \nu_L \oplus e_L)  \oplus 
(\bar{u}_R \oplus \bar{d}_R \oplus \bar{\nu}_R \oplus \bar{e}_R)
=
({\bf 4}, {\bf 2}, {\bf 1}) \oplus (\overline{\bf 4}, {\bf 1}, {\bf 2}),
\eea
written all in the left-handed ($L$) Weyl basis.\footnote{To be clear, 
we have the Weyl spacetime spinor ${\bf 2}_L$ of Spin(1,3) for $({\bf 4}, {\bf 2}, {\bf 1}) \oplus (\overline{\bf 4}, {\bf 1}, {\bf 2})$ of $su(4) \times su(2)_{\rL} \times su(2)_{\rR}$.
Here we use the $L$ and $R$ to specify the left/right-handed spacetime spinor of Spin(1,3).
We use the L and R to specify the left or right internal spinor representation of $su(2)_{\rL} \times su(2)_{\rR}$.
\label{footnote:LR}
}

\subsection{{The Left-Right  $su(3) \times su(2)_{\rL} \times su(2)_{\rR} \times u(1)_{\frac{ \mathbf{B}-  \mathbf{L}}{2}}$ model}}

Two version of internal symmetry groups for 
Senjanovic-Mohapatra's Left-Right (LR) model \cite{SenjanovicMohapatra1975} are,
$$
G_{\LR_{q'}} \equiv \frac{{\SU(3)}_c \times {\SU(2)}_{\rm L} \times  {\SU(2)}_{\rm R} \times \U(1)_{\frac{ \mathbf{B}-  \mathbf{L}}{2}} }{\Z_{3 q'}}
$$ with $q'=1,2$.
The particle excitations of this LR model with 16n Weyl fermions are constrained by the representation of 
$G_{\LR_{q'}}$as (see \Fig{fig:LR}):
\begin{multline}
q_L \oplus l_L \oplus \bar{q}_R \oplus \bar{l}_R  
=(u_L \oplus d_L) \oplus (\nu_L \oplus e_L)  \oplus 
(\bar{u}_R \oplus \bar{d}_R) \oplus (\bar{\nu}_R \oplus \bar{e}_R)\\
=
({\bf 3}, {\bf 2}, {\bf 1})_{\frac{1}{6}} \oplus ({\bf 1}, {\bf 2}, {\bf 1})_{\frac{-1}{2}} \oplus  (\overline{\bf 3}, {\bf 1}, {\bf 2})_{\frac{-1}{6}} \oplus  ({\bf 1}, {\bf 1}, {\bf 2})_{\frac{1}{2}},
\end{multline}
written all in the left-handed ($L$) Weyl basis.

In general, there is a QFT embedding, the PS model ($G_{\PS_{q'}}$) $\supset$ the LR model ($G_{\LR_{q'}}$) $\supset$ the SM ($G_{\SM_{q=3q'}}$) for both $q'=1,2$ 
via the internal symmetry group embedding, see the details in \Sec{sec:Embedding-Web}.

\subsection{The $so(10)$ Grand Unification and a DSpin-structure modification}

Georgi and Fritzsch-Minkowski
\cite{Fritzsch1974nnMinkowskiUnifiedInteractionsofLeptonsandHadrons}
hypothesized the {$so(10)$ GUT} (with a local Lie algebra $so(10)$)
that quarks and leptons as spacetime Weyl fermions (each as a 2-component complex ${\bf 2}_L$ of ${\Spin(1,3)}$) become the irreducible 16-dimensional spinor representation of Spin(10) (see \Fig{fig:so10}):
\bea
\text{${\bf 16}^+$ of $G_{so(10)} \equiv \Spin(10)$ gauge group}. 
\eea 
Thus, the 16n Weyl fermions can interact via the Spin(10) gauge fields at a higher energy.
In this case, the 16th Weyl fermion, previously a sterile neutrino to the SU(5),
is \emph{no longer sterile} to the Spin(10) gauge fields; it also carries a charge 1, thus not sterile, under the gauged center subgroup $Z(\Spin(10))=\Z_4$.

In \Refe{Wang2106.16248}, there is a new sector involving either a new discrete torsion WZW term or a fermion parton theory. 
The new sector can be manifested via the new beyond-standard-model (BSM) {Dirac} fermions (each as a 4-component complex ${\bf 2}_L  \oplus {\bf 2}_R$ of ${\Spin(1,3)}$) in
the 10-dimensional vector representation of $so(10)$ or Spin(10):
\bea
\text{${\bf 10}$ of $G_{so(10)} \equiv \Spin(10)$ gauge group}. 
\eea 
This BSM fermion is \emph{not compatible} with the required spacetime-internal symmetry group structure
$(\Spin  \times_{\Z_2^F}\Spin(10))$ that is necessary to realize the $w_2 w_3$ anomaly. The incompatibility is due to
the spin-charge relation (i.e., the spacetime spin and internal charge relation) imposed by $(\Spin  \times_{\Z_2^F}\Spin(10))$
constrains that the fermions can be in the ${\bf 16}$ but not the ${\bf 10}$ of $\Spin(10)$.

\begin{figure}[h!] %[htbp]
%\begin{center}
  \centering
  \hspace{-.9cm}
 \includegraphics[width=4.in]{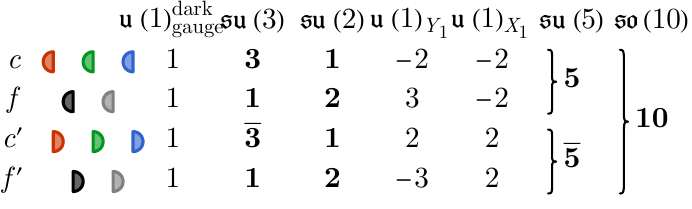}
 \caption{The modified $so(10)$ GUT \cite{Wang2106.16248} with new {Dirac} fermions in the rep
 {${\bf 10}$ of $G_{so(10)} \equiv \Spin(10)$ gauge group}.
 These fermions are called colorons ($c$) and flavorons ($f$).
These fermions can be regarded as the fractionalizations of GUT-Higgs field (see $\Phi^{\rm{bi}}$ in \Sec{sec:QuantumPhaseDiagram})
in terms of the color-flavor separation, 
 similar to the spin-charge separation \cite{Tomonaga1950Remarks,Luttinger1963An-Exactly,Anderson2000Spin-charge}
 in condensed matter phenomenon.
 Their representation
$c \oplus f
 \oplus {c'} \oplus {f'}
 =  ({\bf 3},{\bf 1})_{1,-2,-2}\oplus ({\bf 1},{\bf 2})_{1,3,-2}
 \oplus (\overline{\bf 3},{\bf 1})_{1,2,2}
 \oplus ({\bf 1},{\bf 2})_{1,-3,2} $
of $su(3)_{c}\times su(2)_{\rm L}\times {u(1)'}_{\text{gauge}}^{\text{dark}} \times u(1)_{Y_1}\times u(1)_{X_1}$.
They are
$(c \oplus f)
 \oplus ({c'} \oplus {f'}) = {\bf 5} \oplus \overline{\bf 5}$ of $su(5)$, and ${\bf 10}$ of $so(10)$. 
 }
  \label{fig:so10-parton}
%\end{center}
\end{figure}

The remedy, provided by \Refe{Wang2106.16248}, introduces two different fermion parities, ${\Z_2^F}$ and ${\Z_2^{F'}}$, for the SM fermion ${\bf 16}$ and BSM fermion ${\bf 10}$ respectively,
which together formally forms a double spin structure called DSpin that shares the same spacetime special orthogonal SO rotation: 
\bea
\DSpin \equiv (\Z_2^{F} \times \Z_2^{F'}) \rtimes \SO.
\eea
{Here $\Spin = \Z_2^{F}  \rtimes \SO$ 
and the $\Spin' = \Z_2^{F'} \rtimes \SO$ is another new copy of Spin structure.
}
Thus, the modified $so(10)$ GUT 
in \Refe{Wang2106.16248} asks for a new spacetime-internal structure:
\bea
G^{\text{modified}}_\text{$so(10)$-GUT}\equiv
(\DSpin  \times_{\Z_2^F}\Spin(10)) \times_{\Z_2^{F'}}G_{\text{internal-fermionic-parton}},
\eea
where the internal symmetry of the fermionic parton theory $G_{\text{internal-fermionic-parton}}$ can be implemented via $\U(1)'$ or $\SU(2)'$, etc., 
see \cite{Wang2106.16248}.

Several tables of representation data for the SM, the GG $su(5)$, the PS, and the $so(10)$ models, can already be found in Appendix A of our previous work \cite{Wang2106.16248}.
In comparison, here we provide the flipped $su(5)$ or $u(5)$ model data in Appendix \ref{sec:SM-GUT-table-app}.

\begin{figure}[htbp]
\begin{center}
\includegraphics[width=0.46\columnwidth]{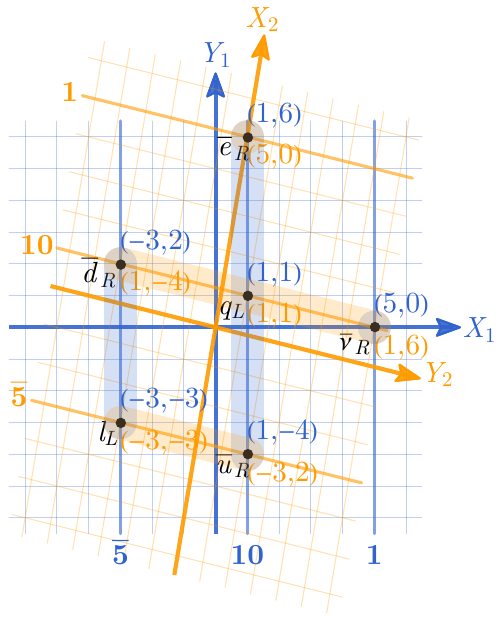}
\caption{Charge lattices of $(X_1,Y_1)$ (in \ccblue{blue} as in Georgi-Glashow [GG] \Fig{fig:GG}) and
$(X_2,Y_2)$ 
(in \ccorange{orange} as in the flipped $u(5)$ GUT \Fig{fig:flip}).
The charge assignments of the SM Weyl fermions 
coincide with the intersection (also the common integer points) of the two sets of charge lattices.
We mark the GG \eq{eq:5-10-1-1st}'s
$(\bar{d}_R \oplus {l}_L ) \oplus (q_L  \oplus \bar{u}_R \oplus   \bar{e}_R) \oplus ( \bar{\nu}_R)
= \overline{\bf 5}_{-3} \oplus {\bf 10}_{1} \oplus {\bf 1}_{5}$  of $su(5)^{\text{1st}} \times u(1)_{X_1}$
highlighted in the three vertical \ccblue{blue} bars.
We mark the \eq{eq:5-10-1-2nd}'s
$(\bar{u}_R \oplus {l}_L ) \oplus (q_L  \oplus \bar{d}_R \oplus   \bar{\nu}_R) \oplus ( \bar{e}_R)
=
\overline{\bf 5}_{-3} \oplus {\bf 10}_{1} \oplus {\bf 1}_{5}$ of $su(5)^{\text{2nd}} \times u(1)_{X_2}$
highlighted in the three tilted horizontal  \ccorange{orange} bars.
}
\label{fig:charge lattice}
\end{center}
\end{figure}

\begin{landscape}
%\pagenumbering{gobble}
\thispagestyle{empty}

\subsection{Representations of SMs and {Five} GUT-like models in a unified Table}
\label{sec:SM-GUT-table}

For readers' convenience to check the quantum numbers of various elementary particles or field quanta of SMs and GUTs,
we combine all the SM, the Georgi-Glashow (GG) or the flipped ($su(5)$ or $u(5)$) models, the Pati-Salam (PS), the left-right (LR), and the $so(10)$ models
in a single \Table{table:fermionAll}. This \Table{table:fermionAll} summarizes the more elaborated Appendix
A of \Refe{Wang2106.16248} and Appendix \ref{sec:SM-GUT-table-app} of present article in a brief but unified way.

 \begin{table}[!h]
$\hspace{-13mm}
  \begin{tabular}{ccccccc c c c c c c  c c  c c  }
    \hline
    $\begin{array}{c}
    \textbf{SM}\\ 
   \textbf{fermion}\\
   \textbf{spinor}\\ 
   \textbf{field}
       \end{array}$
   & ${\SU(3)}$& ${\SU(2)}_{\rm L}$& ${\SU(2)}_{\rm R}$ 
     & $\U(1)_{\frac{ \mathbf{B}-  \mathbf{L}}{2}}$  
   & $\U(1)_{ Y_1}$  
   & $\U(1)_{\tilde Y_R}$   
   & $\U(1)_{\rm{EM}}$ 
   & $\U(1)_{X_1}$
   %$\U(1)_{X}^{\text{1st}}$ 
   &     
    $\Z_{4,X}$
    & $\Z_{2}^F$  &   
    $\U(1)_{X_2}$ & 
    $\U(1)^{}_{Y_2}$ & 
    SU(5)$^{\text{1st}}$ & SU(5)$^{\text{2nd}}$ &  $G_{\PS}$ &  Spin(10)\\
        \hline\\[-4mm]
$u_L$& ${{\mathbf{3}}}$& 
\multicolumn{1}{c}{\multirow{2}{*}{
    {$q_L:\mathbf{2}$} 
     }
     }
&$\mathbf{1}$
& 1/6 
& 1
& 4 & 2/3   
 &1 & 1 &    1 & 1 &1  &
    \multicolumn{2}{c}{\multirow{2}{*}
  {({\bf 3},{\bf 2}) in {\bf 10} }}  
 & 
  \multirow{4}{*}
  {$\begin{array}{c}
     \bf{4},\\
      \bf{2},\\
      1
      \end{array}
     $}
 &  \multirow{8}{*}{
     $\bf{16}
     $}
   \\
 \cline{1-2} \cline{4-13}
%\hline
$d_L$& ${{\mathbf{3}}}$&  &$\mathbf{1}$
& 1/6 
& 1 
& $-2$ &   $-1/3$ 
 &1 & 1 &    1 & 
 1 & 1
   \\
    \cline{1-15}  
 $\nu_L$& $\mathbf{1}$& 
\multicolumn{1}{c}{\multirow{2}{*}{
    {$l_L:\mathbf{2}$} 
    }}
 & $\mathbf{1}$ 
  & $-1/2$
 & $-3$ 
 &  0 & 0   
 & $-3$  &1  &   1 & $-3$ & $-3$ &
    \multicolumn{2}{c}{\multirow{2}{*}
  {({\bf 1},{\bf 2}) in $\bar{\bf 5}$ }} 
    \\
 \cline{1-2} \cline{4-13}   
%\hline 
 $e_L$& $\mathbf{1}$& & $\mathbf{1}$ 
 & $-1/2$
 & $-3$
 &  $-6$  &  $-1$ 
  & $-3$  &1  &   1 & $-3$ & $-3$ \\
 \cline{1-13}  \cline{13-15}         \cline{16-16}   
%%%%%%
%%%%%%
 $\bar{u}_R$& $\bar{\mathbf{3}}$& $\mathbf{1}$ & 
\multicolumn{1}{c}{\multirow{2}{*}{
    {$q_R:\mathbf{2}$} 
     }
     }
     & $-1/6$
& $-4$ & 
$-1$
& $-2/3$   & $1$  &1 &   1 & $-3$ & 2 & in ${\bf 10}$ & in $\bar{\bf 5}$ &  \multirow{4}{*}
  {$\begin{array}{c}
     \bar{\bf{4}},\\
      1,\\
      \bf{2}
      \end{array}
     $}
 \\
     \cline{1-3} \cline{5-13} \cline{14-15}  \\[-3.6mm]  
     %\hline
$\bar{d}_R$& $\bar{\mathbf{3}}$& $\mathbf{1}$ & %\cred{$\mathbf{2}$}  
& $-1/6$
& 2  
& $-1$
& 1/3    & $-3$ &1   &   1 &  1 & $-4$ &  in $\bar{\bf 5}$ & in ${\bf 10}$ 
  \\
 \cline{1-13}   \cline{14-15}  
 $\bar{\nu}_R= {\nu}_L $& $\mathbf{1}$& $\mathbf{1}$& 
\multicolumn{1}{c}{\multirow{2}{*}{
    {$l_R:\mathbf{2}$} 
     }  }
& 1/2 
& 0  &  3 & 0  & 5 &   1 &   1 & 1 &  6 & in ${\bf 1}$ & in ${\bf 10}$ \\
\cline{1-3} \cline{5-13}   \cline{14-15}
%\hline
$\bar{e}_R= e_L^+$& $\mathbf{1}$& $\mathbf{1}$& %\cred{$\mathbf{2}$}  
 & 1/2
& 6  &
3  & 1   &1 &   1 &    1 & 5 & 0 & in ${\bf 10}$ & in ${\bf 1}$ \\
%\hline
%    \hline $\phi_H$ & $\mathbf{1}$ & $\mathbf{2}$& 1/2 & 1 & 0 & -2 &   2 & 3 & 0 & -6  &   -2\\
    \hline\end{tabular}
$
   \caption{Here we list down the Weyl fermion's representations 
   in the $su(3) \times su(2)_{\rL}  \times u(1)$ SM, 
  the left-right ${su(3) \times su(2)_{\rL}  \times su(2)_{\rR}  \times u(1)}$, 
   the Pati-Salam ${su(4) \times su(2)_{\rL}   \times su(2)_{\rR}}$, 
   the Georgi-Glashow $\U(5)_{\hat q=2}^{\text{1st}} \equiv \frac{\SU(5)^{\text{1st}} \times \U(1)_{X=1,\hat q=2} }{\Z_5}$,
   the flipped $\U(5)_{\hat q=2}^{\text{2nd}} \equiv \frac{\SU(5)^{\text{2nd}} \times \U(1)_{X_2,\hat q=2} }{\Z_5}$ model,
   and the $so(10)$ GUT of $\Spin(10)$ group.
  }
 \label{table:fermionAll}
 \end{table}
 
 The $\U(1)_{X_1}$ and $\U(1)_{X_2}$ Lie algebra generators 
in \eq{eq:U5-1st} and \eq{eq:U5-2nd}
are also denoted as the 25th generators out of the 25 generators of the $u(5)^{\text{1st}}$ and $u(5)^{\text{2nd}}$ (the 1st for the GG, and the 2nd for the flipped model).
The $\U(1)_{Y_1}$ and $\U(1)_{Y_2}$ Lie algebra generators are part of the $su(5)^{\text{1st}}$ and $su(5)^{\text{2nd}}$ --- 
they are the 24th generators out of the 25 generators of the $u(5)^{\text{1st}}$ and $u(5)^{\text{2nd}}$. 
In short, we have these relations between different expressions:
 \bea
\left\{
\begin{array}{lcl}
\U(1)_{Y_1} \equiv \U(1)_{\tilde Y} \equiv \U(1)_{6Y} \equiv \U(1)^{\text{1st}}_{T_{24}}. \\
\U(1)_{Y_2} \equiv \U(1)^{\text{2nd}}_{T_{24}}. \\ 
 \U(1)_{X_1} \equiv \U(1)_X \equiv  \U(1)_{5 {({\bf B}-{\bf L})} -4 Y}
\equiv \U(1)_{5 {({\bf B}-{\bf L})} -\frac{2}{3} {Y}_1} \equiv \U(1)^{\text{1st}}_{T_{25}}.\\
\U(1)_{X_2} \equiv \U(1)_{\chi}  \equiv \U(1)^{\text{2nd}}_{T_{25}}. \\
\U(1)_{X_1} \text{ and }  \U(1)_{X_2} \text{ share the same normal subgroup } Z(\Spin(10))=\Z_{4,X} .
\end{array}
\right.
\eea
Here the compactification size of $\U(1)_{\tilde Y}$ is 1/6 of $\U(1)_{Y}$, another way is identifying $\frac{\U(1)_{\tilde Y}}{\Z_6}= \U(1)_{Y}$.
So their charge quantizations are related via $q_{\U(1)_{Y_1}} \equiv q_{\U(1)_{\tilde Y}} \equiv{6}q_{\U(1)_{Y}}$.
We shall simply denote their charge relation as $Y_1 \equiv {\tilde Y} \equiv 6 Y$.
\end{landscape}
%\pagenumbering{arabic}
%
%
These U(1) factors of $X_1,Y_1,X_2$ and $Y_2$ whose corresponding quantized charges span some lattice subspace of a two-dimensional $\R^2$-plane.
The $X_1,Y_1,X_2$ and $Y_2$ charges have the following relations:
\bea
\label{eq:XY-1}
\bpm X_1\\Y_1\epm=\frac{1}{5}\bpm1&4\\6&-1\epm\bpm X_2\\Y_2\epm \equiv M_{\Z_2^{\rm flip} } \bpm X_2\\Y_2\epm &,&\quad \bpm X_2\\Y_2\epm=\frac{1}{5}\bpm1&4\\6&-1\epm\bpm X_1\\Y_1\epm
 \equiv M_{\Z_2^{\rm flip} } \bpm X_1\\Y_1\epm. \quad\\
\label{eq:XY-2}
\bpm X_1\\X_2\epm=\frac{1}{6}\bpm1&5\\5&1\epm\bpm Y_1\\Y_2\epm 
 \equiv M_{XY} \bpm Y_1\\Y_2\epm 
&,& \quad \bpm Y_1\\Y_2 \epm=\frac{1}{4}\bpm-1& 5 \\ 5 &-1\epm\bpm X_1\\X_2\epm
 \equiv M_{YX} \bpm X_1\\ X_2\epm .\;
\quad
\eea
%\begin{itemize}[leftmargin=.mm] 
%\item 
$\bullet$
The \eq{eq:XY-1} shows that $(X_1,Y_1)$ and $(X_2,Y_2)$ can be mapped onto each other via the
$M_{\Z_2^{\rm flip} }
\equiv
\frac{1}{5}(\begin{smallmatrix}1&4\\6&-1\end{smallmatrix})$
where $M_{\Z_2^{\rm flip} }^2=1$, $\det(M_{\Z_2^{\rm flip} })=-1$, and $M_{\Z_2^{\rm flip} }=M_{\Z_2^{\rm flip} }^{-1}$ is itself a {\bf self-inverse}.
The $M_{\Z_2^{\rm flip} }$ is part of the ${\Z_2^{\rm flip}}$ symmetry transformation that 
swaps the GG $u(5)^{\text{1st}}$ model and the flipped $u(5)^{\text{2nd}}$ model.\\
%
%\item 
$\bullet$ The \eq{eq:XY-2} shows $M_{XY} M_{YX}=+1$, $\det(M_{XY})=-2/3$, and 
$\det(M_{YX})=-3/2$; the $M_{XY}$ and $M_{YX}$ are inverse with respect to each other. \\
%
%
%
%\item
$\bullet$ The two sets of charge lattices intersect at integer points that match the charge assignment of the SM Weyl fermions, as shown 
in \Fig{fig:charge lattice}.\footnote{{In \Fig{fig:charge lattice}, {the $(X_1,Y_1) = (X_2,Y_2) =(5,5)$ is also compatible with two integral charge lattices,
but this particle carries a hypercharge $Y_1=5$ and a net EM charge $\frac{5}{6}$ which we do not observe in the experiment.}
Other examples, such as the $(X_1,Y_1)=(5, -5)$ with $(X_2,Y_2) =(-3,7)$ and $(X_1,Y_1)=(-3,7)$ with $(X_2,Y_2) =(5, -5)$, also with nontrivial hypercharges or EM charges, 
that have no evidences yet for their real-world existence.}}
%\end{itemize}
%

Other than the above U(1) factors (i.e., $X_1,Y_1,X_2$ and $Y_2$), we can find the following three U(1) factors of
$T_{3,\rm{L}}$, $T_{3,{\rm R}}$, and  $\frac{ \mathbf{B}-  \mathbf{L}}{2}$. Some comments about these U(1) factors:\\[-6mm]
\begin{enumerate}[leftmargin=.mm] 
\item $T_{3,\rm{L}}$ is generated by the third Lie algebra generator of the ${\SU(2)}_{\rm L}$,
while $T_{3,\rm{R}}$ is generated by the third Lie algebra generator of the ${\SU(2)}_{\rm R}$.
We take its $T_{3,\rm{L/R}}$'s charge ($\pm \frac{1}{2}$) as the coefficient of its Lie algebra generator 
$ \frac{1}{2}\big(\begin{smallmatrix}
1 & 0\\
 0 & - 1
\end{smallmatrix}\big)$.
\item The ${ \mathbf{B}-  \mathbf{L} }$ is the baryon ($\mathbf{B}$) minus lepton ($\mathbf{L}$) number.

\item We have  $T_{3,\rm{L}}+ Y=Q_{\rm{EM}}$,
the Lie algebra linear combination of the third generator of  ${\SU(2)}_{\rm L}$ and  $\U(1)_Y$ gives the $\U(1)_{\rm{EM}}$ charge $Q_{\rm{EM}}$.

\item We have $T_{3,\rm{R}}+ Y=\frac{ \mathbf{B}-  \mathbf{L}}{2}$,
 the Lie algebra linear combination of the third generator of ${\SU(2)}_{\rm R}$ and the $\U(1)_Y$ gives the $\U(1)_{ \mathbf{B}-  \mathbf{L}}$.
We choose the right-handed anti-particle to be in {$\mathbf{2}$} of ${\SU(2)}_{\rm R}$
(so its right-handed particle to be in {$\bar{\mathbf{2}}$} of ${\SU(2)}_{\rm R}$) that makes a specific assignment on the $\pm$ sign of its $T_{3,{\rm R}}$ charge.
So we have the formula, $T_{3,{\rm L}}-T_{3,{\rm R}}=Q_{\rm{EM}}-\frac{ \mathbf{B}-  \mathbf{L}}{2}$.

\item We can introduce a new internal right hypercharge $\U(1)_{{Y}_{\rR}}$ for ${\SU(2)}_{\rm R}$, 
as an analog of the internal left electroweak hypercharge $\U(1)_{{Y}}=\U(1)_{{Y}_{\rL}}$ for ${\SU(2)}_{\rm L}$ sector,
such that 
 $-T_{3,\rm{R}}+ {{Y}_{\rR}}=Q_{\rm{EM}}$
 and  $-T_{3,\rm{L}}+ {{Y}_{\rR}}=\frac{ \mathbf{B}-  \mathbf{L}}{2}$.
    
\item In fact, we can express all aforementioned U(1)
in terms of some linear combinations of three independent generators $T_{3, \rL}$, $T_{3,\rR}$, and  $\frac{ \mathbf{B}-  \mathbf{L}}{2}$.
Here we list down the relations of their charges via the linear combinations of $T_{3, \rL}$, $T_{3,\rR}$, and  $\frac{ \mathbf{B}-  \mathbf{L}}{2}$ charges:
\bea \label{eq:U1-linear-combination}
\left\{
\begin{array}{lcl}
 - T_{3,\rR}+ \frac{ \mathbf{B}-  \mathbf{L}}{2} =  - T_{3,\rL} + Q_{\rm{EM}}&=& Y_{\rL}  \equiv Y \equiv \frac{1}{6} Y_1.\\
  T_{3,\rL} +\frac{ \mathbf{B}-  \mathbf{L}}{2}     = T_{3,\rR}+  Q_{\rm{EM}}   &=& Y_{\rR} \equiv \frac{1}{6}{\tilde Y}_{\rR}.\\
  T_{3,\rL}- T_{3,\rR}+ \frac{ \mathbf{B}-  \mathbf{L}}{2}  &=&Q_{\rm{EM}}.\\
 4 T_{3,\rR}  + 6(\frac{ \mathbf{B}-  \mathbf{L}}{2})   &=& X \equiv X_1 = 5 ({ \mathbf{B}-  \mathbf{L}}) - 4 Y = 5 ({ \mathbf{B}-  \mathbf{L}}) - \frac{2}{3} Y_1.
\end{array}
\right.
\eea
Since the appropriate linear combination of $T_{3,\rR}$ and $(\frac{ \mathbf{B}-  \mathbf{L}}{2})$ in \eq{eq:U1-linear-combination}
contains the $X_1$ and $Y_1$, thus which linear combination as in \eq{eq:XY-1} contains also $X_2$ and $Y_2$.
\item The SM (\Fig{fig:SM}) with a gauge group $G_{\SM_q}$ contains the U(1) Lie algebra generators of 
$T_{3,\rL}$ and $Y_1$, thus also $Q_{\rm{EM}}$.
The SM does \emph{not} contain those of $T_{3,\rR}$,  $\frac{ \mathbf{B}-  \mathbf{L}}{2}$,
$X_1$, $X_2$, or $Y_2$. 
\item The GG (\Fig{fig:GG}) $su(5)$ GUT with a gauge group SU(5) contains the U(1) Lie algebra generators of $T_{3, \rL}$ and $Y_1$, 
thus their linear combinations include $Q_{\rm{EM}}$,
but not the other U(1).

 The GG $u(5)$ GUT with a gauge group $\U(5)_{\hat q=2}$ contains the U(1) Lie algebra generators of 
 $T_{3, \rL}$, $Y_1$ and $X_1$, thus their linear combinations include everything else: $Q_{\rm{EM}}$, $\frac{ \mathbf{B}-  \mathbf{L}}{2}$, $Y_2$ and $X_2$,
also $T_{3, \rR}$ and $Y_{\rR}$.
\item The flipped model (\Fig{fig:flip}) $su(5)$ with only a gauge group SU(5) contains the U(1) Lie algebra generators of $T_{3, \rL}$ and $Y_2$, 
thus their linear combinations does not include $Q_{\rm{EM}}$,
nor the other U(1). Thus the flipped model with only a gauge group SU(5) is not enough to contain the SM gauge group $G_{\SM_q}$.

 The flipped $u(5)$ GUT with a gauge group $\U(5)_{\hat q=2}$ contains the U(1) Lie algebra generators of 
 $T_{3, \rL}$, $Y_2$ and $X_2$, thus their linear combinations include everything else: $Q_{\rm{EM}}$, $\frac{ \mathbf{B}-  \mathbf{L}}{2}$, $Y_1$ and $X_1$,
also $T_{3, \rR}$ and $Y_{\rR}$.
\item The PS model (\Fig{fig:PS}) with a gauge group $G_{\PS_{q'}}$ contains the U(1) Lie algebra generators of 
$T_{3, \rL}$, $T_{3,\rR}$, and  $\frac{ \mathbf{B}-  \mathbf{L}}{2}$.
The LR model contains also all these generators. 
Since the linear combinations of these three generators give all the aforementioned U(1) Lie algebra generators 
in \eq{eq:U1-linear-combination}, the PS and LR models contain all these.
\item The $so(10)$ GUT (\Fig{fig:so10}) with a gauge group Spin(10) 
contains the U(1) Lie algebra generators of 
 $T_{3, \rL}$, $Y_1$ and $X_1$. But it contains \emph{not only} the \emph{discrete} $\Z_{4,X}$ subgroup \emph{but also} the \emph{continuous
Lie group} $\U(1)_{X_1}$. Note that $Z(\Spin(10)) = \Z_{4,X_1} = \Z_{4,X_2} = \Z_{4,X}$.
\end{enumerate}

\newpage
\section{Quantum Landscape as Quantum Phase Diagram (Moduli Space)}
\label{sec:Quantum-Landscape}

We present the internal symmetry group embedding of various SMs and GUTs in
\Sec{sec:Embedding-Web}. Then we use this group embedding structure to explore
a quantum phase diagram containing these SMs and GUTs,
and their quantum criticalities (e.g., critical points or critical regions)
in \Sec{sec:QuantumPhaseDiagram}.
 
\subsection{Embedding Web by Internal Symmetry Groups}
\label{sec:Embedding-Web}

\begin{figure}[h!] %[htbp]
%\begin{center}
  \centering
  \hspace{-.9cm}
 \includegraphics[width=7.in]{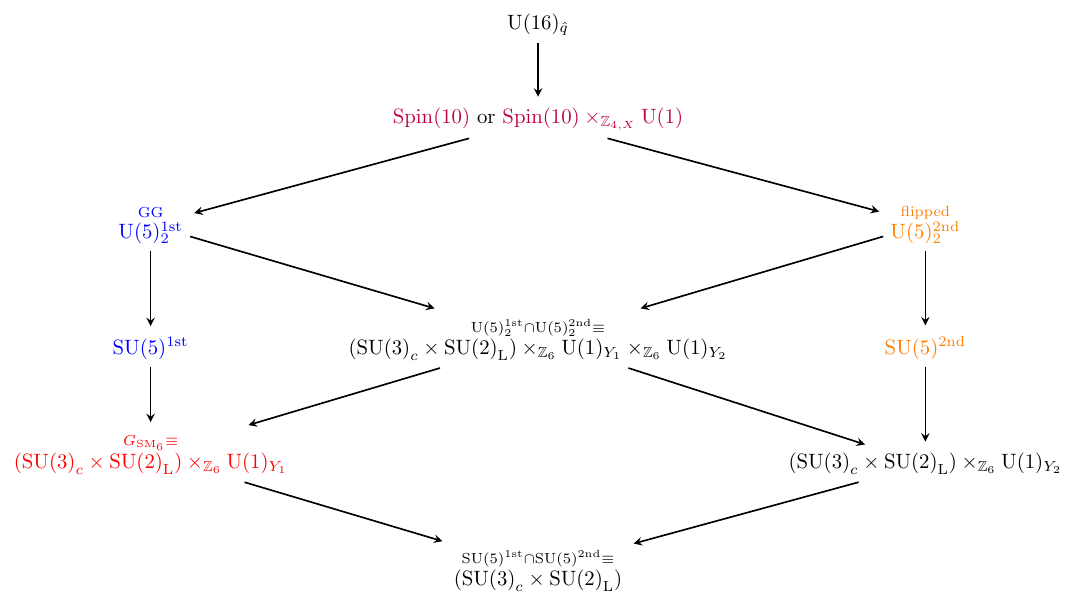}
 \caption{Internal symmetry group embedding web for the $so(10)$ GUT, the Georgi-Glashow (GG) and the flipped $u(5)$ models, and the Standard Model (SM).}
  \label{fig:embed-Lie-group-spacetime-2109-1}
%\end{center}
\end{figure}

For 16 Weyl fermion models, there is a maximal internal symmetry group U(16)
that rotates the 16 flavor of spacetime Weyl spinors in ${\bf 2}_L$. But this U(16) again requires a refined definition, say $\U(16)_{\hat q}$, as we did in \Sec{sec:Refined-U5-group}.  
For our purpose, we just need some appropriate subgroup $\Spin(10)$ or ${{\Spin(10) \times_{\Z_{4,X}} \U(1)}}$
of $\U(16)_{\hat q}$, say ${\hat q}=1$, 
such that 
$\U(16)_{\hat q} \supset \Spin(10)$ or
$\U(16)_{\hat q} \supset {{\Spin(10) \times_{\Z_{4,X}} \U(1)}}$. 
Then we can obtain the following embedding web in \Fig{fig:embed-Lie-group-spacetime-2109-1}
and \Fig{fig:embed-Lie-group-spacetime-2109-2}.
The arrow direction $G_1 \to G_2$ implies that $G_1$ internal symmetry can be broken down to $G_2$ (for example via appropriate scalar Higgs condensation),
this also implies the embedding $G_1 \supset G_2$ and the inclusion $G_1 \hookleftarrow G_2$.

Some comments are in order:
\begin{enumerate}[leftmargin=.mm] 
\item From \eq{eq:U5-embed}, \eq{eq:U5-embed-2}, and \eq{eq:U5-embed-3},
we see that
$\Spin(10) \supset \U(5)_{\hat q=2}$, and 
$\frac{\Spin(10)\times \U(1)}{\Z_4} \supset \U(5)_{\hat q=2}$.
We provide a verification via exponential maps of the Lie algebras into these Lie groups embedding in Appendix \ref{app:Embedding}.
Also there are two versions of $\U(5)_{\hat q=2}$, the 1st for GG and the 2nd for the flipped model (see \Sec{sec:two-u(5)}).
Obviously, we also have $\U(5)_{\hat q} \supset \SU(5)$ for both the 1st  GG and the 2nd the flipped model 

\begin{figure}[h!] %[htbp]
%\begin{center}
  \centering
  \hspace{-.9cm}
 \includegraphics[width=7.in]{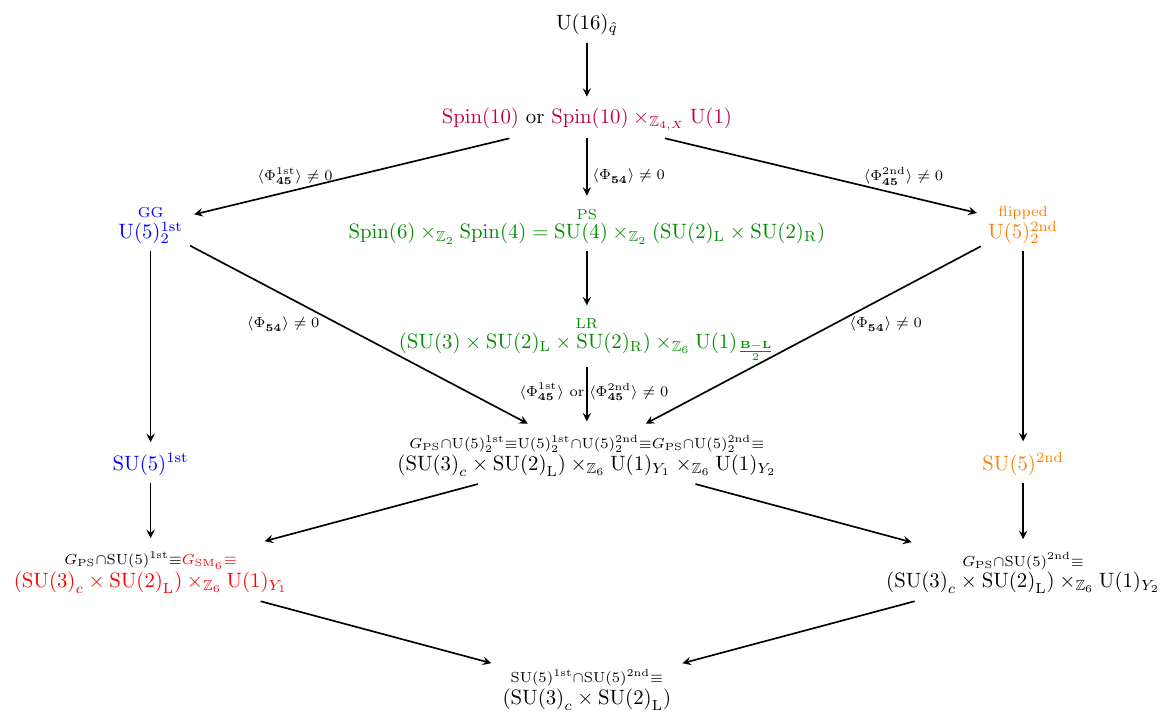}
 \caption{Follow \Fig{fig:embed-Lie-group-spacetime-2109-1},
 an internal symmetry group embedding web for the $so(10)$ GUT, the GG and the flipped $u(5)$ models, 
 and the SM; here we include also the Pati-Salam (PS) and Left-Right (LR) models.
 We can use the GUT-Higgs condensation to trigger the different routes of the symmetry breaking patterns --- 
 we explain the realization of the vacuum expectation value (vev)
 $\langle \Phi_{\bf{45}}^{\text{1st}} \rangle$, $\langle \Phi_{\bf{45}}^{\text{2nd}} \rangle$ and $\langle \Phi_{\bf{54}}^{\text{}} \rangle$ in
 \Sec{sec:QuantumPhaseDiagram}. This figure generalizes the previous studies in \cite{BaezHuerta0904.1556} and in our prior work \cite{Wang2106.16248}.}
  \label{fig:embed-Lie-group-spacetime-2109-2}
%\end{center}
\end{figure}

\item We can explicitly check that the intersection and the union of two Lie groups of GG and the flipped model:
\bea
{\U(5)_{2}^{\text{1st} } \cap \U(5)_{2}^{\text{2nd}} } &=& {({\SU(3)}_c \times {\SU(2)}_{\rm L}) \times_{{\Z_6}} \U(1)_{Y_1}  \times_{{\Z_6}} \U(1)_{Y_2}}.\cr
{\U(5)_{2}^{\text{1st} } \cup \U(5)_{2}^{\text{2nd}} } &=& \Spin(10). \\
{\SU(5)_{2}^{\text{1st} } \cap \SU(5)_{2}^{\text{2nd}} } &=& {({\SU(3)}_c \times {\SU(2)}_{\rm L})}. \nn
\eea

This check is presented in Appendix \ref{sec:two-u5-Intersection-Join} based on the Lie algebra data in \Table{tab:embeddings},
then we can verify the exponential maps of the Lie algebras into these Lie groups 
Also we show
$\SU(5)^{\text{1st}} \supset  G_{\SM_6}$ but $\SU(5)^{\text{2nd}} \not \supset  G_{\SM_6}$, but
$\U(5)^{\text{2nd}}  \supset  G_{\SM_6}$ and
$\SU(5)^{\text{2nd}}  \supset 
{({\SU(3)}_c \times {\SU(2)}_{\rm L}) \times_{{\Z_6}} \U(1)_{Y_2}}$.

\item \Fig{fig:embed-Lie-group-spacetime-2109-2}
follows \Fig{fig:embed-Lie-group-spacetime-2109-1} by adding the PS and LR models.
We have the PS model $\supset$ the LR model $\supset$ the SM for both $q'=1,2$ 
via the internal symmetry group embedding:
\bea
G_{\PS_{q'}} \supset  G_{\LR_{q'}} \supset 
G_{\SM_{q=3q'}}  \equiv \frac{{\SU(3)}_c \times {\SU(2)}_{\rm L} \times \U(1)_{\tilde{Y}}}{\Z_{q=3q'}}.
\eea
Namely, when $q'=1$, we have
$
G_{\PS_{1}} \supset  G_{\LR_{1}} \supset G_{\SM_{3}}.
$
Furthermore, only when $q'=2$, we can have the whole embedded into the Spin(10) for the $so(10)$ GUT:
$
{\Spin(10)} \supset G_{\PS_{2}} \supset  G_{\LR_{2}} \supset G_{\SM_{6}}.
$
\end{enumerate}

\newpage
\subsection{Quantum Phase Diagram and a Mother Effective Field Theory}
\label{sec:QuantumPhaseDiagram}

Now we follow \Refe{Wang2106.16248} on 
the proposed parent effective field theory (EFT) as a modified $so(10)$ GUT (with a Spin(10) gauge group)
with a discrete torsion class of WZW term. In \Refe{Wang2106.16248}, we had proposed that
Georgi-Glashow (GG) $su(5)$ and the Pati-Salam (PS) $su(4) \times su(2) \times su(2)$ models
could manifest different low energy phases of the same parent EFT, but overall all of them share the same quantum phase diagram 
(see Figure 5 and 7 in \cite{Wang2106.16248}). By \emph{quantum phase diagram}, we mean to find 
the governing \emph{zero-temperature} quantum ground states in the parameter spaces 
by tuning the QFT coupling strengths.

In our present work, we further include the GG $u(5)$ and the flipped $u(5)$, and the Left-Right (LR) model into this parent EFT 
(see \Fig{fig:embed-Lie-group-spacetime-2109-2}, and \Fig{fig:phase} below).
The parent EFT is basically the same modified $so(10)$ GUT in \Refe{Wang2106.16248}
(except we need to refine the
vev $\langle \Phi_{\bf{45}}^{\text{1st}} \rangle$, $\langle \Phi_{\bf{45}}^{\text{2nd}} \rangle$ and $\langle \Phi_{\bf{54}}^{\text{}} \rangle$ below in 
\Sec{sec:QuantumPhaseDiagram-Internal-global}),
thus we follow exactly the same notations and conventions there in \cite{Wang2106.16248}.
This parent EFT contains the following actions:
\bea
\label{eq:YM-Weyl}
S_{\text{YM-Weyl}} &=& \int_{M^4} \Tr(F_{} \wedge \star F_{}) 
+ \dd^4 x \big( {\psi}^\dagger_L  (\ii {\bar  \sigma}^\mu {D}_{\mu, A} ) \psi_L \big),\\
\label{eq:Higgs}
S_{\text{Higgs}} &=&  \int_{M^4} \dd^4 x \big(| {D}_{\mu, A}\Phi_{{\mathbf{R}}}  |^2 -{\rm U}(\Phi_{{\mathbf{R}}} )\big),\\
\label{eq:Yukawa}
S_{\text{Yukawa}}  &=& 
 \int_{M^4} \dd^4 x \Big(\frac{1}{2}\phi^\intercal \Phi^{\rm{bi}}  \phi
+ \frac{1}{2}\sum_{a=1}^{5}\big(\psi_L^\intercal\ii\sigma^2(\phi_{2a-1}\Gamma_{2a-1}-\ii\phi_{2a}\Gamma_{2a})\psi_L+\text{h.c.}\big)
  \Big),\\
\label{eq:WZW}
S^\text{WZW} &=& 
\frac{1}{\pi}\int_{M^5} \CB({\Phi}_{\bf 54}) \wedge\dd \CC({\Phi}_{\bf 45}) \Big\vert_{{M^4=\prt {M^5}}} =
 \pi \int_{M^5} B(\tilde{\Phi}^{\rm{bi}})\smile \delta C(\hat{\Phi}^{\rm{bi}}) )  \Big\vert_{{M^4=\prt {M^5}}}.
\eea
In summary, we have:\\
$\bullet$ the action of Yang-Mills field strength 2-form $F = \dd A - \ii { g} A \wedge A$ and Weyl fermion $\psi_L$ coupling term $S_{\text{YM-Weyl}}$ in \eq{eq:YM-Weyl},\\
$\bullet$ the Higgs or GUT-Higgs $\Phi_{{\mathbf{R}}}$ action $S_{\text{Higgs}}$ of some representation ${\mathbf{R}}$ in \eq{eq:Higgs},\\
$\bullet$ the Yukawa-like action \eq{eq:Yukawa} coupling between the fermion $\psi_L$ in {\bf 16} of Spin(10),
the GUT-Higgs $\phi$ in the vector {\bf 10} of Spin(10),
and the $\Phi^{\text{bi}}$ in the bivector ${\bf 100} = {\bf 10} \times {\bf 10}$ of Spin(10).
The $\sigma^2$ matrix acts on the 2-component spacetime Weyl spinor $\psi_L$. 
The $\Gamma_a$ (with $a\in \{1,2,\dots,10\}$) are ten rank-16 matrices satisfying $\{\Gamma_{2a-1},\Gamma_{2b-1}\}=2\delta_{ab}, \{\Gamma_{2a},\Gamma_{2b}\}=2\delta_{ab}, [\Gamma_{2a-1},\Gamma_{2b}]=0$ (for $a,b=1,2,\cdots,5$).
So far all the terms listed above, \eq{eq:YM-Weyl}, \eq{eq:Higgs}, and  \eq{eq:Yukawa},
are within the framework of Landau-Ginzburg type of internal symmetry breaking via the Higgs field.
\\
$\bullet$ the WZW term. In order to go beyond the Landau-Ginzburg paradigm to realize a richer quantum criticality between GG and PS models, 
we require to add a WZW term written on a 4d manifold $M^4$ 
as a 5d $M^5$'s boundary.
The WZW term \eq{eq:WZW} is written in terms of $\R$-valued gauge fields,
2-form $\CB({\Phi}_{\bf 54})$ and 2-form $\CC({\Phi}_{\bf 45})$,
or $\Z_2$-valued cohomology class fields
$B(\tilde{\Phi}^{\rm{bi}})$ and $C(\hat{\Phi}^{\rm{bi}})$.
\Refe{KravecMcGreevySwingle1409.8339} verifies first that the specific $B \dd C$ term matches the $w_2(TM) w_3(TM)$ anomaly.
\Refe{Wang2106.16248} later constructs the $B(\tilde{\Phi}^{\rm{bi}})\smile \delta C(\hat{\Phi}^{\rm{bi}}) ) $ out of the GUT-Higgs fields to
matches the $w_2(TM) w_3(TM) = w_2(V_{\SO(10)})w_3(V_{\SO(10)}) $ anomaly.
An $\SO(10)$ real bivector field $\Phi^{\rm{bi}}\in \R$  is obtained from the tensor product of the two $\phi$,
in the ${\bf 10} \otimes {\bf 10} = {\bf 1}_{\rm S}  \oplus {\bf 45}_{\rm A} \oplus {\bf 54}_{\rm S}$ of $so(10)$ also of Spin(10),
with the anti-symmetric (A) and symmetric (S) parts of tensor product: 
 \be\hspace{-2mm} \label{eq:Phi-bi-phi}
\Phi^{\rm{bi}}_{ab} = \phi_{a}\phi_{b} \text{ includes}\left\{\begin{array}{l}
\Tr\Phi^{\rm{bi}}=\sum_{a}\Phi^{\rm{bi}}_{aa} \text{ gives $\Phi_{{\mathbf{R}}}=\Phi_{{\mathbf{1}}}$ in ${\bf 1}_{\rm S}$}. \\ 
\hat{\Phi}^{\rm{bi}}\equiv
\Phi^{\rm{bi}}_{[a,b]} = \tfrac{1}{2}(\Phi^{\rm{bi}}_{ab}-\Phi^{\rm{bi}}_{ba})=\tfrac{1}{2}(\phi_a\phi_b-\phi_b\phi_a)=\tfrac{1}{2}[\phi_a,\phi_b] \text{ gives $\Phi_{{\mathbf{R}}}=\Phi_{{\mathbf{45}}}$ in ${\bf 45}_{\rm A}$} .\\
\tilde{\Phi}^{\rm{bi}}\equiv
\Phi^{\rm{bi}}_{\{a,b\}} = \tfrac{1}{2}(\Phi^{\rm{bi}}_{ab}+\Phi^{\rm{bi}}_{ba})=\tfrac{1}{2}(\phi_a\phi_b+\phi_b\phi_a)=\tfrac{1}{2}\{\phi_a,\phi_b\} \text{ gives $\Phi_{{\mathbf{R}}}=\Phi_{{\mathbf{54}}}$ in ${\bf 54}_{\rm S}$}.
\end{array}\right.
\ee
We construct this WZW term \eq{eq:WZW} under a precise constraint to match the mod 2 class of 4d global gauge-gravitational anomaly captured by the 5d $w_2 w_3$ term.

\pagebreak

\begin{figure}[!t] %[htbp]
\centering
\includegraphics[width=0.72\textwidth]{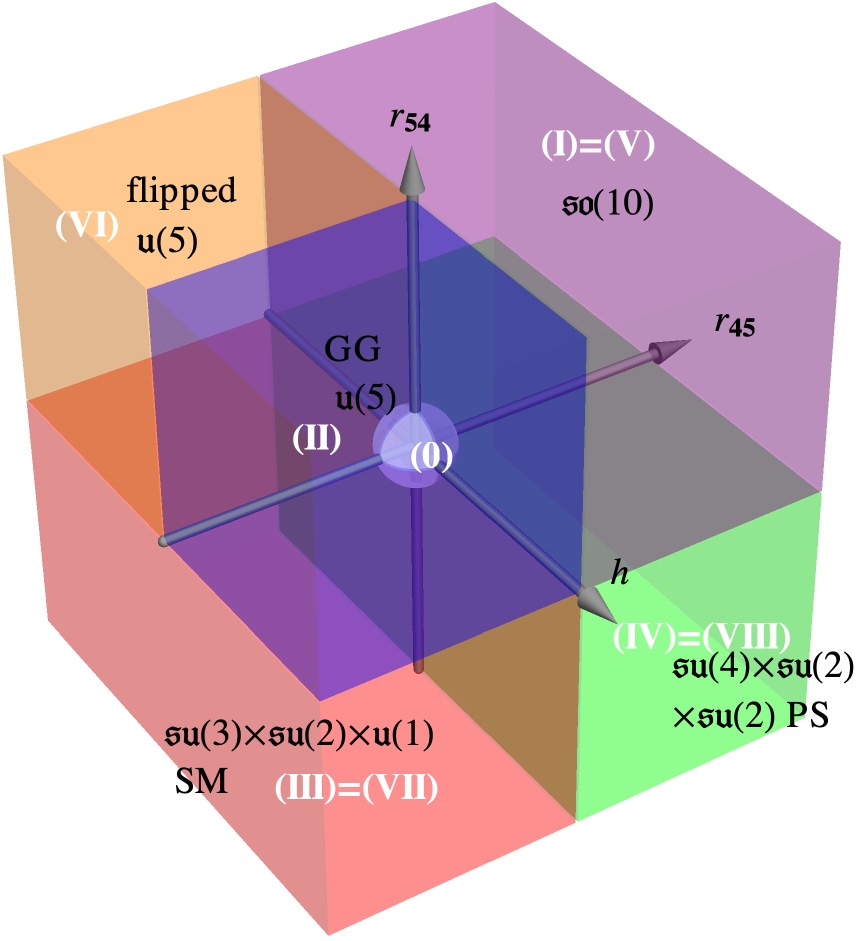}
\caption{The {\bf quantum phase diagram} 
(also the {\bf moduli space} of vacuum expectation values (vevs) of a set of scalar fields)
is separated to eight octants.
The colors of quantum phases are designed to match the colors in \Fig{fig:SM} to \Fig{fig:so10}.
Here the real parameter $r_{\mathbf{R}} \in \R$ denotes the coefficient of the effective quadratic potential 
\eq{eq:U-potential}'s ${\rm U}(\Phi_{{\mathbf{R}}})$
of $\Phi$ field in the representation ${\mathbf{R}}$. 
The corresponding Higgs $\Phi$ field condenses in the representation-${\mathbf{R}}$ if $r_{\mathbf{R}}<0$.
The $\langle \Phi_{\bf{54}}^{\text{}} \rangle \neq 0$ condenses when $r_{\mathbf{54}}<0$.
There are however two distinct $\langle \Phi_{\bf{45}}^{\text{1st}} \rangle$ and $\langle \Phi_{\bf{45}}^{\text{2nd}} \rangle$ condensations for $r_{\mathbf{45}}<0$;
 the two vevs are selected by $h > 0$ and $h<0$ respectively.  
 The quantum phase diagram contains the following phases in the eight octants:
\newline
$\bullet$ {the $so(10)$ GUT (with $\langle \Phi_{\bf{45}} \rangle=\langle \Phi_{\bf{54}} \rangle=0$ 
in the first and fifth octants, labeled by (I) and (V)),}
\newline
$\bullet$ {the Georgi-Glashow $u(5)$ GUT 
(GG, with $\langle \Phi_{\bf{45}}^{\text{1st}} \rangle \neq 0$ but $\langle \Phi_{\bf{54}} \rangle=0$ 
in the second octant (II)),}
\newline
$\bullet$ {the flipped $u(5)$ GUT (with $\langle \Phi_{\bf{45}}^{\text{2nd}} \rangle \neq 0$ but $\langle \Phi_{\bf{54}} \rangle=0$ in the sixth octant (VI)),}
\newline
$\bullet$ {the $su(4) \times su(2)_{\mathrm{L}} \times su(2)_{\mathrm{R}}$ Pati-Salam model (PS, 
with $\langle \Phi_{\bf{54}} \rangle \neq 0$ but $\langle \Phi_{\bf{45}} \rangle=0$ in the fourth and eight octants, (IV) and (VIII)),}
\newline
$\bullet$ {the $su(3)\times su(2)\times u(1)$ Standard Model (SM, 
with both $\langle \Phi_{\bf{45}} \rangle \neq 0$ and $\langle \Phi_{\bf{54}} \rangle \neq 0$,
in the third and seventh octants, (III) and (VII)).}
\newline
$\bullet$ the quantum critical region (around (0) in the white ball region) occurs 
\emph{if} the criticality is enforced by the 4d boundary anomaly on of a 5d $w_2 w_3$ invertible TQFT,
and \emph{if} we had added the WZW term into the modified $so(10)$ GUT,
and \emph{if} %\cred
{the
${\U(1)'}_{\text{gauge}}^{\text{dark}}$ is deconfined, namely its fine structure constant $g'^2$ is below a certain critical value $g_c'^2$,}
{and typically near the origin with small $r_{\mathbf{45}}$ and $r_{\mathbf{54}}$}.
The region outside (0) is shown in \Fig{fig:phase-two} (a), the region (0) inside is shown in \Fig{fig:phase-two} (b).
We summarize the nature of the phase transitions in \Sec{sec:conclude}'s \Table{table:GUT-transition}.
}
\label{fig:phase}
\end{figure}
%
%

%\pagebreak[4]

\clearpage

\begin{figure}[!ht] %[htbp]
\centering
(a)\includegraphics[width=0.4\textwidth]{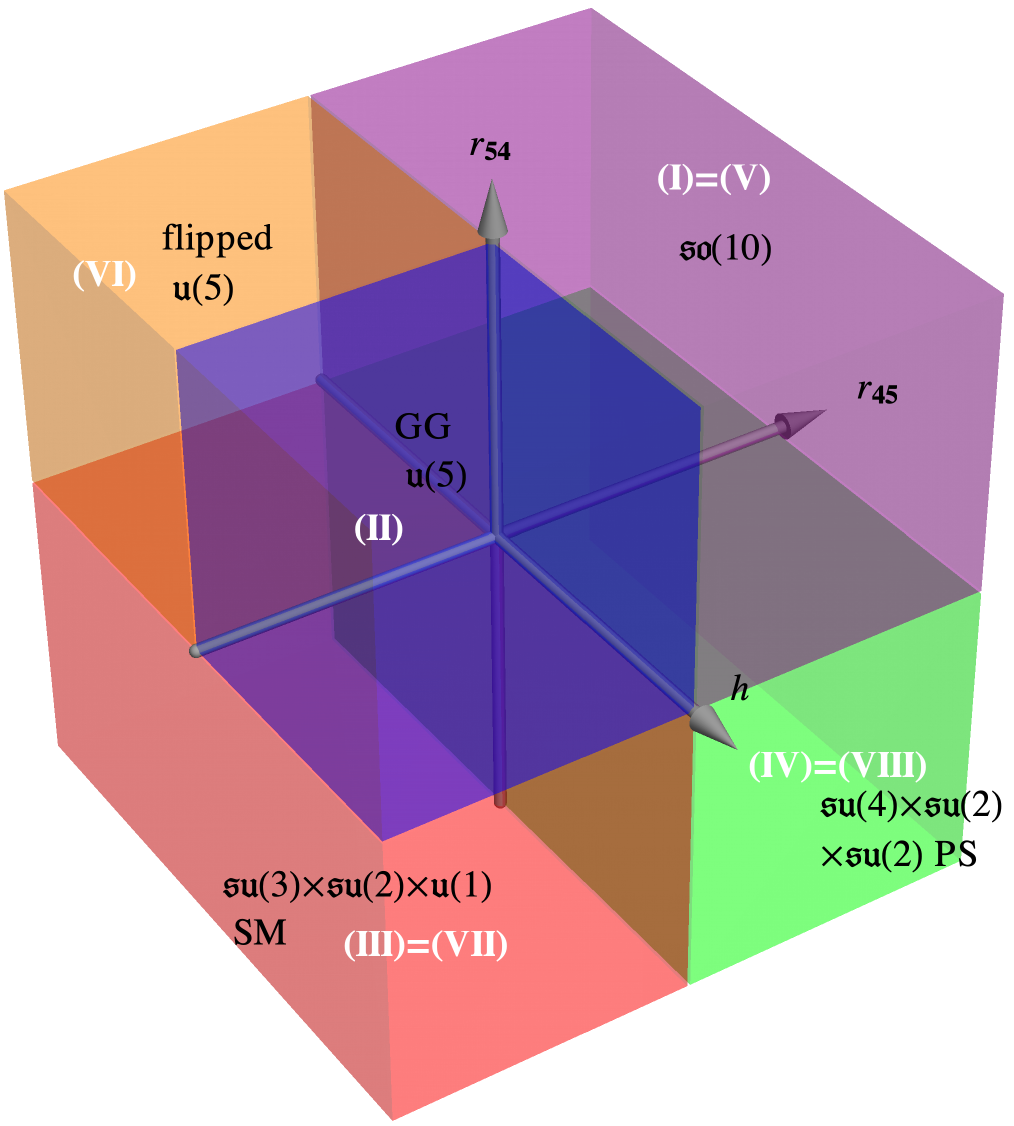}
(b)\includegraphics[width=0.4\textwidth]{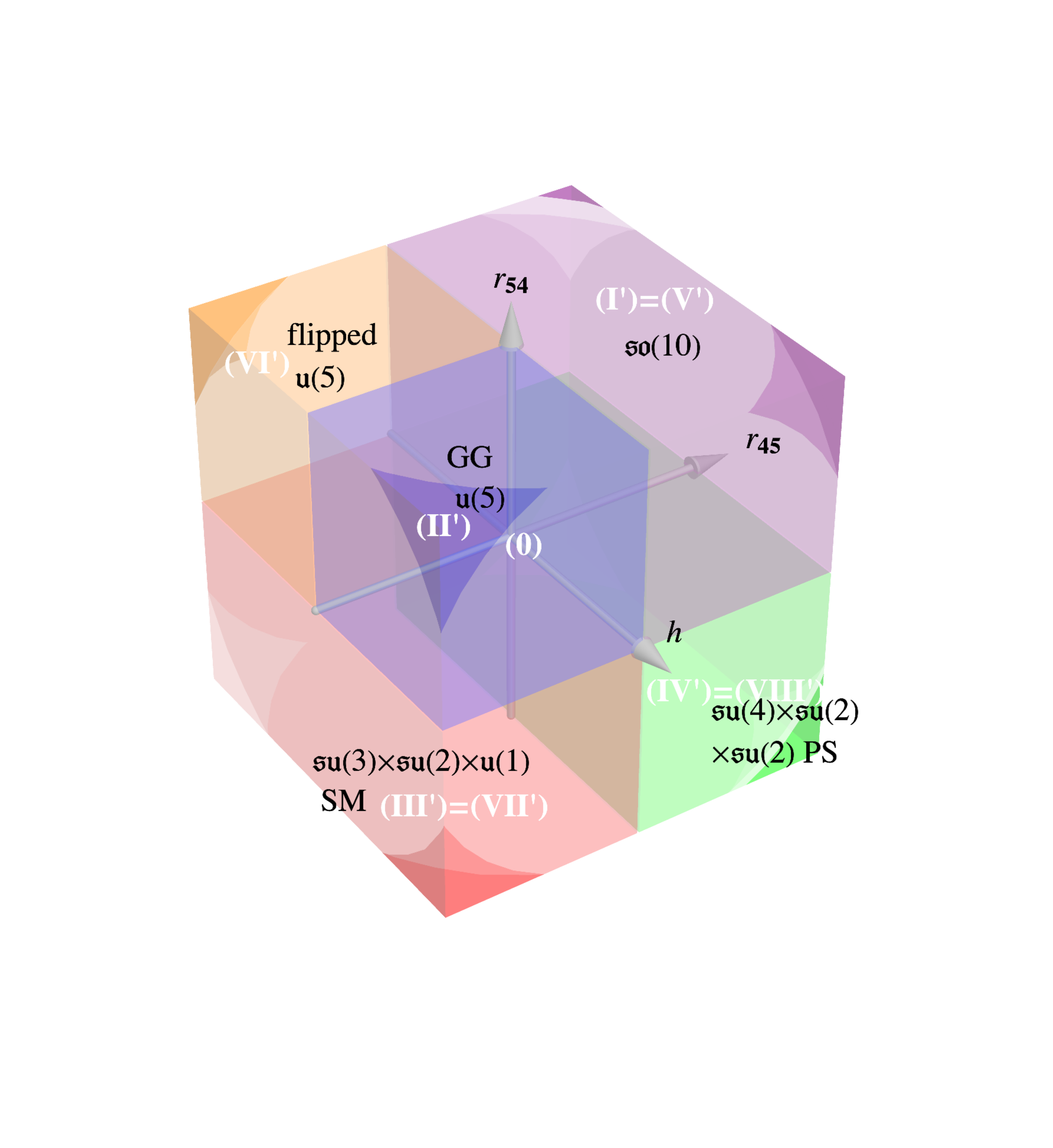}
\caption{{(a) If ${\U(1)'}_{\text{gauge}}^{\text{dark}}$ is confined, namely its fine structure constant $g'^2$ is above a certain critical value,
then we call those eight octants as (I) to (VIII).
(b) If ${\U(1)'}_{\text{gauge}}^{\text{dark}}$ is deconfined with $g'^2$ is below a certain critical value,
then we call those eight octants as (I)$'$ to (VIII)$'$, where they are altogether shown as the white ball region (0) in \Fig{fig:phase},
such that (0)$=$(I)$'+$(II)$'+$(III)$'+$(IV)$'+$(V)$'+$(VI)$'+$(VII)$'+$(VIII)$'$.
We will also summarize the nature of the phase transitions in \Sec{sec:conclude}'s \Table{table:GUT-transition}.}
}
\label{fig:phase-two}
\end{figure}

%\pagebreak

%\newpage 
%\pagebreak[4]

{\bf Manifestation of the WZW term in terms of a fermionic parton theory}: 
Follow \Refe{Wang2106.16248},
we have another realization of WZW term \eq{eq:WZW} by integrating out some massive 5d fermionic parton theory ($|m| \gg 0$,
we denote QED$'$ for it is beyond the ordinary SM's quantum electrodynamics QED sector):
\be \label{eq:QED5}
S^\text{WZW}_\text{QED$'_5$}[\xi,\bar{\xi},a,\Phi,\CB,\CC]
=
\int_{M^5}\bar{\xi}(\ii \tilde \gamma^\mu D'_\mu -m-\tilde \gamma^5\tilde{\Phi}^\text{bi}-\tilde \gamma^6\ii\hat{\Phi}^\text{bi}-\ii\tilde \gamma^5\tilde \gamma^\mu\tilde \gamma^\nu 
\CB_{\mu\nu}-\ii \tilde \gamma^6\tilde \gamma^\mu\tilde \gamma^\nu \CC_{\mu\nu})\xi\;\dd^5 x.
\ee
{The covariant derivative $D'_\mu=\nabla_\mu-\ii a_\mu-\ii g A_\mu$ contains the minimal coupling of the fermionic parton $\xi$ to a new emergent dynamical %\ccred
{${\U(1)'}_{\text{gauge}}^{\text{dark}}$} field $a_\mu$, as well as
the minimal coupling to the $\SO(10)$ gauge field $A_\mu$ %\ccred
{(which is part of the Spin(10) gauge field)}. 
We may treat the gauge field $A_\mu$ as a background field for now, and discuss the dynamically gauged $A_\mu$ later in \Sec{sec:QuantumPhaseDiagram-Internal-gauged}.}
The previously introduced 
two 2-form $\R$ gauge fields $\CB=\CB_{\mu\nu}\dd x^{\mu}\wedge\dd x^{\nu}$ and $\CC=\CC_{\mu\nu}\dd x^{\mu}\wedge\dd x^{\nu}$ couple to the 
{8-component Dirac fermionic parton} $\xi$ (or doubled version of 4-component Dirac fermion) in 5d.
While the 4d interface at $m=0$ appears in between the 5d bulk $m>0$ and $m<0$ phases:
\be \label{eq:QED4}
S^\text{WZW}_{\text{QED}'_4}{[\xi,\bar{\xi},a,\Phi]}=\int_{M^4} \bar{\xi}(\ii\gamma^\mu D'_\mu-\tilde{\Phi}^\text{bi}-\ii\gamma^{\text{FIVE}}\hat{\Phi}^\text{bi})\xi \; \dd^4x,
\ee
now with the {4-component Dirac fermionic parton} $\xi$ in 4d. Some explanations below:

\noindent
$\bullet$ In 4d, 
we can already define five gamma matrices $\gamma^0,\gamma^1,\gamma^2,\gamma^3$, and {$\gamma^{\text{FIVE}} \equiv (\ii \gamma^0\gamma^1\gamma^2\gamma^3)$, 
all are rank-4 matrices.\footnote{Denote $\sigma^{\mu\nu\cdots} \equiv \sigma^\mu\otimes\sigma^\nu\otimes\cdots$ as the direct product of the standard Pauli matrices.
Explicit matrix representation of 4d gamma matrices are
$$\gamma^0=  \sigma^{10},\gamma^1=  \ii \sigma^{21},\gamma^2=  \ii \sigma^{22},\gamma^3=  \ii \sigma^{23}, \gamma^{\text{FIVE}} 
\equiv (\ii \gamma^0\gamma^1\gamma^2\gamma^3)= -\sigma^{30}.$$
\label{ft:4dgamma}} 
The 4d Dirac fermion $\xi$ in \eq{eq:QED4} is a 4-component complex fermion 
${\bf 2}_L  \oplus {\bf 2}_R$ of $\Spin(1,3)$. The $\xi$ is also in the 10-dimensional vector representation of $so(10)$ or Spin(10).
Namely, the 4d Dirac fermion $\xi$ in \eq{eq:QED4} is overall in the following rep:
\bea \label{eq:4d-fermion}
 \text{(${\bf 2}_L  \oplus {\bf 2}_R)$ of ${\Spin(1,3)}$)}  \times \text{(${\bf 10}$ of ${\Spin(10)}$)}. 
\eea

\noindent
$\bullet$ 
However, the 4d parton theory has some extra symmetry that are not presented in the original theory, including:
(i) $\U(1)':\xi\to e^{\ii \theta}\xi$, 
(ii) $\Z_2^{\text{CP}'}:\xi(t,\vec{x})\to\gamma^0\gamma^\text{FIVE}\xi^*(t,-\vec{x})$, 
(iii) $\Z_2^{\text{T}'}:\xi(t,\vec{x})\to\mathcal{K} \gamma^0\gamma^\text{FIVE}\xi(-t,\vec{x})$
{with the complex conjugation operation $\mathcal{K}$ sending $\ii \to - \ii$.} 
Note that the CP$'$ and T$'$ symmetries are for partons from the fractionalization of the WZW term (\eq{eq:WZW} and \eq{eq:QED4}), 
which are unrelated to the CP and T symmetries of the chiral fermions in the GUTs. The massless Dirac fermion are two Weyl fermions:  $\xi = \xi_\text{L} + \xi_\text{R}$.\\
(i) {The $\U(1)'$ symmetry forbids any Majorana mass of the form $\xi_\text{L/R}^{\rm T}\ii\sigma^2\xi_\text{L/R}$ that potentially gaps out the Dirac fermion $\xi$}.\\ 
(ii) Under the CP$'$ symmetry, $\Z_2^{\text{CP}'}:\bar{\xi}\xi\to-\bar{\xi}\xi$. The $\bar{\xi}\xi$ mass is forbidden by the CP$'$ symmetry. \\
(iii) Under the T$'$ symmetry, $\Z_2^{\text{T}'}: \ii \bar{\xi}\gamma^\text{FIVE}\xi\to-\ii \bar{\xi}\gamma^\text{FIVE}\xi$. 
The $\ii \bar{\xi}\gamma^\text{FIVE}\xi$ mass is forbidden by T$'$. \\
Therefore in the presence of these $\U(1)'$, CP$'$, and T$'$ symmetries,
the fermionic partons must {not be gapped by quadratic fermion mass terms (thus the quadratic theory is gapless)}.
{This explains why the Dirac fermionic parton $\xi$ in the vector rep ${\bf 10}$ of $so(10)$ can have a $w_2 w_3$ anomaly \cite{Wang2106.16248} 
(which is checked based on the new SU(2) anomaly trick \cite{WangWenWitten2018qoy1810.00844}) 
---  although it looks that this vector rep can have quadratic mass ($\xi_\text{L/R}\ii\sigma^2\xi_\text{L/R}$, $\bar{\xi}\xi$ or $\ii \bar{\xi}\gamma^\text{FIVE}\xi$)
gapping all $\xi$ out (which seems n\"aively suggests they cannot have any anomaly), the enforced 
 $\U(1)'$, CP$'$, and T$'$ symmetries can actually protect $\xi$ from adding those mass terms.} 
% \ccred{These CP$'$ and T$'$ symmetries also exhibit a C$'$-P$'$-T$'$ fractionalization structure \cite{WangCPT2109.15320} here with a group 
%  $\Z_2^{\text{CP}'} \times \Z_2^{\text{T}'} \times {\Z_2^F}$ or
% $\Z_4^{\text{CP}'} \times_{\Z_2^F} \Z_4^{\text{T}'}$.}
 Because these  $\U(1)'$, CP$'$, and T$'$ symmetries are not physical symmetries of the original WZW term \eq{eq:WZW}, these symmetries must all be dynamically gauged.

\noindent
$\bullet$ In 5d, 
we define five gamma matrices $\tilde \gamma^0, \tilde \gamma^1, \tilde \gamma^2, \tilde \gamma^3$, and $\tilde \gamma^4$.
{However, the 5d gamma matrices have different matrix representations than the 4d gamma matrices --- they are related by the dimensional reduction on the domain wall normal to the $x_1$ direction.}
By doubling the fermion content, we are able to introduce two more gamma matrices, denoted $\tilde \gamma^5$ and $\tilde \gamma^6$, such that all seven gamma matrices $\tilde \gamma^0,\cdots ,\tilde \gamma^6$ 
are rank-8 matrices satisfying the Clifford algebra relation $\{\tilde \gamma^\mu,\tilde \gamma^\nu\}=2\delta^{\mu\nu}$.\footnote{In contrast to footnote \ref{ft:4dgamma}, 
explicit matrix representations of 5d gamma matrices are
$$\tilde \gamma^0=  \sigma^{200}, \tilde \gamma^1=  \ii \sigma^{300}, \tilde \gamma^2=  \ii \sigma^{131}, \tilde \gamma^3=  \ii \sigma^{132}, \tilde \gamma^4=  \ii \sigma^{133},
\tilde \gamma^5=  \ii \sigma^{110}, \tilde \gamma^6 =  \ii \sigma^{120}.$$} 
So the 5d Dirac fermion $\xi$ in \eq{eq:QED5}
is a doubled version of 4-component complex fermion (${\bf 4}$ of ${\Spin(1,4)}$), 
as the 8-component complex fermion. 
The $\xi$ is also in the 10-dimensional vector representation of $so(10)$ or Spin(10).
Namely, the 5d Dirac fermion $\xi$ in \eq{eq:QED5} is in the following rep:
\bea \label{eq:5d-fermion}
2 \times \text{(${\bf 4}$ of ${\Spin(1,4)}$)}  \times \text{(${\bf 10}$ of ${\Spin(10)}$)}. 
\eea
Notice that the 5d bulk fermions \eq{eq:5d-fermion} have doubled components of the 4d interface fermions \eq{eq:4d-fermion}.

\noindent
$\bullet$ For the 5d bulk theory, if we define the $m > 0$ as a trivial gapped vacuum (say at $x>0$), then one can check that 
the $m< 0$ side (say at $x < 0$) might be a nontrivial gapped vacuum with a low energy invertible TQFT 
describing either gapped invertible topological order or gapped symmetry-protected topological states (SPTs) \cite{Chen2011pg1106.4772}
in quantum matter. 
Indeed, the \eq{eq:WZW}'s partition function
$\exp(\ii S^\text{WZW})$ can match with $\exp(\ii \pi \int w_2 w_3)$
 in a closed 5d bulk without a boundary, thus it describes the invertible TQFT $w_2 w_3$.
The WZW term \eq{eq:WZW} also gives a 4d interface description,
that lives on a 4d boundary of a 5d bulk.

Below we explore quantum phases and their criticalities or phase transitions in \Fig{fig:phase} by two aspects:\\
(1) when the {internal symmetries are treated as global symmetries} (as toy models) in \Sec{sec:QuantumPhaseDiagram-Internal-global}, 
and \\
(2) when the {internal symmetries are dynamically gauged} (as they are gauged in our real-world quantum vacuum) in \Sec{sec:QuantumPhaseDiagram-Internal-gauged}.

\subsubsection{Internal symmetries treated as global symmetries}
\label{sec:QuantumPhaseDiagram-Internal-global}

In the limit when we treat their {\bf internal symmetries as global symmetries} (then the Yang-Mills gauge field $A$ is only a non-dynamical background field), 
the GG, the flipped, PS and LR models match the global gauge-gravitational $w_2w_3(TM) = w_2w_3(V_{so(10)})$ anomaly via the internal symmetry-breaking 
from Spin(10) to each individual subgroup (as the breaking pattern in \Fig{fig:embed-Lie-group-spacetime-2109-2}).
The corresponding QFTs in \Fig{fig:embed-Lie-group-spacetime-2109-2} 
again manifest different low energy phases of the same parent EFT, but overall all of them share the same quantum phase diagram.
By sharing the same quantum phase diagram, we mean that they have the same Hilbert space and the same 't Hooft anomaly constraints at a deeper UV.
In other words, they are in the same \emph{deformation class of QFTs}, particularly advocated by 
Seiberg \cite{NSeiberg-Strings-2019-talk}.\footnote{In fact, many works on gapping the anomaly-free chiral fermions are based on the same logic: The chiral fermions
that are free from 't Hooft anomalies of the chiral symmetry $G$, must be gappable without any chiral symmetry breaking in $G$, via the $G$-symmetry preserving interaction deformations.
See a series of work along this direction and references therein: 
Fidkowski-Kitaev \cite{FidkowskifSPT2} in 0+1d, 
Wang-Wen \cite{Wang2013ytaJW1307.7480,Wang2018ugfJW1807.05998} for gapping chiral fermions in 1+1d,
You-He-Xu-Vishwanath \cite{YouHeXuVishwanath1705.09313, YouHeVishwanathXu1711.00863} in 2+1d,
and notable examples in 3+1d by Eichten-Preskill \cite{Eichten1985ftPreskill1986},
Wen \cite{Wen2013ppa1305.1045}, 
You-BenTov-Xu \cite{You2014oaaYouBenTovXu1402.4151, YX14124784}, 
BenTov-Zee \cite{BenTov2015graZee1505.04312},
Kikukawa \cite{Kikukawa2017ngf1710.11618}, 
Wang-Wen \cite{WangWen2018cai1809.11171},
Razamat-Tong \cite{RazamatTong2009.05037, Tong2104.03997}, 
Catterall et al \cite{Catterall2020fep, CatterallTogaButt2101.01026}, etc. The techniques of gapping chiral fermions can be used in gapping the mirror sector.}

\begin{enumerate}[leftmargin=.mm] 
\item
Starting from the $so(10)$ GUT (modified with WZW term or not), 
the condensation of $\Phi_\mathbf{45}$ or/and $\Phi_\mathbf{54}$ will drive the symmetry breaking transitions to various lower energy 
and lower internal symmetry phases,  
summarized in \Fig{fig:embed-Lie-group-spacetime-2109-2}. 
In particular, if $\Phi$ condenses to a specific configuration $\langle\Phi\rangle$, 
the original Lie algebra $\g_\text{large}$ will be broken to its {subalgebra that commutes with} $\langle\Phi\rangle$, 
given by 
\bea
\g_\text{small}=\z_{\g_\text{large}}(\langle\Phi\rangle)\equiv\{{\rm T} \in\g_\text{large}|[{\rm T},\langle\Phi\rangle]=0\}. 
\eea
To realize the symmetry breaking from the Lie algebra $\g_{so(10)} \equiv so(10)$,
\eqs{\g_{so(10)} &\to\g_\text{PS} \equiv su(4) \times su(2)_{\rm L} \times su(2)_{\rm R}, \\
\g_{so(10)}&\to\g_\text{GG} \equiv su(5)^{\text{1st}} \times u(1)_{X_1},\\
\g_{so(10)}&\to\g_\text{flipped} \equiv su(5)^{\text{2nd}} \times u(1)_{X_2},
} 
we must have the relations
\eqs{\z_{so(10)}(\langle\Phi_\mathbf{54}\rangle)&=\g_\text{PS},\\
\z_{so(10)}(\langle\Phi_\mathbf{45}^{\text{1st}}\rangle)&=\g_\text{GG},\\
\z_{so(10)}(\langle\Phi_\mathbf{45}^{\text{2nd}}\rangle)&=\g_\text{flipped},
}
whose solutions reads
\eqs{\label{eq:Phi_config}\langle\Phi_\mathbf{54}\rangle&\propto\smat{-3&&&&\\&-3&&&\\&&2&&\\&&&2&\\&&&&2}\otimes\smat{1&\\&1},\\
\langle\Phi_\mathbf{45}^{\text{1st}}\rangle&\propto \smat{1&&&&\\&1&&&\\&&1&&\\&&&1&\\&&&&1}\otimes\smat{0&1\\-1&0},\\
\langle\Phi_\mathbf{45}^{\text{2nd}}\rangle&\propto \smat{-1&&&&\\&-1&&&\\&&1&&\\&&&1&\\&&&&1}\otimes\smat{0&1\\-1&0}.\\
}
$\bullet$ The $\langle\Phi_\mathbf{54}\rangle$ explicitly distinguishes the first four-dimensional subspace from the last six-dimensional subspace of the $so(10)$ vector, therefore breaking $\g_{so(10)}=so(10)$ down to 
$\g_\text{PS}=so(6)\times so(4)$. \\
$\bullet$ The $\langle\Phi_\mathbf{45}^{\text{1st}}\rangle$ is proportional to the $u(1)_{X_1}$ generator, 
which effectively requires the unbroken generators to commute with $u(1)_{X_1}$ generator. 
This singles out $\g_\text{GG}=\z_{so(10)}(u(1)_{X_1})=su(5)^{\text{1st}}\times u(1)_{X_1}$. \\
$\bullet$ {The $\langle\Phi_\mathbf{45}^{\text{2nd}}\rangle$ can be obtained via the $\Z_2^{\text{flip}}$ transformation on the  $\langle\Phi_\mathbf{45}^{\text{1st}}\rangle$,
where $\Z_2^{\text{flip}}$ is described in \Sec{sec:SM-GUT-table} and Appendix \ref{app:Flipping-Isomorphism}.
The $\langle\Phi_\mathbf{45}^{\text{2nd}}\rangle$ is proportional to the $u(1)_{X_2}$ generator, 
which effectively requires the unbroken generators to commute with $u(1)_{X_2}$ generator. 
This singles out $\g_\text{GG}=\z_{so(10)}(u(1)_{X_2})=su(5)^{\text{2nd}}\times u(1)_{X_2}$.} \\
$\bullet$  Using  \eq{eq:Phi_config}, one can further verify that
\eqs{\z_{\g_\text{PS}}(\langle\Phi_\mathbf{45}^{\text{1st}}\rangle)=
\z_{\g_\text{PS}}(\langle\Phi_\mathbf{45}^{\text{2nd}}\rangle)
=\z_{\g_\text{GG}}(\langle\Phi_\mathbf{54}\rangle)
=\z_{\g_\text{flipped}}(\langle\Phi_\mathbf{54}\rangle)=\g_\text{SM},}
which explicitly confirms that the simultaneous condensation of $\Phi_\mathbf{45}$ (any of $\Phi_\mathbf{45}^{\text{1st}}$ and $\Phi_\mathbf{45}^{\text{2nd}}$)
and $\Phi_\mathbf{54}$ indeed breaks the internal symmetry to $\g_\text{SM}$.
{These results also agree with the representation data and branching rules listed in \cite{Slansky1981, 1912.10969LieART, Yamatsu1511.08771}.}
%\item Manifestation of the WZW term in terms of a fermionic parton theory: \\
 
\item {\bf Quantum criticalities and phase transitions due to GUT-Higgs, with or without WZW term}:\\
In \Fig{fig:phase} and its caption, we have enumerated all the ground states in all the eight octants of the phase diagram
descended from the 4d parent theory. In particular, the 4d phases (in the bulk portion of the phase diagram) maintain regardless of whether we add
the WZW term $S^\text{WZW}$ into the Landau-Ginzburg type of 4d parent theory of the action:
\bea
S_{\text{YM-Weyl}} +S_{\text{Higgs}} +S_{\text{Yukawa}}.  
\eea
The phase transition (between the eight octants in \Fig{fig:phase})
is triggered by the following GUT-Higgs potential
${\rm U}(\Phi_{{\mathbf{R}}})$ appeared in \eq{eq:Higgs}, for example,\footnote{Here we extract
${h  \; \Phi_{\bf{45}} \cdot (\langle \Phi_{\bf{45}}^{\text{1st}} \rangle - \langle \Phi_{\bf{45}}^{\text{2nd}} \rangle)}$
from
 $h' \Big(  (\Phi_{\bf{45}} - \langle \Phi_{\bf{45}}^{\text{1st}}  \rangle)^2 - (\Phi_{\bf{45}} - \langle \Phi_{\bf{45}}^{\text{2nd}} \rangle)^2  \Big) + \dots$.}
\be \label{eq:U-potential}
{\rm U}(\Phi_{{\mathbf{R}}}) =\Big( r_{{\mathbf{45}}}^{} (\Phi_{{\mathbf{45}}}^{})^2 +\lambda_{{\mathbf{45}}}^{} (\Phi_{{\mathbf{45}}}^{})^4\Big)+ 
\Big(r_{{\mathbf{54}}} (\Phi_{{\mathbf{54}}})^2 +\lambda_{{\mathbf{54}}} (\Phi_{{\mathbf{54}}})^4\Big) + 
{h  \; \Phi_{\bf{45}} \cdot (\langle \Phi_{\bf{45}}^{\text{1st}} \rangle - \langle \Phi_{\bf{45}}^{\text{2nd}} \rangle)} + \dots.
\ee

Although the 4d phases in all the eight octants are \emph{not sensitive} to the WZW term,
the {\bf quantum critical region} (labeled as (0)) and the {\bf  phase boundaries} between the eight octants 
are \emph{highly sensitive} to the WZW term. In those critical regions, we must examine the nature of criticality 
by looking into the full action
\bea
S_{\text{YM-Weyl}} +S_{\text{Higgs}} +S_{\text{Yukawa}}+ S^\text{WZW}
=S_{\text{YM-Weyl}} +S_{\text{Higgs}} +S_{\text{Yukawa}}+ S^\text{WZW}_\text{QED$'_5$|QED$'_4$}.  
\eea 
In particular, in the right hand side of equality, 
we focus on a specific scenario that the low energy physics of WZW is manifested by the 5d bulk/4d interface QED$'$ written as
$S^\text{WZW}_\text{QED$'_5$|QED$'_4$}$ from \eq{eq:QED5} and \eq{eq:QED4}, with deconfined emergent dark gauge fields of $\U(1)'$ only near the critical region.

Below in Remark \ref{remark:LG} and \ref{remark:LG-beyond} respectively,
we describe the Landau-Ginzburg type criticalities (not sensitive to WZW term),
and the beyond Landau-Ginzburg type criticalities (sensitive to WZW term and anomaly constraints).

\item How many distinct phase transitions there are in \Fig{fig:phase}? We can enumerate those occur in the $h >0$ side
with the four quadrants (I), (II), (III), and (IV) shown in \Fig{fig:phase-quadrant-1}.
Then we can enumerate those occur in the $h <0$ side
with the last four quadrants (V), (VI), (VII), and (VIII) shown in \Fig{fig:phase-quadrant-2}.

In \Fig{fig:phase-quadrant-1}, there are four phase transitions between these four phases: 
 (I)-(II), (II)-(III), (III)-(IV), and (IV)-(I). 
 There are also another four phase transitions from either of
 these four phases into the quantum critical region in (0):
Namely, (I)-(0), (II)-(0), (III)-(0), and (IV)-(0).
So there are totally eight possible phase transitions in \Fig{fig:phase-quadrant-1}.

In \Fig{fig:phase-quadrant-2},
there are four phase transitions between these four phases: 
 (V)-(IV), (VI)-(VII), (VII)-(VIII), and (VIII)-(V). 
 There are also another four phase transitions from either of
 these four phases into the quantum critical region in (0):
Namely, (V)-(0), (VI)-(0), (VII)-(0), and (VIII)-(0).
So there are also totally eight possible phase transitions in \Fig{fig:phase-quadrant-2}.

%\pagebreak%\newpage

\begin{figure}[!t] %[htbp]
\centering
\includegraphics[width=0.72\textwidth]{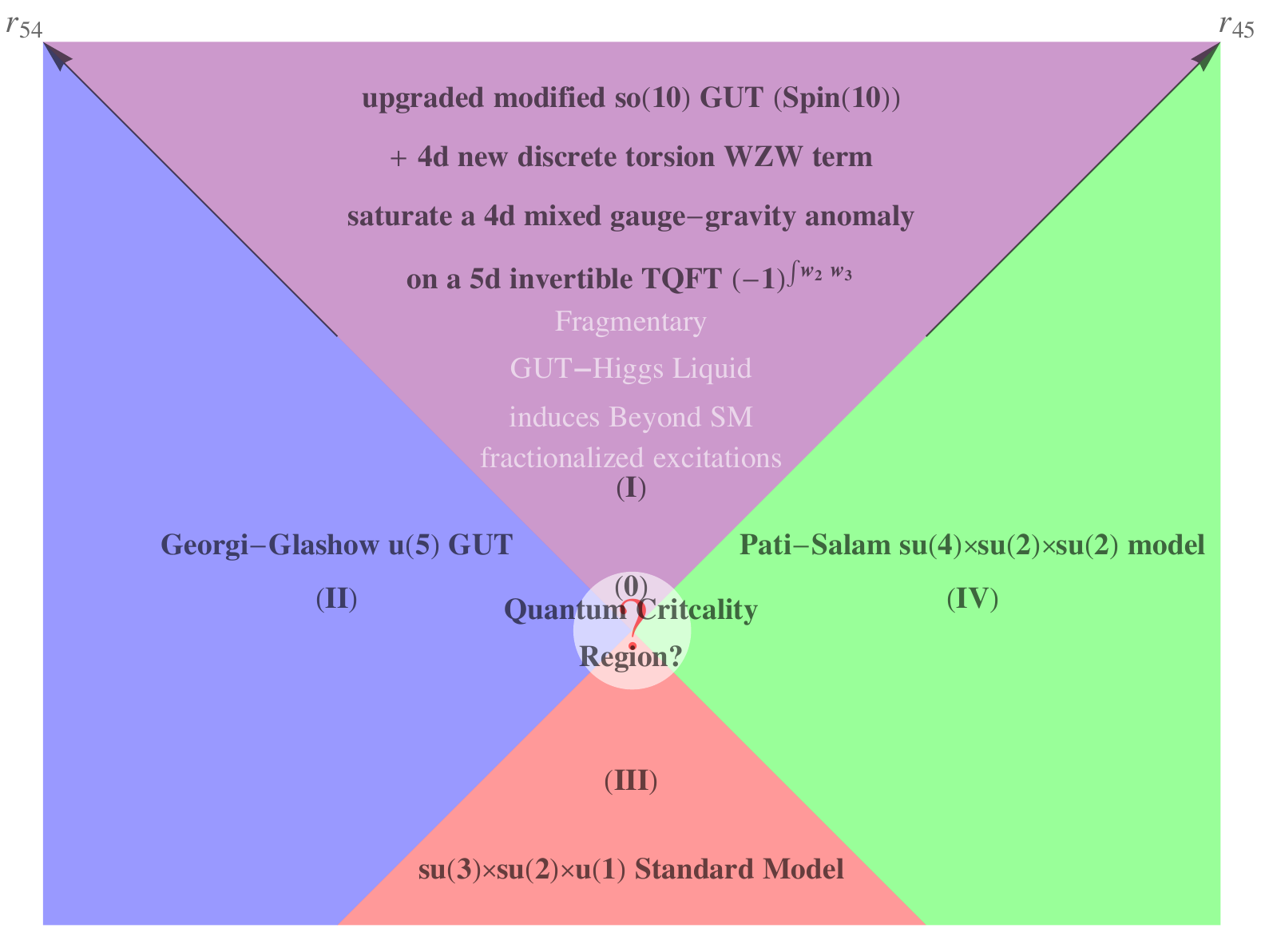} 
%{GEQCP2d-so10-quadrant-GG-corner.pdf}
\caption{A typical $h>0$ slice of \Fig{fig:phase}'s quantum phase diagram. 
The coordinates are potential tuning parameters $r_{{\mathbf{45}}}$ and $r_{{\mathbf{54}}}$ of \eq{eq:U-potential}.
}
\label{fig:phase-quadrant-1}
\end{figure}

\begin{figure}[!t] %[htbp]
\centering
\includegraphics[width=0.72\textwidth]{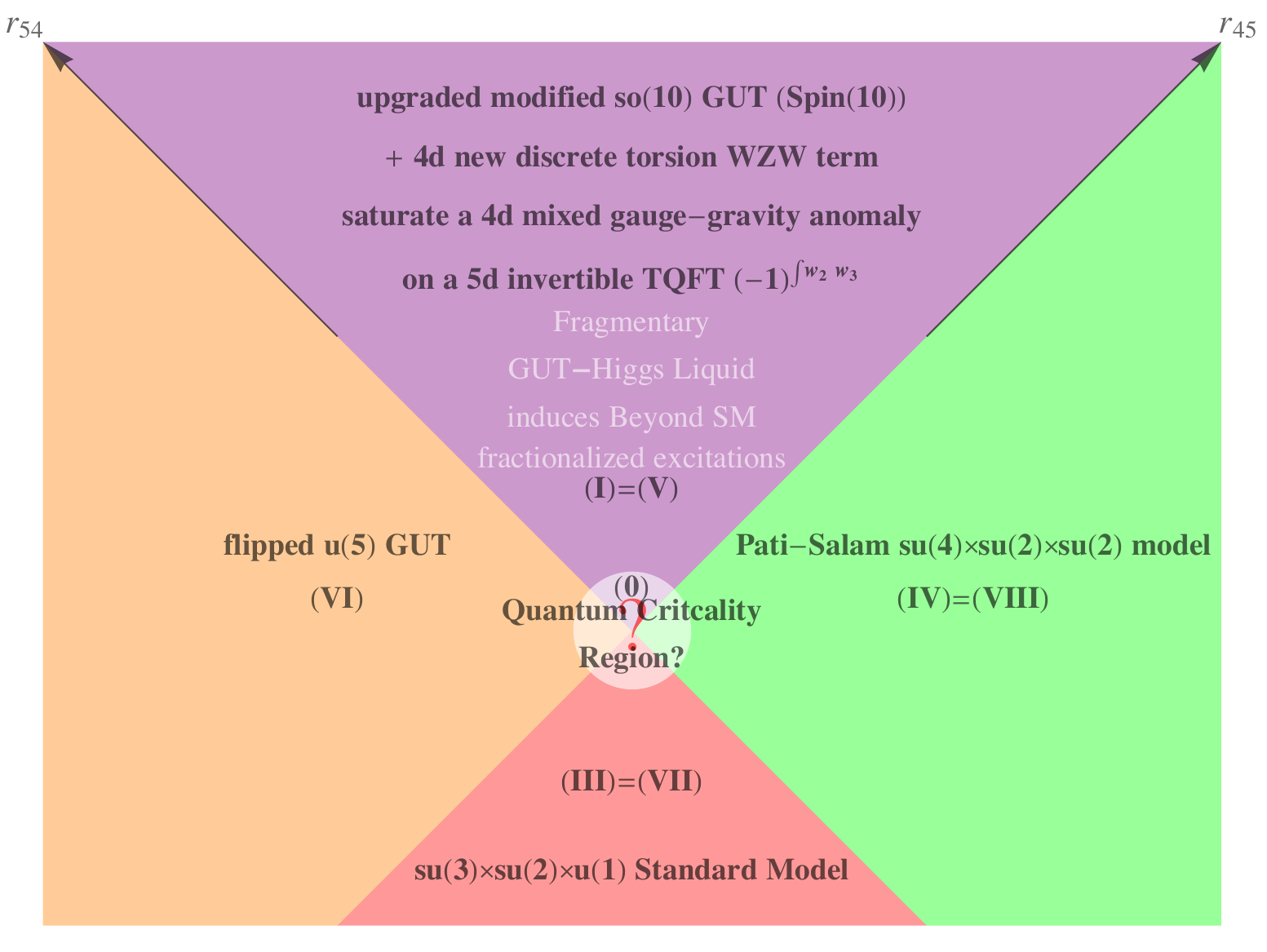} 
\caption{A typical $h<0$ slice of \Fig{fig:phase}'s quantum phase diagram.
The coordinates are potential tuning parameters $r_{{\mathbf{45}}}$ and $r_{{\mathbf{54}}}$ of \eq{eq:U-potential}.
}
\label{fig:phase-quadrant-2}
\end{figure}

Furthermore, the phase structures show many phases are the same: (I)=(V), (III)=(VII), and (IV)=(VIII).
The differences between (II) and (VI) are simply two different Landau-Ginzburg symmetry-breaking vacua.
Thus all phase transitions in \Fig{fig:phase-quadrant-1} have the exactly same nature as those in \Fig{fig:phase-quadrant-2}.
There is only one more phase transition between (II) and (VI), that is not shown in \Fig{fig:phase-quadrant-1} or \Fig{fig:phase-quadrant-2}. 
So {\bf  totally there are nine distinct possible phase transitions in \Fig{fig:phase}} that we will enumerate.

{Even more precisely, the stable quantum critical region (0) can also have distinct symmetry breaking orders,
so we can further precisely denote the dark gauge $\U(1)'$-deconfined
critical region (0) as
\bea \label{eq:critical-region}
\text{(0) $=$ (I)$'+$(II)$'+$(III)$'+$(IV)$'+$(V)$'+$(VI)$'+$(VII)$'+$(VIII)$'$}.
\eea
Each of them has the symmetry breaking orders from the $\U(1)'$-confined region (mentioned previously as (I) to (VIII)).
Thus, we also have the identifications of some deconfined critical phases:
(I)$'=$(V)$'$, (III)$'=$(VII)$'$, and (IV)$'=$(VIII)$'$.}

%\newpage
Here are some terminologies for the type of phase transitions:\\
(1) {\bf \emph{Landau-Ginzburg}} type: Based on the original symmetry group (kinematic) broken down to an unbroken symmetry group (dynamics) via a symmetry-breaking order parameter $\< \CO_{\text LG} \>$. \\
(2) {\bf \emph{Beyond Landau-Ginzburg}} type: Cannot be characterized via merely symmetry-breaking order parameters. e.g., phase transitions involving topological terms (e.g., WZW term), 
SPTs, intrinsic topological orders, or the 't Hooft anomaly matching differently on two sides of phases, etc. \\
(3) {\bf \emph{Wilson-Fisher}} type: Phase transition due to a scalar field condensation $\< \Phi \> \neq 0$, 
especially via $r \Phi^2 + \lambda \Phi^4$ type potential. However, {Wilson-Fisher} fixed point requires the beyond-mean-field 
RG correction to the Gaussian fixed point.\\
(4) {\bf \emph{Gross-Neveu type}} type: Typically the Wilson-Fisher type with additional Yukawa coupling between fermions $\chi$ and scalars $\Phi$ as $\chi^\dagger \Phi \chi$, again a phase transition due to a
condensation $\< \Phi \> \neq 0$.\\
(5) {\bf Order of phase transition}: \\
For a minimal positive $N$, if the $N$th derivative of the free energy (of QFT)
with respect to the driving parameter is discontinuous at the transition, 
it is called the $N$th-order phase transition.\\
$\bullet$ First-order transition is also called a discontinuous transition, 
while the correlation length remains finite and no additional gapless excitations appear at the transition.\\
$\bullet$ Second-order and higher-order is called a continuous transition,
while the correlation length diverges to infinite\footnote{Here the correlation function for Landau-Ginzburg paradigm
is typically the two-point correlator $\< \CO(x_1) \CO(x_2) \>$ of local order parameters of individual spacetime points. 
In contrast, the correlation function for topological order beyond the Landau-Ginzburg paradigm
is the correlator of strings or higher-dimensional extended operators.
} 
and additional gapless excitations (thus called critical, sometimes described by conformal field theory) 
appear at the transition.\\
This above definition is applicable to Landau-Ginzburg as well as beyond Landau-Ginzburg paradigm.

If the transition happens to be within Landau-Ginzburg paradigm, then:\\ 
$\bullet$ First-order means the order parameter has a discontinuous jump at the transition.\\
$\bullet$ Second-order and higher-order means the order parameter changes continuous without a jump at the transition.

\item \label{remark:LG}
{\bf Landau-Ginzburg type {criticalities and phase transitions}}:

$\bullet$ {\bf Phase transition between the octant (II) to (VI)}:

This is the phase transition between the {\bf GG $u(5)$ and flipped $u(5)$ GUTs}.
In both the octant (II) and (VI), we have $r_{{\mathbf{45}}}^{}<0$.
The phase transition between (II) and (VI) is 
triggered by $h>0$ and $h<0$, causing $\langle \Phi_{\bf{45}}^{\text{1st}} \rangle \neq 0$ or 
$\langle \Phi_{\bf{45}}^{\text{2nd}} \rangle \neq 0$. In general, by tuning $h$,
these condensations jump from zero to nonzero, thus it is the 
{\bf  first-order phase transition of traditional Landau-Ginzburg symmetry-breaking type with the order parameter discontinuity}.
The WZW term does not play any role in the phase transition. 
This also means there is no critical gapless mode directly associated with this phase transition.

$\bullet$ {\bf Phase transition between the octant (II) to (III), similar to (VI) to (III)}:

This is the phase transition between the {\bf GG $u(5)$ GUT and SM} 
(similarly, the transition between the {\bf flipped $u(5)$ GUT and SM}).
The phase transition is triggered by tuning $r_{{\mathbf{54}}}^{}>0$ to $r_{{\mathbf{54}}}^{}<0$ 
of $r \Phi^2 + \lambda \Phi^4$ 
in \eq{eq:U-potential}.
It is the  {\bf continuous phase transition of
Wilson-Fisher type Landau-Ginzburg symmetry-breaking type}.

$\bullet$ {\bf Phase transition between the octant (IV) to (III)}:

This is the phase transition between the {\bf PS and SM}.
The phase transition is triggered by tuning $r_{{\mathbf{45}}}^{}>0$ to $r_{{\mathbf{45}}}^{}<0$ 
of $r \Phi^2 + \lambda \Phi^4$ 
in \eq{eq:U-potential}.
It is again the {\bf continuous phase transition of
Wilson-Fisher type Landau-Ginzburg symmetry-breaking type}.

\item  \label{remark:LG-beyond}
{\bf  Beyond Landau-Ginzburg type criticalities and phase transitions}: With the WZW term,
the criticality between the GG and the PS,
and the criticality between the flipped $u(5)$ and the PS,
both of these criticalities are governed by the Beyond Landau-Ginzburg paradigm.
The critical regions are drawn in the white region around the origin in \Fig{fig:phase-quadrant-1} or \Fig{fig:phase-quadrant-2}.

$\bullet$ {\bf Phase transition between the octant (I) to (II), similar to (I) to (VI)}:\\
The WZW term and the anomaly of 5d $w_2w_3$ invertible TQFT play a crucial role in (I).
The 4d phase transition from the modified $so(10)$ (I) to GG (II) (similarly, the $so(10)$ (I) to the flipped (VI))
is a boundary phase transition on the 5d bulk $w_2w_3$.
The phase transition is triggered not merely by tuning $r_{{\mathbf{45}}}^{}>0$ to $r_{{\mathbf{45}}}^{}<0$ 
of $r \Phi^2 + \lambda \Phi^4$ 
in \eq{eq:U-potential},
but also by the symmetry breaking to cancel the anomaly on the 4d boundary of 5d invertible TQFT (when entering from (I) to (II) or to (VI)). 
Overall, it is the {\bf continuous phase transition of
Wilson-Fisher type but Beyond-Landau-Ginzburg paradigm due to the %\cred
{anomaly matching via the symmetry breaking on the 4d boundary of
5d invertible TQFT}}.

$\bullet$ {\bf Phase transition between the octant (I) to (IV)}:\\
The WZW term and the anomaly of 5d $w_2w_3$ invertible TQFT play a crucial role in (I).
The 4d phase transition from the modified $so(10)$ (I) to PS (IV) is a boundary phase transition on the 5d bulk $w_2w_3$.
The phase transition is triggered not merely by tuning $r_{{\mathbf{54}}}^{}>0$ to $r_{{\mathbf{54}}}^{}<0$ 
of $r \Phi^2 + \lambda \Phi^4$ 
in \eq{eq:U-potential},
but also by the symmetry breaking to cancel the anomaly on the 4d boundary of 5d invertible TQFT (when entering from (I) to (IV)). 
Overall, it is the {\bf continuous phase transition of
Wilson-Fisher type but Beyond-Landau-Ginzburg paradigm due to the %\cred
{anomaly matching via the symmetry breaking on the 4d boundary of
5d invertible TQFT}}.

$\bullet$ {\bf Phase transition between the octant (II) to the critical region (0) 
(the situation is similar to phase transitions of (III) to (0), (IV) to (0), and (VI) to (0))}:\footnote{More precisely,
we really mean to specify here the phase transitions from the $\U(1)'$ confined phase of GUT/SM to the corresponding $\U(1)'$ deconfined phase of GUT/SM,
namely (II) to (II)$'$, (III) to (III)$'$, (IV) to (IV)$'$, and (VI) to (VI)$'$. The critical region (0) actually has children sub-phases including those in \Fig{fig:phase-two} (b).} \\
The WZW term and the anomaly of 5d $w_2w_3$ invertible TQFT play a crucial role in the critical region (0).
The 4d phase transitions from the either models of GUT/SM of (II), (III) and (IV) to the critical region (0) 
is a boundary phase transition on the 5d bulk $w_2w_3$.
These 4d phase transitions are Gross-Neveu type
because we also have Yukawa-Higgs interactions $\chi^\dagger \Phi \chi$ in \eq{eq:QED4} (more than Wilson-Fisher of $r \Phi^2 + \lambda \Phi^4$).
Moreover, there are deconfined dark gauge fields of $\U(1)'$ in the critical region (0), but the $\U(1)'$ is confined outside the critical region (0).
Therefore,
%\cred
{overall it is the {\bf continuous phase transition of
deconfined-confined QED$'$-Gross-Neveu type beyond-Landau-Ginzburg paradigm}. 
It is beyond Landau-Ginzburg also due to two effects (1) WZW term and (2) anomaly matching via the symmetry breaking on the 4d boundary of
5d invertible TQFT.}

$\bullet$ {\bf Phase transition between the octant (I) to the critical region (0) (or more precisely (I)$'$)}:\\
The WZW term and the anomaly of 5d $w_2w_3$ invertible TQFT play a crucial role in both the critical region (I)$'$ and the modified $so(10)$ GUT (I).

The 4d phase transitions from the (I) to the critical region (I)$'$ 
is a boundary phase transition that the 5d bulk $w_2w_3$ is always required since the Spin(10) is preserved throughout the transition 
(if we regard the Spin(10) global symmetry is realized locally onsite).

--- If the \emph{deconfined} dark gauge fields $\U(1)'$ in the critical region (I)$'$ becomes \emph{confined} in the region (I),
then overall it is the {\bf continuous deconfined-confined phase transition of
QED$'_4$ Beyond-Landau-Ginzburg paradigm, without any global symmetry-breaking}.
%\cred
{The QED$'_4$ describes the $\U(1)'$ dark gauge field coupled to fermionic partons. 
When the gauge coupling is strong enough, ${g'}^2 > {g_c'}^2$, it is possible to drive a confinement transition, which gaps out all the fermionic partons and removes the $\U(1)'$ photon from the low-energy spectrum. However, this nonperturbative nature of deconfined to confined phase transition of QED$'_4$ 
cannot be captured easily by perturbative renormalization group or Feynman diagram analysis.}\footnote{{\Refe{Kondo9803133} 
studies the deconfined to confined phase transition of QED$'_4$. 
\Refe{Kondo9803133} suggests that its nature is a Berezinskii-Kosterlitz-Thouless (BKT) phase transition, as an infinite order continuous phase transition.}
}

--- If the \emph{deconfined} dark gauge fields $\U(1)'$ in the critical region (I)$'$ remains \emph{deconfined} in the region (I),
then overall there is no phase transition. The critical region (I)$'$ and (I) are smoothly connected as the same critical region.

\end{enumerate}

The critical region (0) is further broken down to the different symmetry-breaking orders from (I)$'$ to (VIII)$'$
shown in \eq{eq:critical-region} with totally 5 refined phases. Hence there are more refined versions of phases transitions 
than what we had discussed above.

Overall, we need to enumerate all the possible phase transitions between these regions:
the ${\U(1)'}_{\text{gauge}}^{\text{dark}}$-confined regions
 (I)$'$=(V)$'$, 
 (II)$'$, 
(III)$'$=(VII)$'$, 
(IV)$'$=(VIII)$'$,
and
(VI)$'$, and 
the ${\U(1)'}_{\text{gauge}}^{\text{dark}}$-deconfined critical regions
 (I)$'$=(V)$'$, 
 (II)$'$, 
(III)$'$=(VII)$'$, 
(IV)$'$=(VIII)$'$,
and (VI)$'$.
We summarize the nature of these phase transitions in \Table{table:GUT-transition}.

\begin{landscape}
\thispagestyle{empty}
%%%%%%%%%%%%%%%%%%%%%%%%%%%% 

\begin{center}
\begin{table}[!h]
%\centering
\hspace{35.8mm}
\finline[\fontsize{9}{9}]{Alpine}{
\noindent
\makebox[\textwidth][c] %[l]
{
	\begin{tabular}{|c c  | c c c c c c c| }
	\hline
		\multicolumn{9}{|c|}{Properties of Quantum Phase Transition}\\
		\hline
		& & \multicolumn{7}{c|}{Internal Global Symmetry}  \\
		\hline
		& &
		$\begin{array}{c}
		\text{Order of}\\ 
		\text{transition}
		\end{array}$
		& $\begin{array}{c}
		\text{Landau-Ginzburg}\\
		\text{order parameter}\\
		\< \CO_{\text LG} \>\\
		\end{array}$
		& $\begin{array}{c}
		\text{Critical}\\ 
		\text{theory}
		%\text{Wilson-Fisher (WF)}\\
		%\text{Gross-Neveu-Yukawa (GNY)}\\
		%\text{QED$'_4$-GNY}
		\end{array}$
		&
		$\begin{array}{c}
		\text{Anomaly}\\
		\text{Sym.Preserved}\\
		\text{Sym.Breaking}
		\end{array}$
		& 
		$\begin{array}{c}
		\text{Fermionic}\\
		\text{parton}
		\end{array}$ 
		&
		$\begin{array}{c}
		\text{Deconfined}\\
		\text{${\U(1)'}_{\text{gauge}}^{\text{dark}}$}
		\end{array}$  
		 &
		 $\begin{array}{c}
		\text{Beyond}\\
		\text{LGW}
		\end{array}$	
		\\
		\hline
	(I)-(II)  & $so(10)$-GG & Cont. &  Condense $\langle \Phi_{\bf{45}}^{\text{1st}}  \rangle \neq 0$  &  \multirow{2}{*}{WF} &  
	            \multirow{2}{*}{$\begin{array}{c}
	           \text{SP to SB}
	           \end{array}$} 
	            & 
	            \multirow{2}{*}{$\begin{array}{c}
	           \text{gapped}\\
	           \text{(confined)}
	           \end{array}$}
	             & 
	             \multirow{2}{*}{$\begin{array}{c}
	           \text{No}
	           \end{array}$}
	             & \multirow{2}{*}{$\begin{array}{c}
	           \text{Yes}\\
	           \text{(anom.)}
	           \end{array}$}
	              \\
	(I)-(VI) & $so(10)$-flipped  & Cont. & Condense $\langle \Phi_{\bf{45}}^{\text{2nd}}  \rangle \neq 0$  &   &  & & & \\
	\hline
	(I)-(IV) & $so(10)$-PS  & Cont. & Condense $\langle \Phi_{\bf{54}}^{}  \rangle \neq 0$  & WF & {$\begin{array}{c}
	           \text{SP to SB} 
	           \end{array}$} &
	           {$\begin{array}{c}
	           \text{gapped}\\
	           \text{(confined)}
	           \end{array}$}
	            & No & {$\begin{array}{c}
	           \text{Yes}\\
	           \text{(anom.)}
	           \end{array}$} \\
	\hline
	(II)-(III) & GG-SM & \multirow{2}{*}{Cont.}  & \multirow{2}{*}{Condense $\langle \Phi_{\bf{54}}^{}  \rangle \neq 0$} &  \multirow{2}{*}{WF} &
	          \multirow{2}{*}{$\begin{array}{c}
	           \text{SB to SB} 
	           \end{array}$}  & 
	           \multirow{2}{*}{$\begin{array}{c}
	           \text{gapped}\\
	           \text{(confined)}
	           \end{array}$}
	            & \multirow{2}{*}{$\begin{array}{c}
	           \text{No}
	           \end{array}$} & \multirow{2}{*}{No}  \\
	(VI)-(III)	 & flipped-SM &     &   & & & & &   \\
	\hline
	(IV)-(III)	& PS-SM & Cont. & Condense $\langle \Phi_{\bf{45}}^{}  \rangle \neq 0$ & WF &  {$\begin{array}{c}
	           \text{SB to SB} 
	           \end{array}$} & 
	           {$\begin{array}{c}
	           \text{gapped}\\
	           \text{(confined)}
	           \end{array}$}
	           & No & No \\
	\hline
	\hline 
	(I)-(III)	& $so(10)$-SM & Cont. & $\begin{array}{c}
	           \text{Condense}\\
	\langle \Phi_{\bf{45}}^{}  \rangle, \langle \Phi_{\bf{54}}^{}  \rangle \neq 0
	 \end{array}$ & WF &  {$\begin{array}{c}
	           \text{SP to SB} 
	           \end{array}$} & 
	           {$\begin{array}{c}
	           \text{gapped}\\
	           \text{(confined)}
	           \end{array}$}
	           & No & {$\begin{array}{c}
	           \text{Yes}\\
	           \text{(anom.)}
	           \end{array}$} \\   
	\hline 
	(II)-(IV)	& GG-PS &  \multirow{2}{*}{Cont.} &  \multirow{2}{*}{$\begin{array}{c}
	           \text{Swap $\langle \Phi_{\bf{45}}  \rangle \neq 0$}\\
	           \text{and $\langle \Phi_{\bf{54}}  \rangle \neq 0$} 
	           \end{array}$ } &  \multirow{2}{*}{WF} &  \multirow{2}{*}{$\begin{array}{c}
	           \text{SB to SB} 
	           \end{array}$} & 
	           \multirow{2}{*}{$\begin{array}{c}
	           \text{gapped}\\
	           \text{(confined)}
	           \end{array}$}
	           & \multirow{2}{*}{$\begin{array}{c}
	           \text{No}
	           \end{array}$} & \multirow{2}{*}{$\begin{array}{c}
	           \text{No}
	           \end{array}$}  \\   
	(VI)-(IV)	& flipped-PS &  & 
	&  &   & & & \\                     
	\hline
	\hline
	(II)-(VI) & GG-flipped  &   1st &
	           $\begin{array}{c}
	           \text{Swap $\langle \Phi_{\bf{45}}^{\text{1st}}  \rangle \neq 0$}\\
	           \text{and $\langle \Phi_{\bf{45}}^{\text{2nd}}  \rangle \neq 0$} 
	           \end{array}$  & No & {$\begin{array}{c}
	           \text{SB to SB} 
	           \end{array}$} & 
	         {$\begin{array}{c}
	           \text{gapped}\\
	           \text{(confined)}
	           \end{array}$}  & No & No \\
	\hline
	\hline
	\hline
	(I)$'$-(II)$'$ & $so(10)'$-GG$'$ &    \multirow{2}{*}{Cont.}  & Condense $\langle \Phi_{\bf{45}}^{\text{1st}}  \rangle \neq 0$  & \multirow{2}{*}{QED$'_4$-GNY} & \multirow{2}{*}{$\begin{array}{c}
	           \text{SP to SB} 
	           \end{array}$} &  \multirow{2}{*}{$\begin{array}{c}
	           \text{gapless}\\
	           \text{(deconfined)}
	           \end{array}$} & 
	           \multirow{2}{*}{$\begin{array}{c}
	           \text{Yes}
	           \end{array}$} & 
	           \multirow{2}{*}{$\begin{array}{c}
	           \text{Yes}\\
	           \text{(an.dcf.)}
	           \end{array}$}
	           \\
	(I)$'$-(VI)$'$ & $so(10)'$-flipped$'$  & & Condense $\langle \Phi_{\bf{45}}^{\text{2nd}}  \rangle \neq 0$ & & & & & \\
	\hline
	(I)$'$-(IV)$'$ & $so(10)'$-PS$'$  & {Cont.} & Condense $\langle \Phi_{\bf{54}}^{}  \rangle \neq 0$ & QED$'_4$-GNY & {SP to SB}  & {$\begin{array}{c}
	           \text{gapless}\\
	           \text{(deconfined)}
	           \end{array}$} & Yes & 
	          {$\begin{array}{c}
	           \text{Yes}\\
	           \text{(an.dcf.)}
	           \end{array}$} \\
	\hline
	(II)$'$-(III)$'$ & GG$'$-SM$'$ & \multirow{2}{*}{Cont.}  & \multirow{2}{*}{Condense $\langle \Phi_{\bf{54}}^{}  \rangle \neq 0$} &  \multirow{2}{*}{WF} &
	          \multirow{2}{*}{$\begin{array}{c}
	           \text{SB to SB} 
	           \end{array}$}  & 
	            \multirow{2}{*}{$\begin{array}{c}
	           \text{gapped}\\
	           \text{(Higgs SSB)}
	           \end{array}$}
	           & \multirow{2}{*}{Yes} &   \multirow{2}{*}{$\begin{array}{c}
	           \text{No}
	           \end{array}$}  \\
	(VI)$'$-(III)$'$	 & flipped$'$-SM$'$ &     &   & & & & &  \\
         \hline
	(IV)$'$-(III)$'$       & PS$'$-SM$'$ & Cont. & Condense $\langle \Phi_{\bf{45}}^{}  \rangle \neq 0$ & WF &  {$\begin{array}{c}
	           \text{SB to SB} 
	           \end{array}$} & 
	           {$\begin{array}{c}
	           \text{gapped}\\
	           \text{(Higgs SSB)}
	           \end{array}$}
	           & Yes & 
	            {$\begin{array}{c}
	           \text{No}          
	           \end{array}$}
	            \\
	\hline
	\hline
	(I)$'$-(III)$'$ & $so(10)'$-SM$'$  & {Cont.} & 
	$\begin{array}{c}
	           \text{Condense}\\
	\langle \Phi_{\bf{45}}^{}  \rangle, \langle \Phi_{\bf{54}}^{}  \rangle \neq 0
	 \end{array}$ & QED$'_4$-GNY & {SP to SB} & {$\begin{array}{c}
	           \text{gapless}\\
	           \text{(deconfined)}
	           \end{array}$} & Yes &
	             {$\begin{array}{c}
	           \text{Yes}\\
	           \text{(an.dcf.)}
	           \end{array}$} 
	           \\
	\hline
	(II)$'$-(IV)$'$	& GG$'$-PS$'$ &  \multirow{2}{*}{Cont.} &  \multirow{2}{*}{$\begin{array}{c}
	           \text{Swap $\langle \Phi_{\bf{45}}  \rangle \neq 0$}\\
	           \text{and $\langle \Phi_{\bf{54}}  \rangle \neq 0$} 
	           \end{array}$ } &  \multirow{2}{*}{QED$'_4$-GNY} &  \multirow{2}{*}{$\begin{array}{c}
	           \text{SB to SB} 
	           \end{array}$} &  \multirow{2}{*}{$\begin{array}{c}
	           \text{gapless}\\
	           \text{(deconfined)}
	           \end{array}$} & \multirow{2}{*}{$\begin{array}{c}
	           \text{Yes}
	           \end{array}$} &   
	           \multirow{2}{*}{$\begin{array}{c}
	           \text{Yes}\\
	           \text{(decf.)}
	           \end{array}$}  \\   
	(VI)$'$-(IV)$'$  & flipped$'$-PS$'$ &  & 
	&  &   & & & \\   
	\hline
	\hline
	(II)$'$-(VI)$'$ & GG$'$-flipped$'$  &   1st &
	           $\begin{array}{c}
	           \text{Swap $\langle \Phi_{\bf{45}}^{\text{1st}}  \rangle \neq 0$}\\
	           \text{and $\langle \Phi_{\bf{45}}^{\text{2nd}}  \rangle \neq 0$} 
	           \end{array}$  & No & {$\begin{array}{c}
	           \text{SB to SB} 
	           \end{array}$} & {$\begin{array}{c}
	           \text{gapped}\\
	           \text{(Higgs SSB)}
	           \end{array}$}
	           & Yes & No  \\                  
	\hline
	\hline
	\hline
	 (I)$'$-(I)& $so(10)'$-$so(10)$ & \multirow{5}{*}{Cont.}  & \multirow{5}{*}{No} & \multirow{5}{*}{$\begin{array}{c}
	           \text{deconfined-}\\
	            \text{confined}
	            \end{array}$}
	            & {SP to SP} & 
	            \multirow{5}{*}{$\begin{array}{c}
	           \text{gapless}\\
	           \text{(deconfined)}
	           \end{array}$}
	            & \multirow{5}{*}{Yes} & \multirow{5}{*}{$\begin{array}{c}
	           \text{Yes}\\
	           \text{(decf.)}
	           \end{array}$} \\
	${(\rm II)}'$-(II)& GG$'$-GG&   &  & 
	            & {SB to SB}  & & & \\
	${(\rm III)}'$-(III)& SM$'$-SM&   &  & 
	            & {SB to SB} & & & \\
	${(\rm IV)}'$-(IV)& PS$'$-PS&   &  & 
	            & {SB to SB} & & & \\
	${(\rm VI)}'$-(VI)& flipped$'$-flipped&   &  & 
	            & {SB to SB} & & &\\
	\hline
	\end{tabular}
}
\caption{Here we summarize the
properties of quantum phase transition
in the limit when the internal symmetries is treated as ungauged global symmetries in \Sec{sec:QuantumPhaseDiagram-Internal-global}.
We abbreviate: {Wilson-Fisher (WF)},
the QED$'_4$ with {Gross-Neveu-Yukawa ({QED$'_4$-GNY})},
Symmetry Preserved (SP),
Symmetry Breaking (SB),
and Landau-Ginzburg-Wilson (LGW).
For beyond LGW, (anom.) means due to anomaly,
(decf) means due to deconfined-confined phase transition.
(an.dcf.) means due to both anomaly and deconfined-confined effects.
}
\label{table:GUT-transition}
}
\end{table}
\end{center}
%%%%%%%%%%%%%%%%%%%%%%%%%%%% 

\end{landscape}

\subsubsection{Internal symmetries are dynamically gauged}
\label{sec:QuantumPhaseDiagram-Internal-gauged}

When the internal symmetries are treated as global symmetries in
\Sec{sec:QuantumPhaseDiagram-Internal-global},
the $\U(1)_{\text{gauge}}^{'\text{dark}}$ is the only dynamical gauge sector,
which becomes deconfined only emergent near the quantum critical region (the white region (0) in \Fig{fig:phase},
in \Fig{fig:phase-two} (b),
and in \Fig{fig:phase-quadrant-1} and \Fig{fig:phase-quadrant-2}).
The emergent deconfined $\U(1)_{\text{gauge}}^{'\text{dark}}$ gauge field near the quantum critical region
is exactly the reason why we name the Gauged Enhanced Quantum Criticality beyond the Standard Model in our prior work
\cite{Wang2106.16248}.

When internal symmetries are dynamically gauged as they are in our quantum vacuum,
then internal symmetry groups become gauge groups.
We have additional gauge sectors such as Spin(10), $\U(5)_{\hat q = 2}$,
$\frac{\SU(4)_c\times(\SU(2)_\rL\times \SU(2)_\rR)}{\mathbb{Z}_{2}}$, and $\frac{{\SU(3)} \times {\SU(2)} \times \U(1)_{\tilde{Y}}}{\Z_6}$, etc.
{We can ask the dynamical fates of these gauge theories: confined or deconfined?
When the number of fermions are comparably small as in our quantum vacuum, 
the RG beta function computation
shows the asymptotic freedom at UV and the confinement at IR  for a non-abelian gauge theory.}
Other abelian gauge sectors, such as $\U(1)_{X_1}$,  $\U(1)_{X_2}$, and  $\U(1)_{\rm EM}$,
can stay deconfined. We summarize their dynamics in \Table{table:GUT-gauged-phase}.

\begin{table}[!h]
\begin{center}
\begin{tabular}{| lcc c |}
\hline
\multicolumn{4}{|c|}{Internal Symmetry Gauged}\\
\hline
& &   confined sectors & deconfined sectors   \\
\hline
& (I)=(V) &  $so(10)$ & \\
\hline
& (II) & $su(5)$  & $u(1)_{X_1}$ \\
\hline
& (III)=(VII) & $su(3) \times su(2)$  & $u(1)_{\text{EM}}$ \\
\hline
& (IV)=(VIII) &  $su(4) \times su(2)_{\rm L} \times su(2)_{\rm R}$ & \\
\hline
& (VI) &  $su(5)$ & $u(1)_{X_2}$ \\
\hline
\hline
\multirow{5}{*}{(0)} & (I)$'$=(V)$'$ & $so(10)$  &  $u(1)_{\text{gauge}}^{'\text{dark}}$ \\
\cline{2-4}
& (II)$'$ &  $su(5)$ & $u(1)_{X_1} \times u(1)_{\text{gauge}}^{'\text{dark}}$ \\
\cline{2-4}
& (III)$'$=(VII)$'$ &  $su(3) \times su(2)$  & $u(1)_{\text{EM}} \times u(1)_{\text{gauge}}^{'\text{dark}}$ \\
\cline{2-4}
& (IV)$'$=(VIII)$'$ & $su(4) \times su(2)_{\rm L} \times su(2)_{\rm R}$  & $u(1)_{\text{gauge}}^{'\text{dark}}$ \\
\cline{2-4}
& (VI)$'$ &  $su(5)$ &  $u(1)_{X_2} \times u(1)_{\text{gauge}}^{'\text{dark}}$ \\
\hline
\end{tabular}
\end{center}
\caption{When internal symmetries are dynamically gauged as in our quantum vacuum,
we enlist the {\bf confined} gauge groups
and {\bf deconfined} gauge groups for each of the phases in \Fig{fig:phase-quadrant-1} and \Fig{fig:phase-quadrant-2}.
For simplicity, we only write down the Lie algebra in the lower case. The full Lie group in the capital case can be read from \Fig{fig:embed-Lie-group-spacetime-2109-2}.
The $u(1)_{\text{gauge}}^{'\text{dark}}$ only emerges and deconfines in the quantum critical region (0), including (I)$'$ to (VIII)$'$.
{The $u(1)_{\text{gauge}}^{'\text{dark}}$ disappears and confines in the (I) to (VIII).}
}
\label{table:GUT-gauged-phase}
\end{table}

%\cred
{When internal symmetries are dynamically gauged,
the phase transitions described in \Table{table:GUT-transition}
also change or upgrade. 
For example, the Wilson-Fisher transitions of scalar fields 
become the Anderson-Higgs transitions of scalar fields interacting with gauge fields.
Some of the QED$'_4$ transitions also need to take into account of the nonabelian gauge fields from Spin(10) or its subgroups,
which lead to various QCD$'_4$ transitions.}

Moreover, once the ordinary internal symmetry (i.e., the 0-symmetry) is dynamically gauged,
the outcome gauge theory can have extended objects (e.g., Wilson or 't Hooft 1d lines) 
which are the charged objects of the generalized global symmetries \cite{Gaiotto2014kfa1412.5148} (i.e., here the 1-symmetry).
We systematically explore these generalized global symmetries of SM and GUT in the next Section \ref{sec:HigherSymmetries}.

\pagebreak
\newpage

\section{Higher Symmetries %and Categorical Symmetries
of Standard Models and Grand Unifications}
\label{sec:HigherSymmetries}

Here we point out that once the internal symmetry of SM and GUT ($G_{\SM_{q=1,2,3,6}}$, $G_{\GG}$, $G_{\PS_{q'=1,2}}$, $G_{so(10)}$, $\dots$) are dynamically gauged (as they are in the dynamical gauge theories),
there are dynamical Wilson or 't Hooft lines as charged objects charged under the generalized global symmetries \cite{Gaiotto2014kfa1412.5148}. 
Pioneer works \cite{AharonyASY2013hdaSeiberg1305.0318, Tong2017oea1705.01853} 
have pointed out the Wilson line spectrum differences between different versions of $G_{\SM_q}$ for $q=1,2,3,6$.
Other prior works also study the higher symmetries for these $G_{\SM_q}$ \cite{Wan2019sooWWZHAHSII1912.13504, MonteroWang, AnberPoppitz2110.02981}.
%In addition, a recent work \cite{AnberPoppitz2110.02981} uses some of these higher-symmetries of SM and their gauging to study the nonperturbative and cosmological effects.
In comparison, our present work will include not only the higher symmetries of SM$_q$ in \cite{Wan2019sooWWZHAHSII1912.13504, MonteroWang, AnberPoppitz2110.02981}, but also other pertinent GUT models.

Recall \cite{Gaiotto2014kfa1412.5148},
when the {\bf \emph{charged objects}} are 1d line operators (say along a closed curve ${\gamma^1}$),
the  {\bf \emph{charge operators}} (also known as the  {\bf \emph{symmetry generators}} or {\bf \emph{symmetry defects}}) are codimension-2 topological operators 
(as 2d surface operators say on a closed surface $\Sigma^2$, in a 4d spacetime). 
As an example, when the gauge group is abelian, the path integral expectation value of the link configuration between the 1d line and 2d surface operators
with a linking number ${\rm Lk}( \Sigma^2, {\gamma^1_{}} )$ evaluated on a closed 4-manifold $M^4$ 
is schematically given by:
\bea
\langle \exp(\ii  \theta \ointint_{ \Sigma^2} B_{\text{charge}}  ) \cdot \exp(\ii  q \oint_{\gamma^1} A_{\text{charged}} )\rangle = 
\e^{\ii q \theta {\rm Lk}( \Sigma^2, {\gamma^1_{}} ) 
} \cdot
\langle \exp(\ii  q \oint_{\gamma^1} A_{\text{charged}} )\rangle
\Big\vert_{M^4}.
\eea
The expectation value is topological independent from the continuous deformation of the codimension-2 topological operator of $B$, as long as it does not cross the charged object of $A$.
Hence this explains the meaning of the name {\bf \emph{topological operator}}: the system and the {topological operator} of $B$
do \emph{not} have to be gapped, but its correlator is the same independent from the topological deformation.
%
%\cred
{However, we need to include $\langle \exp(\ii  q \oint_{\gamma^1} A_{\text{charged}} )\rangle$ on the right hand side whenever this operator $A$ is non-topological.} 
If $A$ describes the 1d Wilson line (or denoted as 1-line), the topological 2d surface operator of $B$ is often called the topological Gukov-Witten 2-surface operator \cite{GukovWitten0612073, GukovWitten0804.1561}.
See further in-depth discussions on the topological operators in \cite{MonteroRudelius2104.07036}.

This expression is also the analogous Ward identity for the {\bf generalized 1-form global symmetry}, or simply denoted as the {\bf 1-symmetry}.
If we normalize the expectation value properly,
the proportionality ($\propto$) becomes the equality ($=$) to the statistical Berry phase $\e^{\ii q \theta {\rm Lk}( \Sigma^2, {\gamma^1_{}} ) }$.
The abelian phase $\e^{\ii q \theta}$ means that we are focusing on the abelian 1-symmetry:\\
$\bullet$ When it is an abelian U(1) 1-symmetry denoted as $\U(1)_{[1]}$, then $q \in \Z$ and $\theta \in [0, 2 \pi)$.\\
$\bullet$ When it is an abelian $\Z_\rN$ 1-symmetry denoted as $\Z_{\rN,[1]}$, then $q \in \Z_\rN$ and $\theta = \frac{2 \pi k}{\rN}$ with $k \in \Z_\rN$.\\
For a gauge theory of gauge group $G_g$, the electric 1-symmetry is associated with the unbroken center subgroup $Z(G_g)$,
the magnetic 1-symmetry is associated with the unbroken Pontryagin dual group of the first homotopy group: $\pi_1(G_g)^{\vee}\equiv\Hom(\pi_1(G_g),\U(1))$.
Here are some familiar examples (results summarized in \Table{table:higher-sym-GUTmodel}):
\begin{enumerate}[leftmargin=.mm] 
\item
For a 4d pure U(1) gauge theory, we have the electric 1-symmetry $\U(1)_{[1]}^e$
 and the magnetic 1-symmetry $\U(1)_{[1]}^m$, whose 1-symmetry measurements are characterized by the following two expectation values:
 \bea
\U(1)_{[1]}^e &:& \langle \exp(\ii  \theta_e \ointint_{ \Sigma^2} \star \dd A_{\; \text{charge}-\U(1)_{[1]}^e}  ) \cdot \exp(\ii  q_e \oint_{\gamma^1} A_{\; \text{charged}-\U(1)_{[1]}^e} )\rangle = 
\e^{\ii q_e \theta_e {\rm Lk}( \Sigma^2, {\gamma^1_{}} ) }
\cdot
\langle \exp(\ii  q_e \oint_{\gamma^1} A )\rangle
\Big\vert_{M^4}.\cr
\U(1)_{[1]}^m &:&\langle 
\exp(\ii  \theta_m \ointint_{ \Sigma^2}  \dd A_{\; \text{charge}-\U(1)_{[1]}^m}    ) \cdot \exp(\ii  q_m \oint_{\gamma^1} V_{\; \text{charged}-\U(1)_{[1]}^m} ) 
\rangle = 
\e^{\ii q_m \theta_m {\rm Lk}( \Sigma^2, {\gamma^1_{}} ) }
\cdot
\langle \exp(\ii  q_m \oint_{\gamma^1} V )\rangle
\Big\vert_{M^4}.\nn
\eea
The Wilson line of a 1-form gauge field $A$ is the $\U(1)_{[1]}^e$ electric charged object,
and  't Hooft line's of a dual 1-form gauge field  $V$ is the $\U(1)_{[1]}^m$ magnetic charged object; they are related by the Hodge dual $\star$ as $\dd A = \star \dd V$.
An open boundary of the magnetic 2-surface $\dd A$ gives rise to 1d object closely related to an \emph{improperly quantized} electric Wilson 1-line of $A$.
Vice versa, 
an open boundary of the electric 2-surface $\dd V$ gives rise to 1d object closely related to an \emph{improperly quantized} magnetic 't Hooft 1-line of $V$.

$\bullet$ The \emph{improperly quantized} operators refer to the continuous unquantized $\theta$, thus these operators live on the boundary of one-higher dimensional operator.\\
$\bullet$ In contrast, the \emph{genuine} operators (e.g., those Wilson or 't Hooft lines specified by the quantized $q_e. q_m \in \Z$) do not require to live on the boundary of one-higher dimensional operator.

\item For a 4d pure SU(N) gauge theory, we have the electric 1-symmetry $\Z_{\rN,[1]}^e$
characterized by
 \be \label{eq:link-We-Ue}
\langle  \exp(\ii \frac{2\pi}{ \rN} \ointint_{\Sigma^2} \Lambda) 
\;
\Tr_{\text{R}}( \text{P} \exp(\ii \oint_{\gamma^1} a))
\rangle =
\exp\left(\frac{\ii 2 \pi}{ \rN}{\text{Lk}({\Sigma^2},{\gamma^1})}\right)
\cdot
\langle  \Tr_{\text{R}}( \text{P} \exp(\ii \oint_{\gamma^1} a)) \rangle,
%(-1)^{\text{Lk}({\gamma^1},{\Sigma^2})}.
 \ee
where gauge field $a$ is Lie algebra su(N) valued. 
The $\text{P} \exp(\ii \oint a)$ specifies a SU(N) group element where P is the path ordering.
Tr is the trace in the  representation R of SU(N).  Here we take the fundamental representation for R.
The $\Lambda \in \H^2(M^4,\Z_{\rN})$ as a $\Z_{\rN}$-cohomology class, tightly related to the
generalized second Stiefel-Whitney class  $w_2(V_{\PSU(\rN)})\in \H^2(M, \Z_\rN)$, 
as the obstruction of promoting the PSU(N) bundle to SU(N) bundle. 
This becomes obvious when we promote the SU(N) 
gauge theory to a U(N) gauge theory with additional constraints, here and below following \cite{Gaiotto2014kfa1412.5148}.
The U(N) gauge theory also has the benefits to go to the PSU(N) gauge theory.
In the U(N) gauge theory, we introduce this constraint to the path integral,
\bea  \label{eq:Lambda-B}
\int [\cD \Lambda] \dots  \exp \big(  \frac{\ii 2 \pi}{ \rN} \int_{M^4} \Lambda \smile (c_1-B )\big)
\eea
with the gauge bundle constraint $c_1=w_2(V_{\PSU(\rN)})=B \mod \rN$
where the first Chern class $c_1 =c_1(V_{\U(\rN)}) \in \Z$ is from the U(1) part of U(N).
By staring at the two expressions in \eq{eq:link-We-Ue} and \eq{eq:Lambda-B},
it becomes also clear that 
an open boundary of the magnetic 2-surface $w_2(V_{\PSU(\rN)})=B$ gives rise to 1d object closely related to the \emph{improperly quantized} electric Wilson 1-line of $a$.
Vice versa, 
an open boundary of the electric 2-surface $\Lambda$ gives rise to 1d object closely related to the \emph{improperly quantized} magnetic 't Hooft 1-line of PSU(N) gauge theory.

\item Thus, for a 4d pure PSU(N) gauge theory, we have the magnetic 1-symmetry $\Z_{\rN,[1]}^m$
characterized by the magnetic 2-surface operator 
$\exp(\ii \frac{2\pi}{ \rN} \ointint_{\Sigma^2} B)$ linking with the magnetic 't Hooft 1-line of PSU(N) gauge theory.

\item For a 4d pure U(N) gauge theory (or the refined U(N)$_{\hat q}$ gauge theory discussed in \Sec{sec:Refined-U5-group}),
the electric 1-symmetry is given by the center $Z(\U(\rN))= \U(1)$,
while the magnetic 1-symmetry is given by the Pontryagin dual group of the first homotopy group 
$\pi_1(\U(\rN))^{\vee}=\Hom(\pi_1(\U(\rN)),\U(1))=\Hom(\Z,\U(1))=\U(1)$.
This means that:\\
$\bullet$ 4d pure U(N) gauge theory without matter kinematically has
${\U(1)}_{[1]}^e$ and ${\U(1)}_{[1]}^m$ 1-symmetries.\\
$\bullet$ 4d pure U(N) (say the refined U(N)$_{\hat q=1}$) gauge theory with the gauge-charged matter 
in the fundamental of SU(N) and in the unit charge of U(1) written as the $(\mathbf{N},1)$ representation,
kinematically reduces the electric 1-symmetry to none --- because the charged Wilson line of ${\U(1)}_{[1]}^e$ of the earlier pure U(N) gauge theory now 
becomes breakable with two open ends attached the gauge-charged matter $(\mathbf{N},1)$ and $(\overline{\mathbf{N}},-1)$, which nullify ${\U(1)}_{[1]}^e$ to zero. 
But the magnetic 1-symmetry ${\U(1)}_{[1]}^m$ maintains.

\end{enumerate}

\begin{table}[!h]
%\centering
	\hspace{-11.8mm}
	\begin{tabular}{|c | c c c c c|}
	\hline
		\multicolumn{6}{|c|}{Higher symmetries of 4d pure gauge theories}\\
		\hline
		QFT & $Z(G_g)$ & $\pi_1(G_g)$. $\pi_1(G_g)^{\vee}$ &
		1-form $e$ sym $G_{[1]}^e$  &
		1-form $m$ sym $G_{[1]}^m$ &
		{ }
		\\
		\hline
		$\U(1)$ & $\U(1)$ & $\Z$. $\U(1)$ &  ${\U(1)}_{[1]}^e$ & ${\U(1)}_{[1]}^m$ & \\
		\hline
		$\SU(\rN)$ & $\Z_\rN$ &  0. 0 &  ${\Z}_{\rN,[1]}^e$ & 0 & \\
		\hline
		$\PSU(\rN)$ & 0 &  $\Z_\rN$. $\Z_\rN$ &  0 & ${\Z}_{\rN,[1]}^m$ & \\
		\hline
		$\U(\rN)$ &  $\U(1)$ &  $\Z$. $\U(1)$ &  ${\U(1)}_{[1]}^e$ & ${\U(1)}_{[1]}^m$  & \\
		\hline
		\hline
		\multicolumn{6}{|c|}{Higher symmetries of 4d SMs or GUTs with SM matters}\\
		\hline
		QFT & $Z(G_g)$ & $\pi_1(G_g)$. $\pi_1(G_g)^{\vee}$ &
		1-form $e$ sym $G_{[1]}^e$  &
		1-form $m$ sym $G_{[1]}^m$ &
		{ }
		\\ 
		\hline
		 $G_{\SM_q} \equiv \frac{{\SU(3)} \times {\SU(2)} \times \U(1)_{\tilde{Y}}}{\Z_q}$ & $\Z_{6/q} \times \U(1)$ & $\Z$. $\U(1)$ &    $\Z_{6/q,[1]}^e$ & ${\U(1)}_{[1]}^m$ & \\
		\hline
		 $G_{\SM_6} \equiv \frac{{\SU(3)} \times {\SU(2)} \times \U(1)_{\tilde{Y}}}{\Z_6}$ & $\U(1)$ & $\Z$. $\U(1)$ &  0 & ${\U(1)}_{[1]}^m$ & \\
		\hline
		SU(5) (GG or flipped) & $\Z_5$ & 0. 0 &  0 & 0 & \\
		\hline
		U(5)$_{\hat q}$ (GG or flipped) & $\U(1)$ & $\Z$. $\U(1)$  &  0 & ${\U(1)}_{[1]}^m$ & \\
		\hline
		$G_{\PS_{q'}}\equiv \frac{\SU(4)\times \SU(2)_\rL \times \SU(2)_\rR}{\Z_{q'}}$ & ${\Z_4 \times_{\Z_{q'}} (\Z_2 \times \Z_2)}$ & $\Z_{q'}$. $\Z_{q'}$  &  $\Z_{2/q',[1]}^e$ &  $\Z_{q',[1]}^m$ & \\ 
		\hline
		$G_{\PS_{2}}\equiv \frac{\SU(4)\times \SU(2)_\rL \times \SU(2)_\rR}{\Z_{2}}$ & ${\Z_4 \times_{\Z_{2}} (\Z_2 \times \Z_2)}$ & $\Z_{2}$. $\Z_{2}$  &  0 & $\Z_{2,[1]}^m$  & \\ 
		\hline
		Spin(10) & $\Z_4$ & 0. 0  &  0 & 0 & \\ 
		\hline
	\end{tabular}
    \caption{For an internal symmetry $G_{\text{internal}}=G_g$ as a gauge group, we list down its center subgroup $Z(G_g)$,
    its first homotopy group  $\pi_1(G_g)$ and its Pontryagin dual $\pi_1(G_g)^{\vee}$. We also list down the
    1-form $e$ sym $G_{[1]}^e$ and
		1-form $m$ sym $G_{[1]}^m$ (without matter for the pure gauge theory, and with SM matter for the SMs and GUTs).
    For $\SM_q$, there is a choice of $q=1,2,3,6$.
    Here
    the SM matters are the 15 of 16 left-handed ($L$) Weyl fermions: 
$(\overline{\bf 3},{\bf 1})_{2,L} \oplus ({\bf 1},{\bf 2})_{-3,L}  
\oplus
 ({\bf 3},{\bf 2})_{1,L} \oplus (\overline{\bf 3},{\bf 1})_{-4,L} \oplus ({\bf 1},{\bf 1})_{6,L} \oplus {({\bf 1},{\bf 1})_{0,L}}$
 of $G_{\SM_q} \equiv \frac{{\SU(3)} \times {\SU(2)} \times \U(1)_{\tilde{Y}}}{\Z_q}$; or
    $\overline{\bf 5} \oplus {\bf 10} \oplus 1$  of SU(5);
    or the $({\bf 4}, {\bf 2}, {\bf 1}) \oplus (\overline{\bf 4}, {\bf 1}, {\bf 2})$ of the $G_{\PS_{q'}}$ with $q'=1,2$;
    or the ${\bf 16}$ of the Spin(10).
    }
\label{table:higher-sym-GUTmodel}   
\end{table}

Now we are ready to provide some systematic applications to SM or GUT (results summarized in \Table{table:higher-sym-GUTmodel}):\footnote{Part of these results presented here 
follow Section 1.5 of \cite{Wan2019sooWWZHAHSII1912.13504} and some unpublished notes of the first author (J.~Wang) with Miguel Montero \cite{MonteroWang}. 
JW thanks Miguel Montero on the related discussions and collaborations. 
}
\begin{enumerate}[leftmargin=.mm] 

\item Higher Symmetry for $G_{\SM_q}$  gauge theory with $q=1,2,3,6$:\\
The electric 1-symmetry is related to the center $Z(G_{\SM_q})=\Z_{6/q} \times \U(1)$,
while the magnetic 1-symmetry is related to the Pontryagin dual group of homotopy group $\pi_1(G_{\SM_q})^{\vee}=\U(1)$.
This means:\\ 
$\bullet$ without the SM fermionic matter of quarks and leptons, we have the corresponding
$\Z_{6/q,[1]}^e \times {\U(1)}_{[1]}^e$ and ${\U(1)}_{[1]}^m$ 1-symmetries.\\
$\bullet$ with the SM fermionic matter for $G_{\SM_6}$, we are left with no electric 1-symmetry, because the gauged charge matter
$({\bf 3},{\bf 2})_{1,L}$ explicitly can open up thus break the minimal charged object Wilson line of ${\U(1)}_{[1]}^e$ with two open ends.
But the magnetic 1-symmetry ${\U(1)}_{[1]}^m$ maintains.\\
$\bullet$ with the SM fermionic matter for $G_{\SM_q}$, 
we are left with electric 1-symmetry $\Z_{6/q,[1]}^e$ and magnetic 1-symmetry ${\U(1)}_{[1]}^m$.\\

\item Higher Symmetry for SU(5) of the Georgi-Glashow (GG) or the Barr's flipped $su(5)$ models:\\
$\bullet$ without the SM fermionic matter, 
the center $Z(\SU(5))=\Z_5$  gives rise to
the electric 1-symmetry $\Z_{5,[1]}^e$, while $\pi_1(\SU(5))= 0$ gives no magnetic 1-symmetry.\\
$\bullet$ with SM fermionic matter such as $\overline{\bf 5}$  of SU(5) breaks the electric 1-symmetry to none.

\item Higher Symmetry for U(5)$_{\hat q=2}$ of the GG or the flipped $su(5)$ models:\\
$\bullet$ without the SM fermionic matter, 
the center $Z(\U(5))=\U(1)$  gives rise to
the electric 1-symmetry ${\U(1)}_{[1]}^e$, while $\pi_1(\U(5))^{\vee}= \U(1)$ gives the magnetic 1-symmetry ${\U(1)}_{[1]}^m$.\\
$\bullet$ with SM fermionic matter such as $\overline{\bf 5}_{-3}$  of U(5)$_{\hat q=2}$ breaks the electric 1-symmetry to none.
But the magnetic 1-symmetry ${\U(1)}_{[1]}^m$ maintains.
\item Higher Symmetry for Pati-Salam $G_{\PS_{q'}}$ gauge theory with $q'=1,2$:\\
$\bullet$ without the SM fermionic matter, the
center $Z(G_{\PS_{q'}})={\Z_4 \times_{\Z_{q'}} (\Z_2 \times \Z_2)}$ gives rise to
the electric 1-symmetry $({\Z_4 \times_{\Z_{q'}} (\Z_2 \times \Z_2)})_{[1]}^e$, while the  
$\pi_1(G_{\PS_{q'}})^{\vee}={\Z_{q'}}$ gives the magnetic 1-symmetry $\Z_{q',[1]}^m$.
\\
$\bullet$ with the matter $({\bf 4}, {\bf 2}, {\bf 1}) \oplus (\overline{\bf 4}, {\bf 1}, {\bf 2})$,
the electric 1-symmetry becomes $\Z_{2/q',[1]}^e$, while the magnetic 1-symmetry $\Z_{q',[1]}^m$ remains.
\item Higher Symmetry for the $so(10)$ GUT and Spin(10) gauge group:\\
$\bullet$ without any SM matter (no ${\bf 16}$ of Spin(10)),
the center $Z(\Spin(10))=\Z_4$  gives rise to
the electric 1-symmetry $\Z_{4,[1]}^e$, while $\pi_1(\Spin(10))^{\vee}= 0$ gives no magnetic 1-symmetry. \\
$\bullet$ with ${\bf 16}$ of Spin(10), there are no electric nor magnetic 1-symmetries left.
\end{enumerate}

\section{Categorical Symmetry and Its Retraction}
\label{sec:CategoricalSymmetries}

\Table{table:higher-sym-GUTmodel} summarizes various higher symmetries of SMs and GUTs. In particular, 
for those embeddable into the Spin(10) group (e.g., only $G_{\SM_q}$ with $q=6$ and $G_{\PS_{q'}}$ with $q'=2$,
listed in \Fig{fig:embed-Lie-group-spacetime-2109-1} and \Fig{fig:embed-Lie-group-spacetime-2109-2}), we only have {1-form symmetries for them}:
\bea
G_{\SM_6} &:& {\U(1)}_{[1]}^m,\cr
\text{GG $su(5)$ GUT with SU(5)} &:& \text{none},\cr
\text{GG or flipped $u(5)$ GUT with $\U(5)_{\hat q =2}$} &:& {\U(1)}_{[1]}^m,\cr
\text{PS model with $G_{\PS_{2}}$} &:& \Z_{2,[1]}^m,\cr 
\text{$so(10)$ or modified $so(10)$ GUT with Spin(10)} &:& \text{none}. \label{eq:1-symmetry}
\eea
Some curious facts are:
\begin{enumerate}[leftmargin=.mm] 
\item All electric 1-symmetries are broken by gauged charged fermionic matter. We are only left with either magnetic 1-symmetries or none in \eq{eq:1-symmetry}.
\item Regardless of which lower energy GUTs that we start with, when we approach to the (modified) $so(10)$ GUT as a mother unified EFT at the deeper UV, all 1-symmetries are gone.
\end{enumerate}
%
%

%\cred
{In \Sec{sec:HigherSymmetries}, we had investigated the higher symmetries that are {\bf invertible global symmetries}.
By invertible global symmetries, we mean that the fusion algebras of symmetry generators (i.e., charge operators) follow the group law.
For any symmetry generator (say $U_1, U_2, \dots$) of an invertible global symmetry, its fusion algebra is a binary operation (say ``$\times$'' for the fusion) 
which must obey:\\ 
(1) the closure,\\
(2) the associativity $U_1 \times (U_2 \times U_3) = (U_1 \times U_2) \times U_3$,\\
(3) the identity operator 1 existence, so $U \times 1 = 1  \times  U = U$,\\
(4) the inverse operator operator $U^{-1}$ existence so that $U \times U^{-1} = 1$.
}

%\cred
{In this Section \ref{sec:CategoricalSymmetries},
we investigate any potential
{\bf non-invertible global symmetries}
 in the SM or GUT models.
The non-invertible global symmetries correspond to
the fusion algebras of symmetry generators (i.e., charge operators) 
do not follow the group law. In particular, we will search for the existence of a symmetry generator $U$ such that the
fusion rule of $U$ with any operator $U'$ can never produce only the identity operator. Namely
\bea
\text{ generally } U \times U' &=&  \sum_j U_j \cr
\text{ or at most } U \times U' &=& 1 + \dots \label{eq:CategoricalSymmetriesfusion}
\eea
where the $\dots$ may be nonzero, including other operators (e.g., $U''$, etc.)
The formal form of fusion algebra sometimes uses ``$\otimes$'' for the fusion, and uses ``$\oplus$'' for the splitting on the right hand side.
Here we simply uses the product ``$\times$'' and the sum ``$+$'' because these relations in \eq{eq:CategoricalSymmetriesfusion} hold in the correlation function
computation, both in the QFT path integral formulation or in the quantum matter lattice regularization formulation. 
Namely, we indeed have this relation holds in the expectation value form
\bea
\< U \times U' \>= \< \sum_j U_j  \>.
\eea
{Non-invertible global symmetries}
have appeared long ago in the 2-dimensional CFTs 
\cite{Verlinde1988sn, Moore1988qv, Frohlich0607247, DavydovKong2010rm1004.4725, BhardwajTachikawa1704.02330, ChangLinShaoWangYin1802.04445}.
They also appeared recently under the name of
\emph{algebraic higher symmetry} or
\emph{categorical symmetry} \cite{KongLanWenZhangZheng2005.14178, KomargodskiOhmoriRoumpedakisSeifnashri2008.07567}, 
and \emph{fusion category symmetry} \cite{ThorngrenWang1912.02817, ThorngrenWang2106.12577}.\footnote{{Note that
Wen et al's usage of categorical symmetry \cite{JiWen1912.13492} is different from other research groups' usage of categorical symmetry.
Instead, Wen et al's usage of {algebraic higher symmetry} \cite{KongLanWenZhangZheng2005.14178} is the same as other research groups' usage of categorical symmetry.}}
From now on, follow the recent development 
\cite{GaiottoKulp2008.05960, NguyenTanizakiUnsal2101.02227, NguyenTanizakiUnsal2104.01824, MonteroRudelius2104.07036, ChoiCordovaHsinLamShao2111.01139, KaidiOhmoriZheng2111.01141},
we shall call these types of symmetries as \emph{non-invertible global symmetries} or \emph{categorical symmetries} interchangeably. 
}

\subsection{Potential Non-Invertible Categorical Symmetry induced by $\Z_2^{\rm flip}$}

Given the SM's $\SU(3)_c \times \SU(2)_{\rm L}$, there are actually two ways to embed it inside an SU(5). 
The first one is the Georgi-Glashow (GG) SU(5) that we denoted $\SU(5)^{\text{1st}}$;
the second one is the Barr's flipped SU(5) that we denoted $\SU(5)^{\text{2nd}}$.
We can define a $\U(5)_{\hat q=2} \equiv \frac{\SU(5) \times \U(1)_{\hat q=2} }{\Z_5}$ for both versions of SU(5), which we call the two versions 
$\U(5)_{\hat q=2}^{\text{1st}}$ and $\U(5)_{\hat q=2}^{\text{2nd}}$.

We denote the GG's $\U(5)$ as the first kind $\U(5)_{\hat q=2}^{\text{1st}}$ embedded inside the Spin(10), where
$$
\U(5)_{\hat q=2}^{\text{1st}} \equiv \frac{\SU(5)^{\text{1st}} \times \U(1)_{X,\hat q=2} }{\Z_5} = \frac{\SU(5)^{\text{1st}} \times \U(1)_{X_1,\hat q=2} }{\Z_5}.
$$
We denote $\U(1)_{X}$ also the $\U(1)_{X_1}$ which is also generated by the 25th Lie algebra generator $T_{25}$ of this $\U(5)_{\hat q=2}^{\text{1st}}$.  
The $\hat q=2$ specifies the identification between the $\SU(5)$'s center and the normal subgroup of $\U(1)$ via our definition in \eq{eq:U5q}.

We denote the Barr's flipped $\U(5)$ as the second kind $\U(5)_{\hat q=2}^{\text{2nd}}$ embedded inside the Spin(10), where
$$
\U(5)_{\hat q=2}^{\text{2nd}} \equiv \frac{\SU(5)^{\text{1st}} \times \U(1)_{\chi,\hat q=2} }{\Z_5} = \frac{\SU(5)^{\text{1st}} \times \U(1)_{X_2,\hat q=2} }{\Z_5}.
$$
We denote $\U(1)_{\chi}$ also the $\U(1)_{X_2}$ which is also generated by the 25th Lie algebra generator $T_{25}$ of this $\U(5)_{\hat q=2}^{\text{2nd}}$.  

%{\bf Potential categorical symmetry and its retraction}: 
There is a magnetic 1-symmetry ${\U(1)}_{[1]}^{m_{X_1}}$ from the
GG's $\pi_1(\U(5)_{\hat q=2}^{\text{1st}})^\vee= \pi_1(  \U(1)_{X_1})^\vee= \U(1)$.
There is a magnetic 1-symmetry ${\U(1)}_{[1]}^{m_{X_2}}$ from the
Barr's flipped model's $\pi_1(\U(5)_{\hat q=2}^{\text{2nd}})^\vee= \pi_1(  \U(1)_{X_2})^\vee= \U(1)$.
There is a $\Z_2^{\rm flip}$ transformation, swapping between $\U(5)_{\hat q=2}^{\text{1st}}$ and $\U(5)_{\hat q=2}^{\text{2nd}}$.
Naively if we study a gauge theory including the union of the gauge group $\U(5)_{\hat q=2}^{\text{1st}}$ and $\U(5)_{\hat q=2}^{\text{2nd}}$,
denoted as $\U(5)_{\hat q=2}^{\text{1st}} \cup \U(5)_{\hat q=2}^{\text{2nd}}$ together with the outer automorphism exchanging these two U(5)s, as
``$(\U(5)_{\hat q=2}^{\text{1st}} \cup \U(5)_{\hat q=2}^{\text{2nd}}) \rtimes \Z_2^{\rm flip}$,''
then we expect to find a potential categorical symmetry for this 4d gauge theory. However, such a potential categorical symmetry is not realized,
for various reasons that we explore in this section.

\subsection{Categorical symmetry retraction from two $\U(5)_{\hat q=2}$ to Spin(10)}
\label{sec:Categorical-symmetry-U5}
\begin{enumerate}[leftmargin=.mm] 
\item {\bf Need Spin(10) to contain both of the two $\U(5)_{\hat q=2}$}:
Dynamically gauging the union $(\U(5)_{\hat q=2}^{\text{1st}} \cup \U(5)_{\hat q=2}^{\text{2nd}})$ already inevitably brings us to the full gauge group Spin(10) 
(checked in Appendix \ref{app:Flipping} and \eq{eq:twoU5-Spin10}).
Thus, needless to say whether we gauge the $\Z_2^{\rm flip}$ or not, when the two U(5)s are gauged, we are already at the full Spin(10).
For a pure Spin(10) gauge theory without matter fields, 
there is only a 1-form electric global symmetry $\Z_{4,[1]}$, and no categorical symmetry. 
 For the Spin(10) gauge theory with fermions in the {\bf 16},
there are neither higher symmetries (according to \Table{table:higher-sym-GUTmodel}) nor categorical symmetries.
\item {\bf GUT-Higgs scale eliminates some electric 1-symmetries}:
We hope to rationalize and characterize why 1-symmetries are gone at the deeper UV in the Spin(10) gauge group,
but there seems to have 1-symmetries in the $\U(5)_{\hat q=2}^{\text{1st}}$ and $\U(5)_{\hat q=2}^{\text{2nd}}$ gauge theories.
The closer look at the Higgs mechanism from the $so(10)$ GUT model with Spin(10) gauge group
to the U(5) gauge theories, we already add GUT-Higgs with the rep ${\bf 45}$ of Spin(10), so we can Higgs Spin(10) down to U(5).
But this rep ${\bf 45}$ has the branching rule from Spin(10) to SU(5)
as
${\bf 45} \sim {\bf 1} \oplus {\bf 10} \oplus {\overline{\bf 10}} \oplus {\bf 24}$ or to the $\U(5)_{\hat q=2}$
${\bf 45} \sim {\bf 1}_0 \oplus {\bf 10}_4 \oplus {\overline{\bf 10}}_{-4} \oplus {\bf 24}_0$.
These branching rules tell us that the electric 1-symmetries, if any, are broken at the GUT-Higgs scale.
Thus, any electric 1-symmetries (such as the ${\Z}_{5,[1]}^e$ for the pure SU(5) gauge theory, or the ${\U(1)}_{[1]}^e$ for the pure U(5) gauge theory)
could be emergent at much lower energy at IR far below the GUT-Higgs scale.
\end{enumerate}

How are the two magnetic 1-symmetries ${\U(1)}_{[1]}^{m_{X_1}}$ and ${\U(1)}_{[1]}^{m_{X_2}}$ disappear when we go to deeper energy from two $\U(5)_{\hat q=2}$ to the Spin(10)?
After all the GUT-Higgs are in the rep of the electric sector not the magnetic sector, so the GUT-Higgs does not remove the magnetic 1-symmetries.
In order to understand how magnetic 1-symmetries disappear or are removed, we study a toy model involving only the two crucial U(1) gauge sector: the $(\U(1)_{X_1} \times_{\Z_{4,X}} \U(1)_{X_2})$ gauge theory.

\subsection{Categorical symmetries in a toy model $\big[ (\U(1)_{X_1} \times_{\Z_{4,X}} \U(1)_{X_2}) \rtimes \Z_2^{\rm flip} \big]$ gauge theory}

We introduce a toy model of $\U(1)_{X_1} \times_{\Z_{4,X}} \U(1)_{X_2}$ gauge theory,
in terms of two U(1) sectors: $\U(1)_{X_1} \equiv \U(1)_{X}^{\text{1st}}$ and $\U(1)_{X_2} \equiv \U(1)_{\chi}^{\text{2nd}}$, 
while $\U(1)_{Y_1} \equiv \U(1)_{\tilde Y}$ and $\U(1)_{Y_2} \equiv \U(1)_{T_{24}}^{\text{2nd}}$, 
we have the following spacetime-internal structures with the spacetime Spin group: 
\bea
&&\hspace{-6mm}
\Spin \times_{\Z_2^\rF} (\U(1)_{X_1} \times_{\Z_{4,X}} \U(1)_{X_2})\cr 
&&\hspace{-6mm}
=\left\{\begin{array}{l}
\Spin \times_{\Z_2^\rF} (\U(1)_{Y_1} \times_{\Z_{5}} \U(1)_{X_1}) = 
(\Spin \times_{\Z_2^\rF} \U(1)_{X_1}) \times_{\Z_{5}} \U(1)_{Y_1}
\supset 
(\Spin \times_{\Z_2^\rF} \Z_{4,X_1}) \times_{\Z_{5}} \U(1)_{Y_1}.
\\
\Spin \times_{\Z_2^\rF} (\U(1)_{Y_2} \times_{\Z_{5}} \U(1)_{X_2}) = 
(\Spin \times_{\Z_2^\rF} \U(1)_{X_2}) \times_{\Z_{5}} \U(1)_{Y_2}
\supset 
(\Spin \times_{\Z_2^\rF} \Z_{4,X_2}) \times_{\Z_{5}} \U(1)_{Y_2}. %\nn
\end{array}
\right.
\quad\quad\quad
\eea
Dynamically gauge the internal symmetry group $(\U(1)_{X_1} \times_{\Z_{4,X}} \U(1)_{X_2})= (\U(1)_{Y_1} \times_{\Z_{5}} \U(1)_{X_1})= (\U(1)_{Y_2} \times_{\Z_{5}} \U(1)_{X_2})$,
we obtain their gauge theory. 

There is also a $\Z_2^{\rm flip}$ as an \emph{outer automorphism} of $(\U(1)_{X_1} \times_{\Z_{4,X}} \U(1)_{X_2})$ 
exchanging the two $\U(1)$ subgroups, which we can define on this spacetime-internal structures
\bea
\Spin \times_{\Z_2^\rF} \big((\U(1)_{X_1} \times_{\Z_{4,X}} \U(1)_{X_2})\rtimes  \Z_2^{\rm flip}\big).
\eea
Several comments about their higher symmetries and categorical symmetries, and their potential obstructions:
\begin{enumerate}[leftmargin=.mm] 

\item {\bf Higher symmetries without gauge charged matter}:

For a pure 4d $G_g=(\U(1)_{X_1} \times_{\Z_{4,X}} \U(1)_{X_2})$ gauge theory without gauge charged matter, we have the center
$$Z(G_g)=(\U(1)_{X_1} \times_{\Z_{4,X}} \U(1)_{X_2})$$ itself inducing the product of the two U(1) electric 1-symmetries, denoted as
${{\U(1)}_{[1]}^{e_{X_1}} \times_{\Z_{4,X[1]}} {\U(1)}_{[1]}^{e_{X_2}}}$
modding out the shared common ${\Z_{4,X}}$ 1-symmetry.
We have 
$$\pi_1(G_g)^{\vee}=\Hom(\pi_1(G_g), \U(1))=\Hom(\Z_{X_1} \times_{} \Z_{X_2}, \U(1)) = 
\U(1)_{X_1} \times_{} \U(1)_{X_2}$$
inducing the magnetic 1-symmetry ${{\U(1)}_{[1]}^{m_{X_1}} \times_{} {\U(1)}_{[1]}^{m_{X_2}}}$. 

\item {\bf $\big[ (\U(1)_{X_1} \times_{\Z_{4,X}} \U(1)_{X_2}) \rtimes \Z_2^{\rm flip} \big]$ gauge theory}:

If we further gauge the $\Z_2^{\rm flip}$ symmetry, we get a nonabelian gauge theory of a gauge group 
$$
\big[ (\U(1)_{X_1} \times_{\Z_{4,X}} \U(1)_{X_2}) \rtimes \Z_2^{\rm flip} \big]. 
$$
If such a 4d gauge theory $\big[ (\U(1)_{X_1} \times_{\Z_{4,X}} \U(1)_{X_2}) \rtimes \Z_2^{\rm flip} \big]$ exists as part of the GUT or BSM physics,
we have a potential categorical symmetry that we elaborate below in the Remark \ref{remark:Potential-categorical}.\footnote{Follow the discussion in 
\cite{MonteroRudelius2104.07036}, 
a similar example is gauging the $\Z_2$ outer automorphism exchanging the two $\rE_8$ in an $(\rE_8 \times \rE_8)$ gauge theory. 
The outcome is  a disconnected compact Lie group gauge theory 
$$(\rE_8 \times \rE_8) \rtimes \Z_2$$
referred to as the gauge symmetry of the
$(\rE_8 \times \rE_8)$ heterotic string theory \cite{DineLeighMacIntire1992}. But the center $Z(\rE_8)=0$ and the $\pi_1(\rE_8)^{\vee}=0$, 
so there are no higher nor categorical symmetries in this gauge theory.}

\item {\bf Electric higher symmetries}: 

For a 4d $G_g=(\U(1)_{X_1} \times_{\Z_{4,X}} \U(1)_{X_2})$ gauge theory with SM gauged electrically charged matter,
the 1-form $e$ symmetry ${{\U(1)}_{[1]}^{e_{X_1}} \times_{\Z_{4,X[1]}} {\U(1)}_{[1]}^{e_{X_2}}}$ 
is explicitly broken to only a subgroup ${\U(1)}_{[1]}^{e}$. 
Furthermore, the remained ${\U(1)}_{[1]}^{e}$ is gone when embedding $(\U(1)_{X_1} \times_{\Z_{4,X}} \U(1)_{X_2})$ into the $\U(5)_{\hat q=2}$ subgroup.
Thus, we are not interested in pursuing the electric higher symmetries further since they are all broken at the GG and flipped $su(5)$ models with matter.

\item {\bf Magnetic higher symmetries}: 

For a 4d $G_g=(\U(1)_{X_1} \times_{\Z_{4,X}} \U(1)_{X_2})$ gauge theory with SM gauged electrically charged matter, 
but we are still left with two 1-form $m$ symmetries preserved: 
${{\U(1)}_{[1]}^{m_{X_1}} \times_{} {\U(1)}_{[1]}^{m_{X_2}}}$, together with the SO structure and the $\Z_2^{\rm flip}$ symmetry.
The full spacetime-internal symmetry for this gauge theory requires at least the structure:
$$
\SO \times \Big( \big( {{\U(1)}_{[1]}^{m_{X_1}} \times_{} {\U(1)}_{[1]}^{m_{X_2}}} \big) \rtimes \Z_2^{\rm flip} \Big)
$$
where the $\Z_2^{\rm flip}$ belongs to an \emph{outer automorphism} of ${{\U(1)}_{[1]}^{m_{X_1}} \times_{} {\U(1)}_{[1]}^{m_{X_2}}}$ 
exchanging the two $\U(1)_{[1]}$.

If we gauge $\Z_2^{\rm flip}$ to obtain the
4d gauge theory $\big[ (\U(1)_{X_1} \times_{\Z_{4,X}} \U(1)_{X_2}) \rtimes \Z_2^{\rm flip} \big]$, naively we only require the 
spacetime-internal symmetry structure:
$$
\SO \times \Big( \big( {{\U(1)}_{[1]}^{m_{X_1}} \times_{} {\U(1)}_{[1]}^{m_{X_2}}} \big) \Big).
$$
But there is a potential categorical symmetry that we elaborate below in the Remark \ref{remark:Potential-categorical}.
\item \label{remark:Potential-categorical}
{\bf Potential categorical symmetry in $\big[ (\U(1)_{X_1} \times_{\Z_{4,X}} \U(1)_{X_2}) \rtimes \Z_2^{\rm flip} \big]$ gauge theory}:

Just like $\O(2)=\U(1) \rtimes \Z_2$ or $\U(1) \rtimes S_N$ gauge theories
has categorical symmetries \cite{MonteroRudelius2104.07036, NguyenTanizakiUnsal2101.02227},
we can look at the potential categorical symmetry of the $\big[ (\U(1)_{X_1} \times_{\Z_{4,X}} \U(1)_{X_2}) \rtimes \Z_2^{\rm flip} \big]$ gauge theory.
As mentioned since the electric higher symmetries are explicitly broken by gauged charged matter, 
we focus on the potential categorical symmetry involving magnetic higher symmetries. 
Each of the $U(1)_{X_1}$ and $U(1)_{X_2}$ has their own 1d 't Hooft line operators and 2d magnetic topological surface operators:
$$
T^{X_j}_{q_{m_j}}  \equiv \exp(\ii  q_{m_j} \oint_{\gamma^1}  V_{X_j}  ), \quad \quad
U^{X_j}_{\theta_{m_j}} \equiv 
\exp(\ii  \theta_{m_j} \ointint_{ \Sigma^2}  \star \dd V_{X_j}  ), 
$$ for $j=1,2$ along some closed 1-curve and some closed 2-surfaces ${ \Sigma^2}$.
They are labeled by magnetic charges $q_{m_j}$ and magnetic angles $\theta_{m_j} \in [0, 2 \pi)$. 
But once the $\Z_2^{\rm flip}$ is dynamically gauged, these operators are not gauge invariant.
Instead, the gauge invariant
't Hooft line is\footnote{It is possible to use the representation theory of certain dual gauge group to describe these 't Hooft lines.
For example, the 't Hooft line in the 4d $\O(2)=\U(1) \rtimes \Z_2$ gauge theory can be understood as 
the Wilson line in the representation of the 4d $\tilde{\O}(2)=\frac{\U(1) \rtimes \Z_4}{\Z_2} = \Pin^-(2)$ gauge theory \cite{MonteroRudelius2104.07036}.
A naive expectation is the following: 
the 't Hooft line in our 4d $\big[ (\U(1)_{X_1} \times_{\Z_{4,X}} \U(1)_{X_2}) \rtimes \Z_2^{\rm flip} \big]$ gauge theory
can be understood as the Wilson line in the representation of the 4d
$\big[\frac{ (\U(1)_{X_1} \times_{\Z_{4,X}} \U(1)_{X_2}) \rtimes \Z_4^{\rm flip} }{\Z_2^{\rm flip}}\big]$ gauge theory.
All these semi-direct product should be defined via proper group extensions.
We leave this subtle topic on the ``S-duality'' of these theories for future work.}
\bea \label{eq:tHooft}
T^{\rm flip}_{q_m} \equiv
T^{X_1}_{q_m} +T^{X_2}_{q_m} \equiv \exp(\ii  q_{m} \oint_{\gamma^1}  V_{X_1}  )  + \exp(\ii  q_{m} \oint_{\gamma^1}   V_{X_2} ). 
\eea
The gauge invariant topological magnetic surface operator is
\bea \label{eq:topological-magnetic-surface}
U^{\rm flip}_{\theta_m} \equiv 
U^{X_1}_{\theta_m}+
U^{X_2}_{\theta_m} \equiv 
\exp(\ii  \theta_m \ointint_{ \Sigma^2}  \star \dd V_{X_1}  )  
+\exp(\ii  \theta_m \ointint_{ \Sigma^2}  \star \dd  V_{X_2}    ). 
\eea
Under the $\Z_2^{\rm flip}$ gauged transformation, the operators labeled by $X_1$ and $X_2$ are swapped.

The 't Hooft lines in \eq{eq:tHooft} are non-topological, in the sense that their deformations introduce propagations of dual photons.
The fusions of these non-topological operators may introduce short-distance singularities and other non-universal contributions in the operator product expansion (OPE).
Nevertheless, there is still a fusion rule following from the topological contribution
\bea
T^{\rm flip}_{q_m} \times T^{\rm flip}_{q'_m}
=T^{\rm flip}_{q_m + q'_m}
+T^{X_1}_{q_m} \times T^{X_2}_{q'_m}
+T^{X_1}_{q'_m} \times T^{X_2}_{q_m}.
\eea
The topological magnetic 2-surface operators are topological of quantum dimension 2, with the following fusion rule: 
\bea
&&U^{\rm flip}_{\theta_m}  \times U^{\rm flip}_{\theta'_m}  = \exp(\ii  {({\theta_m}+ {\theta'_m})} \ointint_{ \Sigma^2}  \star \dd V_{X_1}  )  
+\exp(\ii  {({\theta_m}+ {\theta'_m})} \ointint_{ \Sigma^2}  \star \dd  V_{X_2}    ) \cr 
&&\quad\quad
+\exp(\ii  \theta_m \ointint_{ \Sigma^2}  \star \dd V_{X_1} )  
\exp(\ii  \theta'_m \ointint_{ \Sigma^2}  \star \dd  V_{X_2})    
+\exp(\ii  \theta'_m \ointint_{ \Sigma^2}  \star \dd V_{X_1})   
\exp(\ii \theta_m \ointint_{ \Sigma^2}  \star \dd  V_{X_2})    \quad\quad\quad\quad \cr
&&\quad\quad
=U^{\rm flip}_{\theta_m+ \theta'_m}
+
(U^{X_1}_{\theta_m}
\times U^{X_2}_{\theta'_m} 
+
U^{X_1}_{\theta'_m}
\times U^{X_2}_{\theta_m}).  \label{eq:fusion-2-surface-simple}
\eea
The fusion rule splits, which indicates the magnetic 2-surface operators are non-invertible. 
Only the first operator $U^{\rm flip}_{\theta_m+ \theta'_m}$ of quantum-dimension 2 is 
the same type of magnetic 2-surface operator \eq{eq:topological-magnetic-surface} that we start with.
The remained term is a more general type of 2-surface operator 
\bea
U^{\rm flip}_{\theta_m, \theta'_m} \equiv U^{\rm flip}_{\theta'_m, \theta_m} 
\equiv (U^{X_1}_{\theta_m} \times U^{X_2}_{\theta'_m} +
U^{X_1}_{\theta'_m} \times U^{X_2}_{\theta_m})
\eea 
carrying the dependence of $({\theta_m, \theta'_m})$ 
that still gauge invariant under the $\Z_2^{\rm flip}$'s swapping $X_1 \leftrightarrow X_2$.
The earlier topological surface operator in \eq{eq:topological-magnetic-surface} is the special kind of the more general case: 
$$
U^{\rm flip}_{\theta_m} \equiv U^{\rm flip}_{\theta_m, 0} \equiv U^{\rm flip}_{0, \theta_m}, \quad  U^{\rm flip}_{\theta'_m} \equiv U^{\rm flip}_{\theta'_m, 0} \equiv U^{\rm flip}_{0, \theta'_m}.
$$
The fusion rule between two such general operators also splits
\begin{multline} \label{eq:fusion-2-surface}
U^{\rm flip}_{\theta_m, \theta'_m} \times U^{\rm flip}_{\vartheta_m, \vartheta'_m}
=(U^{X_1}_{\theta_m+\vartheta_m} \times U^{X_2}_{\theta'_m +\vartheta'_m} +
U^{X_1}_{\theta'_m+\vartheta'_m} \times U^{X_2}_{\theta_m +\vartheta_m})
+(U^{X_1}_{\theta_m+\vartheta'_m} \times U^{X_2}_{\theta'_m +\vartheta_m} +
U^{X_1}_{\theta'_m+\vartheta_m} \times U^{X_2}_{\theta_m +\vartheta'_m})\\
=U^{\rm flip}_{{\theta_m+\vartheta_m}, {\theta'_m +\vartheta'_m}} + U^{\rm flip}_{{\theta_m +\vartheta'_m}, {\theta'_m +\vartheta_m}}
\end{multline}
The global symmetry generated by \eq{eq:topological-magnetic-surface} is thus {\bf \emph{non-invertible}}.
It is a {\bf \emph{non-invertible global symmetry}}, or a {\bf \emph{categorical symmetry}}.
\item \label{remark:categorical-retract}
{\bf Categorical symmetry retraction}:
Where does the categorical symmetry  \eq{eq:fusion-2-surface} of the $\big[ (\U(1)_{X_1} \times_{\Z_{4,X}} \U(1)_{X_2}) \rtimes \Z_2^{\rm flip} \big]$ gauge theory
go in the end, after embedding into the GUTs?
\begin{enumerate}[leftmargin=.mm] 
\item Follow \Sec{sec:Categorical-symmetry-U5},
the categorical symmetry is retracted when we embed $\U(1)_{X_1} \subset \U(5)_{\hat q=2}^{\text{1st}}$ 
and $\U(1)_{X_2} \subset \U(5)_{\hat q=2}^{\text{2nd}}$. Because the union of 
gauge group, either ``$(\U(5)_{\hat q=2}^{\text{1st}} \cup \U(5)_{\hat q=2}^{\text{2nd}})$''
or
``$(\U(5)_{\hat q=2}^{\text{1st}} \cup \U(5)_{\hat q=2}^{\text{2nd}}) \rtimes \Z_2^{\rm flip}$,''
require to gauge the full Spin(10) group with Weyl fermions in the {\bf 16}, which has no higher symmetries nor category symmetries.
\item Another possible explanation at IR of categorical symmetry breaking, 
is checking the possible mixed anomaly between the magnetic 1-form symmetries
$\big( {{\U(1)}_{[1]}^{m_{X_1}} \times_{} {\U(1)}_{[1]}^{m_{X_2}}} \big)$ and the $\Z_2^{\rm flip}$-symmetry.
If there is any such mixed anomaly, when the $\Z_2^{\rm flip}$-symmetry is dynamically gauged, then at least part of the magnetic 1-symmetry needs to be broken.
If so, the categorical symmetry must also be broken.

For example, based on the calculation in \cite{WanWang2018bns1812.11967, Wan2019sooWWZHAHSII1912.13504},
the 5th bordism group gives:
\begin{multline} \label{eq:SO-Z2-1-anomaly}
\Omega^{\SO\times\Z_2\times\B\U(1)^{m_{X_1}}\times\B\U(1)^{m_{X_2}} }_5=\Z_2^5 ,  \\
\text{generated by } w_2w_3, \; \tau_5^{m_{X_1}} = \Sq^2 \tau_3^{m_{X_1}}  ,  \;\tau_5^{m_{X_2}} = \Sq^2 \tau_3^{m_{X_2}} ,  \; a^5,  \; aw_2^2.
\end{multline}
\begin{multline} \label{eq:O-Z2-1-anomaly}
\Omega^{\O\times\Z_2\times\B\U(1)^{m_{X_1}}\times\B\U(1)^{m_{X_2}} }_5=\Z_2^{11},  \\ 
\text{generated by }  
w_2w_3,a^5,a^3w_1^2,aw_1^4,aw_2^2,a^2\tau_3^{m_{X_1}},a^2\tau_3^{m_{X_2}},w_1^2\tau_3^{m_{X_1}},w_1^2\tau_3^{m_{X_2}},\tau_5^{m_{X_1}},\tau_5^{m_{X_2}}.
\end{multline}
The particular cobordism classification of 4d anomalies that fits our theory is this \cite{HAHSIV}:
\bea\label{eq:SO-Z2-1-twist-anomaly}
\Omega^{\SO\times\Z_2\ltimes(\B\U(1)^{m_{X_1}}\times\B\U(1)^{m_{X_2}}) }_5.
\eea
However, the twisted cobordism calculation is more difficult, \Refe{HAHSIV} predicts that the 
$\Omega^{\SO\times\Z_2\ltimes(\B\U(1)^{m_{X_1}}\times\B\U(1)^{m_{X_2}}) }_5$ is either the same as $\Z_2^5$ in \eq{eq:SO-Z2-1-anomaly},
or as $\Z_2^4$
where the two generators $\tau_5^{m_{X_1}} = \Sq^2 \tau_3^{m_{X_1}}$ and $\tau_5^{m_{X_2}} = \Sq^2 \tau_3^{m_{X_2}}$ in $\Omega^{\SO\times\Z_2\times\B\U(1)^{m_{X_1}}\times\B\U(1)^{m_{X_2}} }_5$
reduce to a single one
$ \tau_5^{m_{X_1}} + \tau_5^{m_{X_2}}$ in $\Omega^{\SO\times\Z_2\ltimes(\B\U(1)^{m_{X_1}}\times\B\U(1)^{m_{X_2}}) }_5$.
Thus we can use the data from \eq{eq:SO-Z2-1-anomaly} and \eq{eq:O-Z2-1-anomaly}
to deduce whether our $ ( {{\U(1)}_{[1]}^{m_{X_1}} \times_{} {\U(1)}_{[1]}^{m_{X_2}}}) \rtimes \Z_2^{\rm flip}$ symmetry has any 't Hooft anomaly or not.
Several comments are in order:
\begin{enumerate}[leftmargin=.mm] 
\item  %$\bullet$ 
Here $w_j \equiv w_j(TM)$ is the $j$-th Stiefel-Whitney (SW) characteristic class of spacetime tangent bundle $TM$ of manifold $M$.\\
\item  %$\bullet$ 
The $\tilde \tau_3$ is the generator of $\H^3(\B^2\U(1),\U(1))=\Z$
and
$\tau_3 = (\tilde \tau_3 \mod 2)$ 
is the generator of $\H^3(\B^2\U(1),\Z_2)=\Z_2$. The $ \Sq^2$ is from the Steenrod square, here mapping the 3rd cohomology group to the 5th cohomology group. \\
\item  %$\bullet$ 
The upper labels ${m_{X_1}}$ and ${m_{X_2}}$ are for specifying magnetic 1-symmetries from either sector of ${{\U(1)}_{[1]}^{m_{X_1}} \times_{} {\U(1)}_{[1]}^{m_{X_2}}}$.\\
\item  %$\bullet$ 
The $a$ is the generator of $\H^1(\B\Z_2^{\rm flip},\Z_2)=\Z_2$.\\
\item  %$\bullet$ 
In particular, the classification in \eq{eq:O-Z2-1-anomaly} indicates the 
$a^2\tau_3^{m_{X_j}}=\Sq^1 (a\tau_3^{m_{X_j}})= w_1 a\tau_3^{m_{X_j}}$ term, which specifies the  potential mixed anomalies in 4d
between the $\Z_2^{\rm flip}$ symmetry and any of the magnetic U(1) 1-symmetries (for both $j=1,2$).\\
\item  %$\bullet$ 
To check whether this $a^2\tau_3^{m_{X_j}}$ anomaly is present in our theory,
we can couple the magnetic 1-symmetry ${\U(1)}_{[1]}^{m_{X_j}}$ to the 2-form magnetic background $B^{m_{X_j}}$ field,
and couple the $\Z_2^{\rm flip}$ symmetry to the 1-form or 1-cochain gauge field $a$.
Suppose in the presence of $B^{m_{X_j}}$ field, under the larger gauge transformation of $\Z_2^{\rm flip}$ symmetry,
the partition function is not fully invariant but obtains only a $(-1)$ phase, then it signals that the theory has a mod 2 global anomaly.\\
\item  %$\bullet$ 
Under the $\Z_2^{\rm flip}$ symmetry,
the $\U(1)_{X_1}$ and $\U(1)_{X_2}$ are swapped.
Thus their associated gauged charges 
of the quarks and leptons are also swapped.
Out of the 16 Weyl fermions per generation,
the conventional left-handed Weyl spinors 
($u_L$, $d_L$, $\nu_L$, and $e_L$, those also coupled to the SU(2)$_{\rm L}$) maintain their $\U(1)_{X_1}=\U(1)_{X_2}$ gauged charges;
but only the conventional right-handed Weyl spinors 
($\bar{u}_R$, $\bar{d}_R$, $\bar{\nu}_R$, and $\bar{e}_R$,
those can be coupled to the SU(2)$_{\rm R}$, which we also flip them to the left-handed particle as the right-handed anti-particle)
swap their $\U(1)_{X_1}$ and $\U(1)_{X_2}$ gauged charges:
\bea\label{eq:Z2flip-change}
\begin{tabular}{lcc c}
\hline
&  $\U(1)_{X_1} \equiv \U(1)_{X}^{\text{1st}}$ & $\U(1)_{X_2} \equiv \U(1)_{\chi}^{\text{2nd}}$ & \\
\hline
 $\bar{u}_R$&  1 & $-3$\\
\hline
$\bar{d}_R$& $-3$ & 1 \\
\hline
$\bar{\nu}_R= {\nu}_L $& 5 & 1\\
\hline
$\bar{e}_R= e_L^+$& 1 & 5\\
\hline
\end{tabular}.
\eea
Although we do not yet know the full classification of anomalies from \eq{eq:O-Z2-1-anomaly},
but we have enough informations to deduce that actually our $\big[ (\U(1)_{X_1} \times_{\Z_{4,X}} \U(1)_{X_2}) \rtimes \Z_2^{\rm flip} \big]$ gauge theory is: 
$$
\text{{\bf mixed anomaly free} within $\Z_2^{\rm flip}$  and magnetic 1-symmetries
$\big( {{\U(1)}_{[1]}^{m_{X_1}} \times_{} {\U(1)}_{[1]}^{m_{X_2}}} \big) \rtimes \Z_2^{\rm flip}$}.
$$
The reasoning is that the 4d anomaly is captured by the large gauge transformation of 
$\Z_2^{\rm flip}$-background field $a \in \H^1(\B\Z_2^{\rm flip},\Z_2)=\Z_2$ and 
the ${\U(1)}_{[1]}^{m}$-background field 
$\tau_3 \in \H^3(\B^2\U(1),\Z_2)=\Z_2$.
To trigger a non-vanishing 4d anomaly, we must turn on the 3-dimensional background field $\tau_3$,
while the possible anomalies term (involving 1-symmetries) based on dimensional analysis counting into a 5d iTQFT can be:
$a^2\tau_3^{m_{X_j}}$ or $w_1 a\tau_3^{m_{X_j}}$ or $\tau_5^{m_{X_j}} = \Sq^2 \tau_3^{m_{X_j}} $.
Only $a^2\tau_3^{m_{X_j}}$ or $w_1 a\tau_3^{m_{X_j}}$ or some linear form of $a$ as $\tilde{w} a\tau_3^{m_{X_j}}$ 
(where $\tilde{w}$ can be some twisted 1-cochain specified by the structure of \eq{eq:O-Z2-1-anomaly})
involve the desired mixed anomalies.
But the anomaly coefficients 
seem to be zero by any consideration.\footnote{If the anomaly coefficient depends on the number of fermions are swapped under $\Z_2^{\rm flip}$,
then it is the two fermions swapped with the other two fermions in \eq{eq:Z2flip-change},
then the anomaly coefficient seems to be 0 mod 2.
If the anomaly coefficient depends on
their $\U(1)_{X_1}- \U(1)_{X_2}$ charge differences,
we still have $( (-3 -1)+ (1-(-3)) + (1-5) + (5-1)) =0$
%$$\text{ anomaly coefficient of } w_1 a\tau_3^{m_{X_j}} \text{ or } \tilde{w} a\tau_3^{m_{X_j}} \propto ( (-3 -1)+ (1-(-3)) + (1-5) + (5-1)) =0. $$
and $( (-3 -1)^2+ (1-(-3))^2 + (1-5)^2 + (5-1)^2)= 4^3 = 0 \mod 64.$
%$$\text{ anomaly coefficient of } a^2\tau_3^{m_{X_j}} \propto ( (-3 -1)^2+ (1-(-3))^2 + (1-5)^2 + (5-1)^2)= 4^3 = 0 \mod 64.$$
}
Given this data in \eq{eq:Z2flip-change}, these anomaly coefficients suggest that it is highly impossible to have any mod 2 global anomaly.
\item  %$\bullet$ 
Thus we propose the categorical symmetry  \eq{eq:fusion-2-surface} of the $\big[ (\U(1)_{X_1} \times_{\Z_{4,X}} \U(1)_{X_2}) \rtimes \Z_2^{\rm flip} \big]$ gauge theory
actually is sensible and well-defined kinematically, as long as if we do not embed the theory into the full combined theory of GG and flipped $su(5)$ GUTs. 
The categorical symmetry in \eq{eq:fusion-2-surface}  is only retracted after embedding into the GUTs.
\end{enumerate}

\end{enumerate}

\end{enumerate}
%
%
%

%\newpage

\section{Conclusion and Future Directions}
%{Conclusion: BSM Gauge Enhanced Quantum Criticality}
\label{sec:conclude}

Here are some concluding remarks and future directions:\\[-8mm]
\begin{enumerate}[leftmargin=2.mm, label=\textcolor{blue}{\arabic*}., ref={\arabic*}] 
\item {\bf Three generations of SM quarks and leptons}: Our work so far had focused on 1 generation of SM quarks and leptons.
We can ask whether the number n of generations of quarks and leptons modify any story we had discussed previously.

In the conventional $so(10)$ GUT without a WZW term, there is no constraint on n generation since we can take n copies of theories.

In the modified $so(10)$ GUT with a WZW term,
the mod 2 class $w_2w_3$  anomaly  of \eq{eq:w2w3-2} 
is matched by the sector of GUT-Higgs fields (or their fractionalized partons) and their 4d WZW term alone,
we just need to ensure the anomaly index from GUT-Higgs WZW sector contributes 1 mod 2. 

$\bullet$ If each generation of 16 SM Weyl fermions associates with its own GUT-Higgs field,
then the generation number n times of 16 SM Weyl fermions with n GUT-Higgs WZW sector requires a constraint 
${\rm{n}} = 1 \mod 2$ to match the $w_2 w_3$ anomaly, where ${\rm{n}} = 3$ generation indeed works. 

$\bullet$ However, in general, we can just introduce a single (or any odd number) of GUT-Higgs WZW sector
to match the 1 mod 2 class of $w_2 w_3$ anomaly. 
After all, we may only need a single GUT-Higgs to achieve the gauge symmetry-breaking from Spin(10) to other subgroups, regardless of the number of n.
In this case, our discussion on quantum criticalities can be applied to any n of SM or GUT.

Of course, an open question is which model fits the best to the HEP phenomenology and experiment.  

\item {\bf What is mass?}

We come back to a philosophical or metaphysical question: What is mass? From a modern quantum matter perspective,
we can address this issue with a physical and mathematical answer.
The mass more generally describes a massive or gapped energy spectrum -- gapped with respect to the ground state(s).
Here the mass is defined as the correlation function (of the corresponding operators/excitations/states)
decaying exponentially.
Then we can further address what are the known mechanisms for providing a massive or gapped spectrum.

$\bullet$ Traditionally, the {free/single-particle/mean-field} mechanism includes
Anderson-Higgs {\cite{Anderson1963, EnglertBrout1964, Higgs1964}}
or chiral symmetry breaking {\cite{NambuJonaLasinio1961}}
with a bilinear quadratic mass term turned on in a Lagrangian or Hamiltonian theory.

$\bullet$ Modern quantum systems provide other many-body interacting mechanisms to generate a non-quadratic non-mean-field mass,
see \Table{table:mass}.

In our work \cite{Wang2106.16248}, 
and in Ultra Unification \cite{JW2006.16996, JW2008.06499, JW2012.15860},
these many-body interacting mechanisms ((4)-(6)) are used for the new BSM physics.
\begin{center}
\begin{table}[!h]
\hspace{-19mm}
\begin{tabular}{| lcc c |}
%\hline
%\multicolumn{4}{|c|}{}\\
\hline
Mass mechanism &   
$\begin{array}{c}
		\text{Symmetry Property}\\ 
		\end{array}$
 &
 $\begin{array}{c}
\text{Topological Order}\\ 
\text{with low energy TQFT} 
\end{array}$
  &  
   $\begin{array}{c}
   \text{Description:}\\
\text{Free/Single-Particle/Mean-Field}\\ 
\text{or Many-Body Interacting}\\ 
\text{(examples in references)}
\end{array}$
  \\
\hline
(1) Anderson-Higgs &  Symmetry Breaking & \xmark  & 
$\begin{array}{c}
\text{Mean-Field}\\ 
\text{\cite{Anderson1963, EnglertBrout1964, Higgs1964}}
\end{array}$ 
\\
\hline
 (2) $\begin{array}{c}
   \text{Confinement:}\\
\text{Chiral SB} 
\end{array}$ & Symmetry Breaking   & \xmark & 
$\begin{array}{c}
\text{Mean-Field}\\ 
\text{\cite{NambuJonaLasinio1961}}
\end{array}$ 
\\
\hline
 (3) $\begin{array}{c}
   \text{Confinement:}\\
\text{s confinement} 
\end{array}$ & 
$\begin{array}{c}
\text{Symmetry Preserving} 
\end{array}$ 
 & \xmark & Many-Body Interacting \\
\hline
 (4) $\begin{array}{c}
   \text{Symmetric}\\
\text{Mass}\\ 
\text{Generation}\\
 \text{(Anomaly-Free)}
\end{array}$
&  
$\begin{array}{c}
\text{Symmetry Preserving} 
\end{array}$  
&\xmark & 
$\begin{array}{c}
\text{Many-Body Interacting}\\ 
\text{\cite{FidkowskifSPT2,Wang2013ytaJW1307.7480,Wang2018ugfJW1807.05998,
YouHeXuVishwanath1705.09313, YouHeVishwanathXu1711.00863,
Eichten1985ftPreskill1986, Wen2013ppa1305.1045,
You2014oaaYouBenTovXu1402.4151, YX14124784,BenTov2015graZee1505.04312,
Kikukawa2017ngf1710.11618,WangWen2018cai1809.11171,
RazamatTong2009.05037, Tong2104.03997,
Catterall2020fep, CatterallTogaButt2101.01026}}
\end{array}$   
\\
\hline
(5) $\begin{array}{c}
   \text{Symmetric}\\
\text{Gapped}\\ 
\text{Topological Order}\\
 \text{(Anomalous)}
\end{array}$
& 
$\begin{array}{c}
\text{Symmetry Preserving} 
\end{array}$ 
 & \cmark & 
$\begin{array}{c}
\text{Many-Body Interacting}\\ 
\text{\cite{VishwanathSenthil1209.3058, Senthil1405.4015, Wang2017locWWW1705.06728}}
\end{array}$   
  \\
\hline
(6) $\begin{array}{c}
   \text{Symmetry}\\
\text{Extension}\\
\text{Gapped}\\
 \text{(Anomaly}\\
   \text{Trivialized)}
\end{array}$
&
$\begin{array}{c}
\text{Symmetry Extension}\\
\text{$K \to \tilde{G} \overset{\iota}{\to} G$} 
\end{array}$  
& 
$\begin{array}{c}
   \text{\cmark: TO/TQFT if $K$ is gauged,}\\
   \text{and if spacetime dim $d \geq 3$}\\[2mm]
\text{\xmark: no TO/TQFT if $\tilde{G}$}\\ 
\text{remains ungauged.} 
\end{array}$ 
& 
$\begin{array}{c}
\text{Many-Body Interacting}\\ 
\text{\cite{Wang2017locWWW1705.06728}}
\end{array}$  
   \\
\hline
%\hline
\end{tabular}
\caption{Some mechanisms to generate the mass or energy gap, both in the traditional
{free/single-particle/mean-field} level or the quantum many-body interacting systems.
{Chiral SB} for the chiral symmetry breaking.
The s confinement is the smooth confinement without symmetry breaking.
{\cmark} for Yes and {\xmark} for no.
The symmetry-extension mechanism and its short exact sequence or fibration {$K \to \tilde{G} \overset{\iota}{\to} G$}
to construct $G$-symmetric gapped phase, are summarized in {\cite{Wang2017locWWW1705.06728}}.
Symmetry-extension method extends the $G$-symmetry to $\tilde{G}$-symmetry so to
trivialize the anomaly by a gapped system, which can subsequently produce
Symmetric Gapped Topological Order (TO) if we gauge $K$ while still preserves $G$ in a spacetime dimension $d \geq 3$.
}
\label{table:mass}
\end{table}
\end{center}
%%%%%%%%%%%%%%%%%%%%%%%%%%%% 
%
%
%\newpage
\item {\bf What is criticality? What is a phase transition? }

Following footnote \ref{ft:criticality-phase-transition},
now we can revisit the terminology on
{\bf criticalities} vs {\bf phase transitions} with examples in \Fig{fig:phase}-\Fig{fig:phase-quadrant-2}. \\
$\bullet$ The {\bf criticality} means the system with gapless excitations (gapless thus critical, sometimes conformal) and with an infinite correlation length, 
it can be either (i) a {\bf continuous phase transition} (many examples in \Table{table:GUT-transition})
as an unstable critical point/line/etc. as an unstable renormalization group (RG) fixed point which has at least one relevant perturbation in the phase diagram, 
or (ii) a {\bf critical phase} (the white region (0) in \Fig{fig:phase}-\Fig{fig:phase-quadrant-2})
as a stable critical region controlled by a stable RG fixed point which does not have any relevant perturbation
in the phase diagram.\\
$\bullet$ The {\bf phase transition} \cite{subirsachdev2011book} means the phase interface between two (or more) bulk phases in the phase diagram.
The phase transition can be a {\bf continuous phase transition} (second order or higher order, with gapless modes; many examples in \Table{table:GUT-transition})
or a {\bf discontinuous phase transition} (first-order, without gapless modes, and with a finite correlation length; e.g., (II)-(VI) and (II)$'$-(VI)$'$
\Fig{fig:phase}-\ref{fig:phase-two}).\\
\item {\bf Boundary criticality or phase transition}: The modified $so(10)$ GUT with a 4d WZW term lives on a boundary of 5d invertible TQFT $w_2w_3(TM)=w_2w_3(V_{\SO(10)})$ of \eq{eq:w2w3-2}.
When the internal Spin(10) symmetry is treated as a global symmetry (not yet dynamically gauged),
the modern condensed matter viewpoint is that

(1) If we impose any regularization (e.g., lattice)
such that the internal Spin(10) symmetry acts \emph{onsite} (i.e., the global symmetry acts on a local site),
then we must realize the 4d criticality as a boundary criticality on the 4d boundary of the 5d bulk invertible TQFT.
Thus the 't Hooft anomaly in 4d is manifested by the boundary physics of 5d invertible TQFT via the anomaly inflow \cite{1984saCallanHarvey}.
The internal onsite Spin(10) symmetry can be dynamically gauged if we gauge altogether the full 4d boundary-5d bulk coupled system 
(e.g., the \emph{long-range entangled} gauged boundary-bulk coupled systems in Sec.~7 of \cite{Wang1801.05416}).

(2) In contrast, if we are allowed to realize the internal Spin(10) symmetry acts \emph{non-onsite}, then
the 't Hooft anomaly in 4d is realized as the obstruction to gauge this non-onsite symmetry in 4d.

Since we aim to gauge the internal symmetry at the end, we shall take the viewpoint (1), and 
we can interpret our criticalities or phase transitions 
in \Table{table:GUT-transition} as 4d boundary criticalities or phase transitions 
of a 5d gapped bulk.

\item {\bf Bulk criticality or phase transition}:
When the internal Spin(10) symmetry is dynamically gauged,
the 5d bulk turns from a gapped \eq{eq:w2w3-2} to a gapless system 
(because at the Gaussian fixed point where the pure Yang-Mills kinetic term $(\dd A)^2$ in 5d is marginal, the $A^2 \dd A$ and $A^4$ become weak and irrelevant at IR, 
hence the confined gauge fields in 4d becomes deconfined and gapless in the 5d bulk).
The purple regions of the phases with the gauged Spin(10) in
\Fig{fig:phase-quadrant-1} and \Fig{fig:phase-quadrant-2} thus have a 5d gapless bulk.

$\bullet$ The Spin(10) preserving system (purple regions) can be regarded as the GUT-Higgs field \emph{disordered} phase, 
which has the GUT-Higgs field \emph{disordered} both in the 4d boundary and 5d bulk.

$\bullet$ The Spin(10) breaking system (regions other than purple) can be regarded as the GUT-Higgs field \emph{ordered} phase on the 4d boundary,
but the GUT-Higgs field can be either \emph{ordered} (breaking) or \emph{disordered} (preserving) in the 5d bulk 
(e.g., see a recent study \cite{Metlitski2020Boundary} in condensed matter, and references therein). 

So the 5d bulk phases (in \Fig{fig:phase-quadrant-1} and \Fig{fig:phase-quadrant-2})
moving from the purple region to non-purple regions,
can go through either ---

$\bullet$ 5d bulk phase transition:
a \emph{disordered} (Spin(10) preserving and gauge boson \emph{gapless}) phase to an \emph{ordered} (Spin(10) breaking to subgroups and gauge boson \emph{partially gapped}) phase, which happens together with the boundary disorder-order transition. The presence of Spin(10) anomaly on the 4d boundary (and the SPT order in the 5d bulk) could modify the 
critical exponents
and
scaling dimensions of the critical GUT-Higgs fields at the transition, which was discussed as gapless SPT states in condensed matter literature \cite{Zhang2017Unconventional,Scaffidi2017Gapless,Parker2018Topological,Xu2020Topological,Wu2020Boundary,Zhu2021Surface,Duque2021Topological,Ma2021Edge}.

$\bullet$ 5d bulk no phase transition:
maintain a \emph{disordered} (Spin(10) preserving and gauge boson gapless) phase. In this case, the \emph{disorder}-\emph{order} transition is only a boundary transition that happens only in the 4d boundary \cite{Metlitski2020Boundary}.

The 5d bulk phases (in \Fig{fig:phase-quadrant-1} and \Fig{fig:phase-quadrant-2})
moving between different non-purple regions, can introduce another possibility ---

$\bullet$ 5d bulk phase transition:
between different \emph{ordered} (Spin(10) breaking to subgroups and gauge boson \emph{partially gapped}) phases.

\item {\bf 16n vs 15n Weyl fermions, and topological criticality or topological phase transition}:

The SM and GG $su(5)$ models can have either choice of 15 or 16 Weyl fermions for each generation. 
In contrast, in order to be consistent with the SM data constraint,
the PS, the GG $u(5)$, the flipped $u(5)$, and the modified $so(10)$ have 16 Weyl fermions per generation.
Thus, in all the figures from \Fig{fig:phase}-\Fig{fig:phase-two} and \Fig{fig:phase-quadrant-1}-\Fig{fig:phase-quadrant-2}, we see that the hyperplane/line 
set at $r_{\mathbf{45}} =0$ separates
one side $r_{\mathbf{45}} >0$ which has models with 16n Weyl fermions, 
while the other side  $r_{\mathbf{45}} <0$
can have models with either choice of 15n or 16n Weyl fermions
(except the flipped $u(5)$ can only have 16n Weyl fermions).

%\cred
{Overall, we could view the QFTs are governed by a deformation class of QFTs 
by replacing \eq{eq:w2w3-2}
to
\begin{multline}\hspace{-4mm}
\label{eq:Z2Z16}
{{\bf Z}_{{\text{$5$d-iTQFT}}}^{({\rm p}, \upnu)} }
\equiv
%(-1)^{{\rm p} \int_{M^5} w_2 w_3} 
\exp(\ii \pi\cdot {\rm p}\cdot \int_{M^5} w_2 w_3) 
\cdot
\exp(\frac{2\pi \ii}{16} \cdot\upnu \cdot \eta(\text{PD}(\CA_{{\Z_4}} \mod 2)) \bigg\rvert_{M^5}), \\ 
\text{ with } 
{\rm p} \in \Z_2, \quad
\text{a 4d Atiyah-Patodi-Singer $\eta$ invariant} \equiv \eta_{\text{Pin}^+} \in \Z_{16}, \quad \upnu\in \Z_{16}
.\;
\end{multline}
The QFTs with the ungauged internal global symmetries can be deformed to each other, as long as they are matched by 
the 4d boundary 't Hooft anomaly of the 5d cobordism class \eq{eq:Z2Z16}:
First, a potential global $\Z_2$ anomaly, the $w_2 w_3$ anomaly for our 4d WZW term.
Second, the $\Z_{16}$ global anomaly captured by a 5d version of Atiyah-Patodi-Singer (APS) eta invariant for the $\Spin \times_{\Z_2^F} \Z_{4,X}$-structure
 from $\TP_5(\Spin \times_{\Z_2^F} \Z_{4,X})=\Z_{16}$. We can write that 5d APS invariant in terms of the 4d APS invariant of $\Pin^+$-structure from $\TP_4(\Pin^+)=\Z_{16}$.
 The $\CA_{{\Z_4}} \in \H^1(M, \Z_{4,X})$ is a cohomology class discrete gauge field of the $\Z_{4,X}$-symmetry.
The two combined invertible TQFTs, labeled by ${\rm p} \in \Z_2$ and $\upnu\in \Z_{16}$, have a partition function ${{\bf Z}_{{\text{$5$d-iTQFT}}}^{({\rm p}, \upnu)} }$ on $M^5$.
}

$\bullet$ On the GG $u(5)$ model, if we are able to break down the continuous $\U(1)_{X_1}$ variant of 
the baryon minus lepton number $X_1 \equiv 5({ \mathbf{B}-  \mathbf{L}})- \frac{2}{3}Y_1$ symmetry
down to a discrete variant of $\Z_{4,X_1} = \Z_{4,X}$ around the hyperplane at $r_{\mathbf{45}} =0$,
then the $\Z_{4,X}$-symmetry gives rise to a $\Z_{16}$ class global anomaly 
\cite{GarciaEtxebarriaMontero2018ajm1808.00009, WW2019fxh1910.14668}, 
such that we can introduce new sectors in addition to 15n Weyl fermions to go beyond the SM: 
via adding gapped TQFTs or gapless CFTs to cancel the global anomaly (known as Ultra Unification \cite{JW2006.16996, JW2008.06499, JW2012.15860}).

$\bullet$ On the flipped $u(5)$ model, there is also a continuous $\U(1)_{X_2}$ symmetry,
where $X_2 \equiv \frac{1}{5} X_1 + \frac{4}{5} Y_1 =
({ \mathbf{B}-  \mathbf{L}}) + \frac{2}{3}Y_1$, but this $\U(1)_{X_2}$ \emph{cannot} be broken down to $\Z_{4,X_2} = \Z_{4,X}$
if we want to preserve the SM's $\U(1)_{Y_1}$. 
So the $\Z_{16}$ class global anomaly of $\Z_{4,X}$-symmetry does not occur to constrain the flipped $u(5)$ model.

So over the 8 octants in \Fig{fig:phase}-\Fig{fig:phase-quadrant-2}, 
only the regions of the blue (II) and the red (III)=(VII) can have 
the 15n Weyl fermion scenario with new BSM TQFT/CFT sectors.
Also their ${\U(1)'}_{\text{gauge}}^{\text{dark}}$-deconfined analogs 
(II)$'$ and (III)$'$=(VII)$'$ can have 
${\U(1)'}_{\text{gauge}}^{\text{dark}}$-\emph{deconfined criticality}
as well as \emph{topological criticality} simultaneously. 

An interesting future direction is that: 
Under what mechanism without fine-tuning, 
can the 16n to 15n Weyl fermion \emph{topological phase transition} (involving with 4d TQFTs/CFTs on the 15n fermion side)
occur as \emph{the same phase transition simultaneously} 
as those phase transitions described previously in \Table{table:GUT-transition}?

\item {\bf Completeness Hypothesis vs Absence of Global Symmetries vs Absence of Topological Operators}:

We have found that the potential categorical symmetry in the low energy sector of $\big[ (\U(1)_{X_1} \times_{\Z_{4,X}} \U(1)_{X_2}) \rtimes \Z_2^{\rm flip} \big]$-gauge theory
(presumably if we Higgs down the $so(10)$ GUT's Spin(10) down to this restricted subgroup 
 of the two U(5) GUTs: GG and flipped models). 
But the categorical symmetry in fact disappears to none, 
even when we try to embed this $\big[ (\U(1)_{X_1} \times_{\Z_{4,X}} \U(1)_{X_2}) \rtimes \Z_2^{\rm flip} \big]$ to 
the dynamically gauged union of $(\U(5)_{\hat q=2}^{\text{1st}} \cup \U(5)_{\hat q=2}^{\text{2nd}})$, 
which already inevitably requires the full gauge group Spin(10) 
(see \eq{eq:twoU5-Spin10}).

Although our present work only studies QFT and GUT models (not quantum gravity),
we find the \emph{absence} of generalized global symmetries (neither higher symmetries nor categorical symmetries) in the UV model.
It will be interesting to know whether this phenomenon is relatable to 
several conjectures on the universal features of quantum gravity (QG): \\
$\bullet$ {\bf Completeness Hypothesis} about the spectrum \cite{Polchinski2003bq0304042, Banks2010zn1011.5120, HarlowOoguri1810.05338}: every representation of any gauge group must be occupied by particle states.\\ 
$\bullet$ {\bf Absence of Global Symmetries} \cite{Hawking1975ParticleCreation, BanksDixon1988, Coleman1988, GiddingsStrominger1988, KalloshKLLS9502069, ArkaniHamed0601001}: All global symmetries (including the generalization of higher symmetries of extended objects, or non-invertible categorical symmetries, etc.) 
in QG must be either explicitly broken or dynamically gauged. \\ 
$\bullet$ {\bf Absence of Topological Operators}:
Take examples in 4d (or general $d$d) following \cite{MonteroRudelius2104.07036, RudeliusShao2006.10052}, \\
--- the completeness of gauge theory particle state spectrum is equivalent to no topological Gukov-Witten 2-surface (or more general $(d-2)$d)
operators (if and only if Wilson 1-line operators are endable with the 0d particles);\\
---
the completeness of twist vortices (cosmic strings, or string states) is equivalent to no topological Wilson 1-line operators 
(if and only if Gukov-Witten 2-surface operators [or generally $(d-2)$d]
are endable with the twist vortices [or generally $(d-3)$d]).\\
---
the completeness of magnetic monopole
 is equivalent to no topological magnetic 2-surface operators 
(if and only if 't Hooft 1-line operators [or generally $(d-3)$d]
are endable with the 0d magnetic monopoles [or generally $(d-4)$d]).\\
---
the completeness of magnetic twist vortices is equivalent to no topological 't Hooft 1-line operators [or generally $(d-3)$d] 
(if and only if magnetic 2-surface operators 
are endable with the 1d magnetic twist vortices).

It will be illuminating to learn whether the absence of higher and categorical symmetries in our model 
has implications on the completion of the spectrum.
It will be interesting to understand how our model on the modified $so(10)$ GUT 
has any interpretations/connections to the above QG conjectures.

\end{enumerate}

%%%%%%%%%%%%%
%%%%%%%%%%%%%
\section{Acknowledgements} 
JW deeply appreciates Pierre Deligne, Jun Hou Fung, Yuta Hamada, Jacob McNamara, Miguel Montero, 
Shu-Heng Shao, Ryan Thorngren, Cumrun Vafa, Zheyan Wan,
and Yunqin Zheng for helpful comments.
We thank Zheyan Wan for the collaborations on upcoming works.
JW thanks Miguel Montero et al for other related discussions \cite{MonteroWang}.
YZY is supported by a startup fund at UCSD.
This work is supported by 
NSF Grant DMS-1607871 ``Analysis, Geometry and Mathematical Physics'' 
and Center for Mathematical Sciences and Applications at Harvard University.

%\newpage
\appendix

\section{Quantum Numbers and Representations of SMs and GUTs}
\label{sec:SM-GUT-table-app}

Follow the setup of Appendix A of \Refe{Wang2106.16248}, 
here we summarize the representations of ``elementary'' chiral fermionic particles of quarks and leptons of SMs and GUTs in Tables.
In particular, we had already discussed the representations of the GG $su(5)$ GUT and the PS model, and their embedding into the $so(10)$ GUT in 
 Appendix A.1 and A.2 of \Refe{Wang2106.16248};  so we shall skip those.
 We focus on the flipped $su(5)$ GUT.

\paragraph{Spacetime symmetry representation}
{Weyl fermions are spacetime Weyl spinors, which we prefer to write all Weyl fermions as
$
{\bf 2}_L \text{ of } {\Spin(1,3)}=\rm{SL}(2,\C)
$
with a complex representation in the 4d Lorentz signature.
On the other hand,
the Weyl spinor is 
$
{\bf 2}_L \text{ of } \Spin(4)=\SU(2)_L \times \SU(2)_R
$
with a {pseudoreal representation}
in the 4d Euclidean signature.}

\paragraph{Internal symmetry representation}

Below we provide Table \ref{table:SMfermion-flip}
to organize the internal symmetry representations of particle contents of
the flipped $su(5)$ GUT with the $\U(5)_{\hat q=2}^{\text{2nd}}$ gauge group, 
embedding into the $so(10)$ GUT with the Spin(10) gauge group.

\subsection{Embed the SM into the flipped $u(5)$ GUT with $\U(5)_{\hat q=2}^{\text{2nd}}$, then into the $so(10)$ GUT}
\label{sec:SM-GUT-table-flipped}

Let us compare the GG and the flipped $su(5)$ GUT embedding:
\begin{enumerate}[leftmargin=.mm] 
\item
The gauge theory embedding $so(10)$ GUT $\supset$ $\U(5)_{\hat q=2}^{\text{1st}}$ GUT $\supset$ the $su(5)$ GUT $\supset$ the SM$_6$ only contains the  $G_{\SM_{q=6}}$ 
via an internal symmetry group embedding: 
\bea
{\Spin(10)} \supset \U(5)_{\hat q=2}^{\text{1st}} \supset G_{\GG}\equiv \SU(5) \supset   G_{\SM_{6}} \equiv \frac{{\SU(3)}_c \times {\SU(2)}_{\rm L} \times \U(1)_{\tilde{Y}}}{\Z_6}.
\eea
The representations of quarks and leptons for these models are organized in Table 2 of \Refe{Wang2106.16248}.
Here the $\U(5)_{\hat q=2}^{\text{1st}} \equiv \frac{\SU(5)^{\text{1st}} \times \U(1)_{X,\hat q=2} }{\Z_5} = \frac{\SU(5)^{\text{1st}} \times \U(1)_{X_1,\hat q=2} }{\Z_5}$
contains the $\U(1)_{X} \equiv \U(1)_{X_1}$ subgroup.
\item
The gauge theory embedding $so(10)$ GUT $\supset$ $\U(5)_{\hat q=2}^{\text{2nd}}$ GUT $\supset$ the SM$_6$, only contains the  $G_{\SM_{q=6}}$ 
via an internal symmetry group embedding: 
\bea
{\Spin(10)} \supset \U(5)_{\hat q=2}^{\text{2nd}} \supset \underset{\equiv\U(5)_{\hat q=2}^{\text{1st} } \cap \U(5)_{\hat q=2}^{\text{2nd}} }{
{\Big( ({\SU(3)}_c \times {\SU(2)}_{\rm L}) \times_{{\Z_6}} \U(1)^{\text{1st}}_{\tilde{Y}}  \times_{{\Z_6}}  \U(1)^{\text{2nd}}_{T_{24}} \Big)}
}   
\supset   G_{\SM_{6}}.
\eea
In contrast, if we break $\U(5)_{\hat q=2}^{\text{2nd}}$ to $\SU(5)^{\text{2nd}}$,
this route does not contain $G_{\SM_{6}}$:
\bea
{\Spin(10)} \supset \U(5)_{\hat q=2}^{\text{2nd}} \supset  \SU(5)^{\text{2nd}}  
\supset
({\SU(3)}_c \times {\SU(2)}_{\rm L}) \times_{{\Z_6}} \U(1)^{\text{2nd}}_{T_{24}} 
\supset  \underset{\equiv\SU(5)_{}^{\text{1st} } \cap \SU(5)_{}^{\text{2nd}} }{  ({\SU(3)}_c \times {\SU(2)}_{\rm L})}.
\eea
The representations of quarks and leptons for these models are organized in \Table{table:SMfermion-flip}.
Here the $\U(5)_{\hat q=2}^{\text{2nd}} \equiv \frac{\SU(5)^{\text{2nd}} \times \U(1)_{\chi,\hat q=2} }{\Z_5} = \frac{\SU(5)^{\text{1st}} \times \U(1)_{X_2,\hat q=2} }{\Z_5}$
contains the $\U(1)_{\chi} \equiv \U(1)_{X_2}$ subgroup.

\end{enumerate}

%the charge of $\U(1)_{Y}$ is $\frac{1}{6}$ of the charge of  $\U(1)_{\tilde Y }$.

%
%
%
%
\begin{table}[!h]
$\hspace{-20mm}
  \begin{tabular}{c c c c c c c c c c c c c c}
    \hline
    $\begin{array}{c}
    \textbf{SM}\\ 
   \textbf{fermion}\\
   \textbf{spinor}\\ 
   \textbf{field}
       \end{array}$
   & ${\SU(3)}$& ${\SU(2)}$ & $\U(1)_{X_2}$  & $\U(1)_{Y_2}$  & $\underset{\frac{6 X_2}{5} - \frac{Y_2}{5} }{\U(1)_{{Y_1=}}}$  
   & $\U(1)_{\rm{EM}}$ 
    & $\U(1)_{{ \mathbf{B}-  \mathbf{L}}}$  & $\U(1)_{X_1}$
     & 
    $\Z_{5,X}$
 & 
    $\Z_{4,X}$
    & $\Z_{2}^F$  & SU(5)   \\
        \hline\\[-4mm]
        $\bar{u}_R$& $\overline{\mathbf{3}}$& $\mathbf{1}$& $-3$ & 2 &  $-4$  & $\frac{-2}{3}$ & $-1/3$ &1 & $1$  &1 &   1 &
 \multirow{2}{*}{
     $\overline{\bf{5}}$}  &   \\
\cline{1-11} $l_L$& $\mathbf{1}$&  {${\mathbf{2}}$}& $-3$ & $-3$ & $-3$   & 0 or $-1$ & $-1$ & $-3$& $-3$  &1  &   1 \\
\cline{1-12}  $q_L$& ${{\mathbf{3}}}$& $\mathbf{2}$& 1 & 1 & $1$   & $\frac{2}{3}$ or $\frac{-1}{3}$ &  1/3  &1&1 & 1 &    1 & \multirow{3}{*}{
     ${\bf{10}}$} \\
\cline{1-11}\\[-3.8mm] 
 $\bar{d}_R$& $\overline{\mathbf{3}}$& $\mathbf{1}$ & 1 & $-4$ & $2$   & $\frac{1}{3}$ & $-1/3$ & $-3$ & $-3$ &1   &   1 &  \\
\cline{1-11} $\bar{\nu}_R= {\nu}_L $& $\mathbf{1}$&  $\mathbf{1}$ & 1 & 6 & $ 0$   & 0 & 1  & 5 &0 &   1 &   1 & \\
\cline{1-12} $\bar{e}_R= e_L^+$& $\mathbf{1}$&   $\mathbf{1}$ & 5 & 0 & $6$  & 1 & 1  &1 &1 &   1 &    1 &  \;\;${\bf{1}}$\\
%\hline
%    \hline $\phi_H$ & $\mathbf{1}$ & $\mathbf{2}$& 1/2 & 1 & 0 & -2 &   2 & 3 & 0 & -6  &   -2\\
    \hline\end{tabular}
$
   \caption{   The $G_{\SM_{q=6}}$ cannot be embedded into $\SU(5)^{\text{2nd}}$ given the way their representations are chosen as \eq{eq:5-10-1-2nd}.
   The $G_{\SM_{q=6}}$ can be embedded into 
   the flipped $u(5)$ GUT only if we include the full 
   $\U(5)_{\hat q=2}^{\text{2nd}} \equiv \frac{\SU(5)^{\text{2nd}} \times \U(1)_{\chi,\hat q=2} }{\Z_5} = \frac{\SU(5)^{\text{1st}} \times \U(1)_{X_2,\hat q=2} }{\Z_5}$ gauge group. 
   Then we can embed this into the $so(10)$ GUT with a Spin(10) gauge group.
 The $\SU(5)^{\text{2nd}} \not \supset  \U(1)_Y$ or $\U(1)_{Y_1}$,
 but we have $\U(5)_{\hat q=2}^{\text{2nd}}  \supset  \U(1)_{Y}$ or $\U(1)_{Y_1}$.
}
 \label{table:SMfermion-flip}
\end{table}

\section{Embedding}
\label{app:Embedding}
\subsection{Embedding Strategy}
The strategy to embed 
$\SU(5)$ 
in $\Spin(10)$ is to first embed $\SU(5)$ in $\SO(10)$ 
by identifying $\C^5$ with $\R^{10}$ via the complex-to-real mapping $x+\ii y\xrightarrow{r} (x,y)$, 
and then lift the embedding from $\SO(10)$ to $\Spin(10)$ (from $r$ to $r'$ in the following diagram). 

Furthermore, we ask ``how to embed $\U(5)$, 
or more precisely $\U(5)_{\hat q}$ 
(several non-isomorphic versions $\U(5)_{\hat q} \equiv \frac{\SU(5) \times_{\hat q} \U(1) }{\Z_5}$
defined in \eq{eq:U5q},
via the identification $\U(5)_{\hat q} \equiv 
\{ (g, \e^{\ii \theta}) \in \SU(5) \times \U(1) \big\vert  ( \e^{\ii \frac{2 \pi n}{5}} \mathbb{I}, 1) \sim ( \mathbb{I}, \e^{\ii \frac{2 \pi n {\hat q}}{5}} ), n\in \Z_5 \}$),
into $\SO(10)$ or $\Spin(10)$?'' As we shall show, it turns out that 
$\U(5)_{\hat q} \subset \SO(10)$ for ${\hat q}=1,4$,
while $\U(5)_{\hat q} \subset \Spin(10)$ for ${\hat q}=2,3$.

\be\label{eq:embedding}
\begin{tikzcd}
	{\Z_5} & {\tilde\U(1)_{}} \\
	{\SU(5)} & {\U(5)_{\hat q}} \equiv \frac{\SU(5) \times_{\hat q} \U(1) }{\Z_5} & {\U(1)'} \\
	{\Z_2} & {\Spin(10)} & {\SO(10)}
	\arrow[hook, from=3-1, to=3-2]
	\arrow["p", from=3-2, to=3-3]
	\arrow["r", from=2-2, to=3-3]
	\arrow[hook, from=2-1, to=2-2]
	\arrow["\det", from=2-2, to=2-3]
	\arrow["{r'}", from=2-2, to=3-2]
	\arrow[hook, from=1-1, to=2-1]
	\arrow["{z=\e^{\frac{2\pi\ii n}{5}}}",hook, from=1-1, to=1-2]
	\arrow["{z^{5 \hat q}}", from=1-2, to=2-3]
	\arrow[hook, from=1-2, to=2-2]
\end{tikzcd}
\ee
\begin{enumerate}[leftmargin=.mm] 

\item Here we carefully distinguish several different U(1):
(i) the $\U(1)$ in
$\U(5)_{\hat q} \equiv \frac{\SU(5) \times_{\hat q} \U(1) }{\Z_5}$,
(ii) the $\tilde\U(1)$ as the center of ${\U(5)_{\hat q}}$ such that 
$\tilde\U(1) = Z({\U(5)_{\hat q}}) \supset Z({\SU(5)})=\Z_5$,
(iii) the $\U(1)'$ as the quotient $\frac{{\U(5)_{\hat q}}}{\SU(5)}$.
All these three groups are all isomorphic to any ordinary $S^1$,
but they have different roles that we explain below.

\item 
$\bullet$ The ``$\hookrightarrow$'' means the inclusion, the former group can be embedded into the later group.
The ``$\to$'' means the group homomorphism.

$\bullet$ Here ${\SU(5)} \hookrightarrow {\U(5)_{\hat q}} \overset{\det}{\to} {\U(1)'}$ 
is part of the short exact sequence
$1 {\to} {\SU(5)} {\to} {\U(5)_{\hat q}} \overset{\det}{\to} {\U(1)'} {\to} 1$ where
${\U(1)'} \equiv \frac{\U(1)}{\Z_{5 {\hat q} }} $
is related to modding out ${\Z_{5 {\hat q} }}$ of the $\U(1)$ defined in ${\U(5)_{\hat q}}$ in \eq{eq:U5q}.  
The inclusion ${\SU(5)} \hookrightarrow {\U(5)_{\hat q}}$
indicates that ${\SU(5)}$ is a normal subgroup of ${\U(5)_{\hat q}}$, while the ${\U(1)'}$ is a quotient group of ${\U(5)_{\hat q}}$. %(not necessarily a subgroup of ${\U(5)_{\hat q}}$).
So 
$$
{\U(1)'} = \frac{\U(5)_{\hat q}}{\SU(5)} =  \frac{\U(1)}{\Z_{5 {\hat q} }} .
$$

$\bullet$ Here ${\Z_2} \hookrightarrow {\Spin(10)} \overset{p}{\to} {\SO(10)}$
is part of the short exact sequence
$1\to {\Z_2} \to {\Spin(10)} \to {\SO(10)} \to 1$,
so ${\Spin(10)}/{\SO(10)}= {\Z_2}$.

$\bullet$ Here ${\Z_5} \overset{\e^{\frac{2\pi\ii n}{5}}}{\hookrightarrow} {\tilde \U(1)_{}} \overset{z^{5 \hat q}}{\to} {\U(1)'}$
is part of the short exact sequence 
$1 \to {\Z_5} \to {\tilde \U(1)_{}} \overset{z^{5 \hat q}}{\to} {\U(1)'} \to 1$.

$\bullet$ There are no other short exact sequences in this \eq{eq:embedding}, other than the above relations that we report.

\item For ${\Z_5} \overset{z=\e^{\frac{2\pi\ii n}{5}}}{\hookrightarrow} {\tilde\U(1)_{}} \overset{z^{5 \hat q}}{\to} {\U(1)'}$,
the first map ${\Z_5} \overset{z=\e^{\frac{2\pi\ii n}{5}}}{\hookrightarrow} {\tilde\U(1)_{}}$ inclusion
specifies the identification of the center $Z(\SU(5))={\Z_5}
\equiv \{ {z=\e^{\frac{2\pi\ii n}{5}}} | n \in \{0,1,2,3, 4\} \}$
and its map to the total group $\tilde \U(1) \equiv\{  z= \e^{\ii \theta} |  \theta \in [0, 2 \pi) \}$.
The second map ${\tilde\U(1)_{}} \overset{z^{5 \hat q}}{\to} {\U(1)'} = \frac{\U(5)_{\hat q}}{\SU(5)} =  \frac{\U(1)}{\Z_{5 {\hat q} }} $
says that the $z \in {\tilde\U(1)_{}}$ maps to the ${z^{5 \hat q}} \in {\U(1)'}$,
thus also says that the $\e^{\frac{2\pi\ii n}{5}} \in {\tilde\U(1)_{}}$ becomes the kernel of the map $\tilde \U(1) \overset{z^{5 \hat q}}{\to} \U(1)$
 that maps to the identity $\e^{{2\pi\ii {\hat q}}} = 1 \in {\U(1)}$.

All these facts coincide with our definition of
$\U(5)_{\hat q}$ in \eq{eq:U5q},
which also justifies
\eq{eq:embedding}'s second line: ${\SU(5)} \hookrightarrow {\U(5)_{\hat q}} \overset{\det}{\to} {\U(1)'}$.

\item The ${\U(5)_{\hat q}} \overset{r}{\to} {\SO(10)}$ 
and the ${\U(5)_{\hat q}} \overset{r'}{\to} {\Spin(10)}$
only imply that the maps are group homomorphisms. 
The map may or may not be inclusion or embedding, which depends on the value of ${\hat q}$.

If ${\U(5)_{\hat q}} \subset \SO(10)$ (in particular ${\hat q}=1$),
then a non-trivial fact is that the lifting is only possible, if the lifted 
${\U(5)_{\hat q'}} \subset \Spin(10)$
has the $\SU(5)$ fundamental representation carries an ${\hat q'}=2 q \quad (\text{mod } 5)$ charge under $\U(1)$.
Because the isomorphism \eq{eq:U5-isomorphism},
we can deduce that 
$$\text{ ${\U(5)_{\hat q=1,4}} \subset \SO(10)$
while ${\U(5)_{\hat q'=2,3}} \subset \Spin(10)$.}$$
In the following subsections, we provide some nontrivial checks of these facts.

\end{enumerate}

\subsection{Definition of Lie Groups}

\subsubsection{Lie algebra and Lie Group of $\U(5)_{\hat q}$}

Let $\{\ket{\psi_i}|i=1,2,\cdots,5\}$ be a set of orthonormal basis of $\C^5$, i.e.~$\braket{\psi_i}{\psi_j}=\delta_{ij}$ given the dual basis $\bra{\psi_i}=\ket{\psi_i}^\dagger$. The $\U(5)$ group is the group of isometries of $\C^5$, which is canonically identified with the group of $5\times 5$ unitary matrices. They are generated by 25 generators $T_a$ ($a=1,2,\cdots,25$):
\be
U\in\U(5):U=\exp\Big(\sum_{a=1}^{25}\ii\theta_a T_a\Big).
\ee
The generators $T_a$ are basis of $5\times 5$ Hermitian matrices, and can be chosen as
\be
\begin{array}{l}
\begin{array}{lll}
T_{1} =\ket{\psi_1}\bra{\psi_2}+\ket{\psi_2}\bra{\psi_1}, & 
T_{11} =\ii\ket{\psi_1}\bra{\psi_2}-\ii\ket{\psi_2}\bra{\psi_1},  \\

T_{2} =\ket{\psi_1}\bra{\psi_3}+\ket{\psi_3}\bra{\psi_1}, & 
T_{12} =\ii\ket{\psi_1}\bra{\psi_3}-\ii\ket{\psi_3}\bra{\psi_1},  \\ 

T_{3} =\ket{\psi_1}\bra{\psi_4}+\ket{\psi_4}\bra{\psi_1}, & 
T_{13} =\ii\ket{\psi_1}\bra{\psi_4}-\ii\ket{\psi_4}\bra{\psi_1},  \\ 

T_{4} =\ket{\psi_1}\bra{\psi_5}+\ket{\psi_5}\bra{\psi_1}, & 
T_{14} =\ii\ket{\psi_1}\bra{\psi_5}-\ii\ket{\psi_5}\bra{\psi_1},  \\ 

T_{5} =\ket{\psi_2}\bra{\psi_3}+\ket{\psi_3}\bra{\psi_2}, & 
T_{15} =\ii\ket{\psi_2}\bra{\psi_3}-\ii\ket{\psi_3}\bra{\psi_2}, \\ 

T_{6} =\ket{\psi_2}\bra{\psi_4}+\ket{\psi_4}\bra{\psi_2}, & 
T_{16} =\ii\ket{\psi_2}\bra{\psi_4}-\ii\ket{\psi_4}\bra{\psi_2},  \\

T_{7} =\ket{\psi_2}\bra{\psi_5}+\ket{\psi_5}\bra{\psi_2}, & 
T_{17} =\ii\ket{\psi_2}\bra{\psi_5}-\ii\ket{\psi_5}\bra{\psi_2}, \\

T_{8} =\ket{\psi_3}\bra{\psi_4}+\ket{\psi_4}\bra{\psi_3}, & 
T_{18} =\ii\ket{\psi_3}\bra{\psi_4}-\ii\ket{\psi_4}\bra{\psi_3}, \\

T_{9} =\ket{\psi_3}\bra{\psi_5}+\ket{\psi_5}\bra{\psi_3}, & 
T_{19} =\ii\ket{\psi_3}\bra{\psi_5}-\ii\ket{\psi_5}\bra{\psi_3}, \\

T_{10} =\ket{\psi_4}\bra{\psi_5}+\ket{\psi_5}\bra{\psi_4}, & 
T_{20} =\ii\ket{\psi_4}\bra{\psi_5}-\ii\ket{\psi_5}\bra{\psi_4}, \\
\end{array}\\ \\
\begin{array}{l}
T_{21}=\ket{\psi_1}\bra{\psi_1}-\ket{\psi_2}\bra{\psi_2}, \\
T_{22}=2\ket{\psi_3}\bra{\psi_3}-\ket{\psi_4}\bra{\psi_4}-\ket{\psi_5}\bra{\psi_5}, \\
T_{23}=\ket{\psi_4}\bra{\psi_4}-\ket{\psi_5}\bra{\psi_5}, \\
T_{24}=-3\ket{\psi_1}\bra{\psi_1}-3\ket{\psi_2}\bra{\psi_2} +2\ket{\psi_3}\bra{\psi_3} +2\ket{\psi_4}\bra{\psi_4}+2\ket{\psi_5}\bra{\psi_5},\\
T_{25}=\hat q(\ket{\psi_1}\bra{\psi_1}+\ket{\psi_2}\bra{\psi_2} +\ket{\psi_3}\bra{\psi_3} +\ket{\psi_4}\bra{\psi_4} +\ket{\psi_5}\bra{\psi_5}). \\
\end{array}
\end{array}
\ee
In particular, the last five generators $T_{21},T_{22},\cdots,T_{25}$ are the Cartan generators. Among them, the last one $T_{25}$ is the $\U(1)$ subgroup generator. The overall coefficient $\hat q=1,2,3,4$ in $T_{25}$ labels the $\U(1)$ charge carried by the $\SU(5)$ fundamental representation, which is to be determined later. 

The $\U(1)$ subgroup contains a $\Z_5$ subgroup that can be shared with the center of $\SU(5)$,
\be
\Z_5=\langle\e^{\frac{2\pi\ii}{5}T_{25}}\rangle = \{\e^{\frac{2\pi\ii}{5}n T_{25}}|n=0,1,2,3,4\}.
\ee
Assigning $T_{25}$ with different charge numbers $\hat q$ will lead to different embeddings of the $\Z_5$ subgroup in $\SU(5)$, which lead to different 
$\U(5)_{\hat q}=\SU(5)\times_{\Z_5,{\hat q}}\U(1)$ group structures,
\be\label{eq:T25_vs_q}
\e^{\frac{2\pi\ii}{5}T_{25}}=\left\{
\begin{array}{ll}
\e^{\frac{2\pi\ii}{5}\frac{1}{2}(T_{24}-5T_{21})} & \hat q=1,\\
\e^{\frac{2\pi\ii}{5}T_{24}} & \hat q=2,\\
\e^{-\frac{2\pi\ii}{5}T_{24}} & \hat q=3,\\
\e^{-\frac{2\pi\ii}{5}\frac{1}{2}(T_{24}-5T_{21})} & \hat q=4.\\
\end{array}
\right.
\ee
The difference will become important when embedding $\U(5)_{\hat q}$ into $\SO(10)$ or $\Spin(10)$.

\subsubsection{Lie algebra and Lie Group of $\SO(10)$}

Let $\{\ket{e_i}|i=1,2,\cdots,10\}$ be a set of orthonormal basis of $\R^{10}$, i.e.~$\braket{e_i}{e_j}=\delta_{ij}$ 
given by the dual basis $\bra{e_i}=\ket{e_i}^\intercal$. 
The $\SO(10)$ group is the group of rotations in $\R^{10}$, which is canonically identified with the group of $10\times 10$ orthogonal matrices. They are generated by 45 generators:
\be\label{eq:def_SO(10)}
O\in\SO(10):O=\exp\bigg(\sum_{\{i,j\}\subset\{1,\cdots,10\}}\ii\theta_{i\wedge j} \; e_{i\wedge j}\bigg).
\ee
The generators $e_{i\wedge j}$ are basis of $10\times 10$ pure-imaginary anti-symmetric Hermitian matrices, given by
\be
e_{i\wedge j}=\ii(\ket{e_i}\bra{e_j}-\ket{e_j}\bra{e_i})
\ee 
for all 2-subsets $\{i,j\}$ in $\{1,2,\cdots,10\}$. Note that the basis is anti-symmetric $e_{i\wedge j}= -e_{j\wedge i}$ by definition.

\subsubsection{Lie algebra and Lie Group of $\Spin(10)$}

The construction of the $\Spin(10)$ group starts with the Clifford algebra $\Cl(10)=M_{16}(\mathbb{H})$, which is isomorphic to the algebra of $16\times 16$ matrices over the quaternion field $\mathbb{H}$. Let $\Gamma_1,\Gamma_2,\cdots,\Gamma_{10}$ be the generators of $\Cl(10)$, satisfying $\{\Gamma_i,\Gamma_j\}=2\delta_{ij}$. For the purpose of spelling out the matrix representations, these $16\times 16$ quaternion matrices can be written as $32\times 32$ complex matrices, by embedding $M_{16}(\mathbb{H})$ into $M_{32}(\C)$. One explicit choice of the complex matrix representation can be
\be\label{eq:Gamma_i}
\begin{array}{ll}
\Gamma_{1}\bumpeq\sigma^{11000}, & \Gamma_{2}\bumpeq\sigma^{30100},\\ \Gamma_{3}\bumpeq\sigma^{13000}, & \Gamma_{4}\bumpeq\sigma^{30300},\\ \Gamma_{5}\bumpeq\sigma^{12012}, & \Gamma_{6}\bumpeq\sigma^{30221},\\ \Gamma_{7}\bumpeq\sigma^{12020}, & \Gamma_{8}\bumpeq\sigma^{30202},\\ \Gamma_{9}\bumpeq\sigma^{12032}, & \Gamma_{10}\bumpeq\sigma^{30223},
\end{array}
\ee
where $\sigma^{\mu\nu\cdots} \equiv \sigma^\mu\otimes\sigma^\nu\otimes\cdots$ denotes the direct product of Pauli matrices. The symbol $\bumpeq$ (reads ``can be represented as'') indicates that the equality is only a basis dependent statement.

The $\Spin(10)$ group is a Lie group, whose Lie algebra corresponds to the grade-2 subspace of $\Cl(10)$, which is spanned by 45 basis elements (as Lie group generators),
\be\label{eq:Gamma_ij}
\Gamma_{i\wedge j} = \frac{\ii}{2}[\Gamma_i,\Gamma_j],
\ee
such that the Lie group elements are generated by
\be\label{eq:def_Spin(10)}
\mathsf{O}\in\Spin(10):\mathsf{O}=\exp\bigg(\sum_{\{i,j\}\subset\{1,\cdots,10\}}\frac{\ii\theta_{i\wedge j}}{2}\Gamma_{i\wedge j}\bigg).
\ee
The Spin group is fully contained in the even-graded subspace $\Cl^\text{even}=\Cl^0\oplus\Cl^2\oplus\Cl^4\oplus\cdots$ of the associated Clifford algebra. For $\Cl(10)=M_{16}(\mathbb{H})$, the even-graded subspace $\Cl^\text{even}(10)=M_{16}(\C)\oplus M_{16}(\C)$ splits into two independent algebras of $16\times 16$ matrices over the complex field $\C$. The two $M_{16}(\C)$ subspaces are  specified by the following projection operators (projected down from $M_{32}(\C)$),
\be
P_{\pm}=\frac{1\pm\ii\prod_{j=1}^{10}\Gamma_j}{2}.
\ee
This means that the irreducible fundamental representation of $\Spin(10)$ is only 16-dimensional, even though the associated $\Cl(10)$ requires a 32-dimensional representation. By choosing one of the $M_{16}(\C)$ subspace (say the subspace of that survives the $P_{+}$ projection), $\Gamma_{i\wedge j}$ can be represented as $16\times 16$ matrices as follows (substituting \Eqn{eq:Gamma_i} into \Eqn{eq:Gamma_ij} followed by the projection $\Gamma_{i\wedge j}\to P_{+}\Gamma_{i\wedge j} P_{+}$)
\be\label{eq:Gamma45}
\begin{array}{lllll}
\Gamma_{1\wedge2}\bumpeq\sigma ^{1100}, & \Gamma_{1\wedge3}\bumpeq\sigma ^{2000}, & \Gamma_{1\wedge4}\bumpeq\sigma ^{1300}, & \Gamma_{1\wedge5}\bumpeq-\sigma ^{3012}, & \Gamma_{1\wedge6}\bumpeq\sigma ^{1221},\\ 
\Gamma_{1\wedge7}\bumpeq-\sigma ^{3020}, & \Gamma_{1\wedge8}\bumpeq\sigma ^{1202}, & \Gamma_{1\wedge9}\bumpeq-\sigma ^{3032}, & \Gamma_{1\wedge10}\bumpeq\sigma ^{1223}, & \Gamma_{2\wedge3}\bumpeq-\sigma ^{3100},\\ 
\Gamma_{2\wedge4}\bumpeq\sigma ^{0200}, & \Gamma_{2\wedge5}\bumpeq-\sigma ^{2112}, & \Gamma_{2\wedge6}\bumpeq-\sigma ^{0321}, & \Gamma_{2\wedge7}\bumpeq-\sigma ^{2120}, & \Gamma_{2\wedge8}\bumpeq-\sigma ^{0302},\\ 
\Gamma_{2\wedge9}\bumpeq-\sigma ^{2132}, & \Gamma_{2\wedge10}\bumpeq-\sigma ^{0323}, & \Gamma_{3\wedge4}\bumpeq\sigma ^{3300}, & \Gamma_{3\wedge5}\bumpeq\sigma ^{1012}, & \Gamma_{3\wedge6}\bumpeq\sigma ^{3221},\\ 
\Gamma_{3\wedge7}\bumpeq\sigma ^{1020}, & \Gamma_{3\wedge8}\bumpeq\sigma ^{3202}, & \Gamma_{3\wedge9}\bumpeq\sigma ^{1032}, & \Gamma_{3\wedge10}\bumpeq\sigma ^{3223}, & \Gamma_{4\wedge5}\bumpeq-\sigma ^{2312},\\ 
\Gamma_{4\wedge6}\bumpeq\sigma ^{0121}, & \Gamma_{4\wedge7}\bumpeq-\sigma ^{2320}, & \Gamma_{4\wedge8}\bumpeq\sigma ^{0102}, & \Gamma_{4\wedge9}\bumpeq-\sigma ^{2332}, & \Gamma_{4\wedge10}\bumpeq\sigma ^{0123},\\ 
\Gamma_{5\wedge6}\bumpeq\sigma ^{2233}, & \Gamma_{5\wedge7}\bumpeq-\sigma ^{0032}, & \Gamma_{5\wedge8}\bumpeq\sigma ^{2210}, & \Gamma_{5\wedge9}\bumpeq\sigma ^{0020}, & \Gamma_{5\wedge10}\bumpeq-\sigma ^{2231},\\ 
\Gamma_{6\wedge7}\bumpeq-\sigma ^{2201}, & \Gamma_{6\wedge8}\bumpeq-\sigma ^{0023}, & \Gamma_{6\wedge9}\bumpeq\sigma ^{2213}, & \Gamma_{6\wedge10}\bumpeq\sigma ^{0002}, & \Gamma_{7\wedge8}\bumpeq\sigma ^{2222},\\ 
\Gamma_{7\wedge9}\bumpeq-\sigma ^{0012}, & \Gamma_{7\wedge10}\bumpeq\sigma ^{2203}, & \Gamma_{8\wedge9}\bumpeq-\sigma ^{2230}, & \Gamma_{8\wedge10}\bumpeq-\sigma ^{0021}, & \Gamma_{9\wedge10}\bumpeq\sigma ^{2211}.
\end{array}
\ee
The representation space is $\C^{16}$.

\subsection{Embedding and Projection Maps}

As laid out in \Eqn{eq:embedding}, some of the $\U(5)_{\hat q}$ group can be embedded into $\SO(10)$ by an embedding map $r$, and the $\Spin(10)$ group can be projected to $\SO(10)$ by a projection map $p$. This potentially enables us to lift the embedding $r$ to $r'$, then $r'$ will tell us how to embed which $\U(5)_{\hat q'}$ in $\Spin(10)$. 
However, we will show that the lifting is not always possible. It crucially depends on the choice of the charge number 
$\hat q$ or $\hat q'$.

\subsubsection{Embedding Map $r$}

The embedding map $r:\C^5\to\R^{10}$ sends the representation space $\C^5$ to $\R^{10}$ by
\be
r:\sum_{i=1}^{5}z_i\ket{\psi_i}\mapsto\sum_{i=1}^{5} (\Re z_i\ket{e_{2i-1}}+\Im z_i\ket{e_{2i}}).
\ee
It induces a functorial map $r:\U(5)\to\SO(10)$ by
\be
\begin{tikzcd}
	{\C^5} & {\R^{10}} \\
	{\C^5} & {\R^{10}}
	\arrow["r", from=1-1, to=1-2]
	\arrow["r", from=2-1, to=2-2]
	\arrow[""{name=0, anchor=center, inner sep=0}, "{\SO(10)}"', from=2-2, to=1-2]
	\arrow[""{name=1, anchor=center, inner sep=0}, "{\U(5)}", from=2-1, to=1-1]
	\arrow["r", color={blue}, shorten <=6pt, shorten >=6pt, Rightarrow, from=1, to=0]
\end{tikzcd},
\ee
which is the embedding map $r$. The embedding map between Lie groups also implies the embedding map of the corresponding Lie algebras, as $r:\mathfrak{u}(5)\to\mathfrak{so}(10)$, which allows us to establish the relation among generators:
\be\label{eq:r_map}
\begin{array}{l}
\begin{array}{ll}
r(T_{1}) = e_{1\wedge4} - e_{2\wedge3}, & r(T_{11}) = e_{1\wedge3} + e_{2\wedge4},\\ 
r(T_{2}) = e_{1\wedge6} - e_{2\wedge5}, & r(T_{12}) = e_{1\wedge5} + e_{2\wedge6},\\ 
r(T_{3}) = e_{1\wedge8} - e_{2\wedge7}, & r(T_{13}) = e_{1\wedge7} + e_{2\wedge8},\\ 
r(T_{4}) = e_{1\wedge10} - e_{2\wedge9}, & r(T_{14}) = e_{1\wedge9} + e_{2\wedge10},\\ 
r(T_{5}) = e_{3\wedge6} - e_{4\wedge5}, & r(T_{15}) = e_{3\wedge5} + e_{4\wedge6},\\ 
r(T_{6}) = e_{3\wedge8} - e_{4\wedge7}, & r(T_{16}) = e_{3\wedge7} + e_{4\wedge8},\\ 
r(T_{7}) = e_{3\wedge10} - e_{4\wedge9}, & r(T_{17}) = e_{3\wedge9} + e_{4\wedge10},\\ 
r(T_{8}) = e_{5\wedge8} - e_{6\wedge7}, & r(T_{18}) = e_{5\wedge7} + e_{6\wedge8},\\ 
r(T_{9}) = e_{5\wedge10} - e_{6\wedge9}, & r(T_{19}) = e_{5\wedge9} + e_{6\wedge10},\\ 
r(T_{10}) = e_{7\wedge10} - e_{8\wedge9}, & r(T_{20}) = e_{7\wedge9} + e_{8\wedge10},
\end{array}\\ \\
\begin{array}{l}
r(T_{21}) = e_{1\wedge2} - e_{3\wedge4},\\ 
r(T_{22}) = 2 e_{5\wedge6} - e_{7\wedge8} - e_{9\wedge10},\\ 
r(T_{23}) = e_{7\wedge8} - e_{9\wedge10},\\ 
r(T_{24}) = -3 e_{1\wedge2} - 3 e_{3\wedge4} + 2 e_{5\wedge6} + 2 e_{7\wedge8} + 2 e_{9\wedge10},\\ 
r(T_{25}) = \hat{q}(e_{1\wedge2} + e_{3\wedge4} + e_{5\wedge6} + e_{7\wedge8} + e_{9\wedge10}).
\end{array}
\end{array}
\ee
Given that $r$ is a linear map (i.e. $r(\theta_aT_a)=\theta_a r(T_a)$), \Eqn{eq:r_map} automatically specifies the map $r$ for any element in the Lie algebra.

\subsubsection{Projection Map $p$}

The projection map $p:\Spin(10)\to\SO(10)$ is defined by the short exact sequence
\be
1\to{\Z_2}\hookrightarrow{\Spin(10)}\xrightarrow{p}{\SO(10)}\to 1.
\ee
By comparing \Eqn{eq:def_Spin(10)} and \Eqn{eq:def_SO(10)}, the projection map simply identifies the Lie group generators
\be
p\Big(\frac{\Gamma_{i\wedge j}}{2}\Big) = e_{i\wedge j}
\ee
for all pairs $\{i,j\}\subset\{1,2,\cdots,10\}$. The mapping $p:\mathfrak{spin}(10)\xrightarrow{\cong}\mathfrak{so}(10)$ is one-to-one (hence invertible) on the Lie algebra level (but not on the Lie group level), which allows us to define $p^{-1}:\mathfrak{so}(10)\xrightarrow{\cong}\mathfrak{spin}(10)$,
\be\label{eq:p^-1_map}
p^{-1}(e_{i\wedge j})=\frac{\Gamma_{i\wedge j}}{2}.
\ee

\subsubsection{Embedding Map $r'$}

The invertibility of $p$ on the Lie algebra level allows us to lift the map $r$ to $r'$ on the Lie algebra level by defining $r'= p^{-1}\circ r$, i.e.~$r'(T_i)=p^{-1}(r(T_i))$.
\be
\begin{tikzcd}
	{\mathfrak{u}(5)} \\
	{\mathfrak{spin}(10)} & {\mathfrak{so}(10)}
	\arrow["r", hook, from=1-1, to=2-2]
	\arrow["{r'}", hook, from=1-1, to=2-1]
	\arrow["p", shift left=0.5, from=2-1, to=2-2]
	\arrow["{p^{-1}}", shift left=0.5, from=2-2, to=2-1]
\end{tikzcd}
\ee
Using \Eqn{eq:r_map} and \Eqn{eq:p^-1_map}, we obtain
\be\label{eq:r'_map}
\begin{array}{l}
\begin{array}{ll}
r'(T_{1}) = \frac{1}{2}(\Gamma_{1\wedge4} - \Gamma_{2\wedge3}), & r'(T_{11}) = \frac{1}{2}(\Gamma_{1\wedge3} + \Gamma_{2\wedge4}),\\ 
r'(T_{2}) = \frac{1}{2}(\Gamma_{1\wedge6} - \Gamma_{2\wedge5}), & r'(T_{12}) = \frac{1}{2}(\Gamma_{1\wedge5} + \Gamma_{2\wedge6}),\\ 
r'(T_{3}) = \frac{1}{2}(\Gamma_{1\wedge8} - \Gamma_{2\wedge7}), & r'(T_{13}) = \frac{1}{2}(\Gamma_{1\wedge7} + \Gamma_{2\wedge8}),\\ 
r'(T_{4}) = \frac{1}{2}(\Gamma_{1\wedge10} - \Gamma_{2\wedge9}), & r'(T_{14}) = \frac{1}{2}(\Gamma_{1\wedge9} + \Gamma_{2\wedge10}),\\ 
r'(T_{5}) = \frac{1}{2}(\Gamma_{3\wedge6} - \Gamma_{4\wedge5}), & r'(T_{15}) = \frac{1}{2}(\Gamma_{3\wedge5} + \Gamma_{4\wedge6}),\\ 
r'(T_{6}) = \frac{1}{2}(\Gamma_{3\wedge8} - \Gamma_{4\wedge7}), & r'(T_{16}) = \frac{1}{2}(\Gamma_{3\wedge7} + \Gamma_{4\wedge8}),\\ 
r'(T_{7}) = \frac{1}{2}(\Gamma_{3\wedge10} - \Gamma_{4\wedge9}), & r'(T_{17}) = \frac{1}{2}(\Gamma_{3\wedge9} + \Gamma_{4\wedge10}),\\ 
r'(T_{8}) = \frac{1}{2}(\Gamma_{5\wedge8} - \Gamma_{6\wedge7}), & r'(T_{18}) = \frac{1}{2}(\Gamma_{5\wedge7} + \Gamma_{6\wedge8}),\\ 
r'(T_{9}) = \frac{1}{2}(\Gamma_{5\wedge10} - \Gamma_{6\wedge9}), & r'(T_{19}) = \frac{1}{2}(\Gamma_{5\wedge9} + \Gamma_{6\wedge10}),\\ 
r'(T_{10}) = \frac{1}{2}(\Gamma_{7\wedge10} - \Gamma_{8\wedge9}), & r'(T_{20}) = \frac{1}{2}(\Gamma_{7\wedge9} + \Gamma_{8\wedge10}),
\end{array}\\ \\
\begin{array}{l}
r'(T_{21}) = \frac{1}{2}(\Gamma_{1\wedge2} - \Gamma_{3\wedge4}),\\ 
r'(T_{22}) = \frac{1}{2}(2 \Gamma_{5\wedge6} - \Gamma_{7\wedge8} - \Gamma_{9\wedge10}),\\ 
r'(T_{23}) = \frac{1}{2}(\Gamma_{7\wedge8} - \Gamma_{9\wedge10}),\\ 
r'(T_{24}) = \frac{1}{2}(-3 \Gamma_{1\wedge2} - 3 \Gamma_{3\wedge4} + 2 \Gamma_{5\wedge6} + 2 \Gamma_{7\wedge8} + 2 \Gamma_{9\wedge10}),\\ 
r'(T_{25}) = \frac{1}{2}{\hat q}(\Gamma_{1\wedge2} + \Gamma_{3\wedge4} + \Gamma_{5\wedge6} + \Gamma_{7\wedge8} + \Gamma_{9\wedge10}),
\end{array}
\end{array}
\ee
which applies to the whole Lie algebra given that $r'$ is a linear map (i.e. $r'(\theta_aT_a)=\theta_a r'(T_a)$).
However, there could be obstruction to further lift the Lie algebra embedding $r'$ to the Lie group level. The key relies on whether the $\Z_5$ subgroup shared between $\SU(5)$ and $\U(1)$ can be consistently defined after lifting $\U(5)_{\hat q}$ to $\Spin(10)$. This depends on the choice of the charge number $\hat q$.

\subsection{Obstruction to Lifting: Compatibility of $\Z_5$ Center}

Recall \Eqn{eq:T25_vs_q} that the charge number $\hat q$ affects how the $\Z_5$ generator $\e^{\frac{2\pi\ii}{5}T_{25}}\in\U(1)$ is identified with a group element inside $\SU(5)$. The key is to check if any of these identifications will be broken under the embedding map $r'$. Since \Eqn{eq:T25_vs_q} only involves three generators $T_{21}$, $T_{24}$ and $T_{25}$, we will only focus on their images under the $r'$ map. Further more, because $T_{21}$, $T_{24}$ and $T_{25}$ are commuting Cartan generators (so will be $r'(T_{21})$, $r'(T_{24})$ and $r'(T_{25})$), we can find a good basis of $\C^{16}$ to simultaneously diagonalize $r'(T_{21})$, $r'(T_{24})$ and $r'(T_{25})$. By explicit calculation (by substituting \Eqn{eq:Gamma45} to \Eqn{eq:r'_map} and diagonalize the $16\times 16$ matrices), the eigenvalues of $r'(T_{21})$, $r'(T_{24})$ and $r'(T_{25})$ in their common eigenbasis are listed in \Table{tab:Tvalues}.

\begin{table}[htp]
\caption{Eigenvalues of $r'(T_{21})$, $r'(T_{24})$ and $r'(T_{25})$ in their common eigenbasis.}
\begin{center}
\begin{tabular}{cccc}
\hline
$\SU(5)$ & $r'(T_{21})$ & $r'(T_{24})$ & $r'(T_{25})$ \\
\hline
\multirow{5}{*}{$\overline{\mathbf{5}}$}
 & $-1$ & $-3$ & $-3\hat q/2$ \\ 
 & $1$ & $-3$ & $-3\hat q/2$ \\ 
 & $0$ & $2$ & $-3\hat q/2$ \\ 
 & $0$ & $2$ & $-3\hat q/2$ \\ 
 & $0$ & $2$ & $-3\hat q/2$ \\ 
\hline
\multirow{10}{*}{$\mathbf{10}$}
 & $0$ & $-4$ & $\hat q/2$ \\ 
 & $0$ & $-4$ & $\hat q/2$ \\ 
 & $0$ & $-4$ & $\hat q/2$ \\ 
 & $-1$ & $1$ & $\hat q/2$ \\ 
 & $-1$ & $1$ & $\hat q/2$ \\ 
 & $-1$ & $1$ & $\hat q/2$ \\ 
 & $1$ & $1$ & $\hat q/2$ \\ 
 & $1$ & $1$ & $\hat q/2$ \\ 
 & $1$ & $1$ & $\hat q/2$ \\ 
 & $0$ & $6$ & $\hat q/2$ \\ 
\hline
\multirow{1}{*}{$\mathbf{1}$}
 & $0$ & $0$ & $5\hat q/2$ \\  
 \hline
\end{tabular}
\end{center}
\label{tab:Tvalues}
\end{table}
Because of the group isomorphism given in
\eq{eq:U5-isomorphism}, we will only have to check two cases,
$\hat q=1$ and $\hat q=2$:
\begin{itemize}
\item For $\hat q=1$, $\e^{\frac{2\pi\ii}{5}T_{25}}$ is identified with $\e^{\frac{2\pi\ii}{5}(T_{24}-5T_{21})/2}$ in $\U(5)$, but $\e^{\frac{2\pi\ii}{5}r'(T_{25})}$ can not be identified with $\e^{\frac{2\pi\ii}{5}r'((T_{24}-5T_{21})/2)}$ in $\Spin(10)$, because
\be
\begin{split}
r'((T_{24}-5T_{21})/2)-r'(T_{25}|_{\hat q=1})&\bumpeq\diag(\tfrac{5}{2}, -\tfrac{5}{2}, \tfrac{5}{2}, \tfrac{5}{2}, \tfrac{5}{2}, -\tfrac{5}{2}, -\tfrac{5}{2}, -\tfrac{5}{2}, \tfrac{5}{2}, \tfrac{5}{2}, \tfrac{5}{2}, -\tfrac{5}{2}, -\tfrac{5}{2}, -\tfrac{5}{2}, \tfrac{5}{2}, -\tfrac{5}{2})\\
&\overset{\mod 5}{=}\tfrac{5}{2}\mathbb{1},
\end{split}
\ee
meaning $\e^{\frac{2\pi\ii}{5}r'((T_{24}-5T_{21})/2)}=-\e^{\frac{2\pi\ii}{5}r'(T_{25})}$, which posts an obstruction to lift $r'$ out of the exponent to the Lie group level. 

\item For $\hat q=2$, $\e^{\frac{2\pi\ii}{5}T_{25}}$ is identified with $\e^{\frac{2\pi\ii}{5}T_{24}}$ in $\U(5)$, while $\e^{\frac{2\pi\ii}{5}r'(T_{25})}$ can also be identified with $\e^{\frac{2\pi\ii}{5}r'(T_{24})}$ in $\Spin(10)$ consistently, because
\be
\begin{split}
r'(T_{24})-r'(T_{25}|_{\hat q=2})&\bumpeq\diag(0, 0, 5, 5, 5, -5, -5, -5, 0, 0, 0, 0, 0, 0, 5, -5)\\
&\overset{\mod 5}{=}0,
\end{split}
\ee
meaning $\e^{\frac{2\pi\ii}{5}r'(T_{24})}=\e^{\frac{2\pi\ii}{5}r'(T_{25})}$, which posts no obstruction to lift $r'$ out of the exponent to the Lie group level. 
\end{itemize}

In conclusion, the $\Z_5$ center can be compatibly defined and shared between $\SU(5)$ and $\U(1)$, if and only if $\hat q=2$. 
With $\hat q=2$, there is no obstruction to further lift $r'$ to the Lie group level and hence the embedding map $r':\U(5)\to\Spin(10)$ is well-defined.

\section{Flipping}
\label{app:Flipping}

\subsection{Flipping Isomorphism}
\label{app:Flipping-Isomorphism}

The standard GG $\U(5)$ and flipped $\U(5)$ (denoted as $\U(5)_\text{1st}$ and $\U(5)_\text{2nd}$ in the following) can be both embedded in the same $\Spin(10)$ group as long as their charge numbers are taken to be the same even integer. We will implicitly assume the $q=2$ case in the following discussion. $\U(5)_\text{1st}$ and $\U(5)_\text{2nd}$ are related by the flipping isomorphism, denoted as $f$, which is  an \emph{outer} automorphism of $\U(5)$ and also an \emph{inner} automorphism of $\SO(10)$ as well as $\Spin(10)$, as shown in the diagram below.

\be
\begin{tikzcd}
	{\U(5)_\text{1st}} \\
	{\Spin(10)} & {\SO(10)} \\
	{\U(5)_\text{2nd}}
	\arrow["r'_1", hook, from=1-1, to=2-1]
	\arrow["r_1", hook, from=1-1, to=2-2]
	\arrow["p", from=2-1, to=2-2]
	\arrow["r'_2"', hook', from=3-1, to=2-1]
	\arrow["r_2"', hook', from=3-1, to=2-2]
	\arrow["f"', <->, dd, bend right=60, shift right=3, from=1-1, to=3-1]
	\arrow["f"', <->, loop, distance=2em, in=215, out=145, from=2-1, to=2-1]
	\arrow["f", <->, loop, distance=2em, in=325, out=35, from=2-2, to=2-2]
\end{tikzcd}
\ee

The flipping isomorphism $f$ can be specified as an inner automorphism of $f:\SO(10)\to\SO(10)$, such that for $O\in\SO(10)$: $f(O)=F^{-1} O F$ with $F=\e^{\ii \pi e_{2\wedge 4}}$. This can also be interpreted as a basis transformation of $\R^{10}$,
\be
f(\ket{e_i})=F\ket{e_i}=\left\{
\begin{array}{ll}
-\ket{e_i} & i\in\{2,4\}\\
\ket{e_i} & i\notin\{2,4\}
\end{array}
\right.
\equiv(-1)^{\delta_{i\in\{2,4\}}}\ket{e_i},
\ee
where $\delta_{i\in\{2,4\}}$ is the Kronecker delta symbol that equals 1 when $i\in\{2,4\}$ and equals 0 when $i\notin\{2,4\}$. The flipping isomorphism is a duality, as $F^2=1$, or $f\circ f=\id$.

The flipping isomorphism can be lifted to $f:\Spin(10)\to\Spin(10)$, such that for $\mathsf{O}\in\Spin(10)$: $f(\mathsf{O})=\mathsf{F}^{-1} \mathsf{O} \mathsf{F}$ with $\mathsf{F}=\pm\e^{\ii (\pi/2) \Gamma_{2\wedge 4}}=\pm\ii\Gamma_{2\wedge 4}=\mp\Gamma_2\Gamma_4$. The $\pm$ sign ambiguity arise from the $\Z_2$ subgroup freedom when embedding $\SO(10)$ in $\Spin(10)$. Nevertheless, this sign ambiguity does not affect the definition of the flipping isomorphism $f(\mathsf{O})$ because $\mathsf{F}$ always appears twice. The flipping morphism can also be translated to an inner automorphism of $\Cl(10)$, as
\be
f(\Gamma_i)=\mathsf{F}^{-1}\Gamma_i\mathsf{F}=(-1)^{\delta_{i\in\{2,4\}}}\Gamma_i
\ee
which applies to the $\Spin(10)$ generators (as grade-2 elements of $\Cl(10)$) as
\be
f(\Gamma_{i\wedge j})=\mathsf{F}^{-1}\Gamma_{i\wedge j}\mathsf{F}=(-1)^{\delta_{i\in\{2,4\}}+\delta_{j\in\{2,4\}}}\Gamma_{i\wedge j}.
\ee
Again, one can see $f\circ f=\id$.

Given the embedding map $r'_1:\U(5)_\text{1st}\to\Spin(10)$ in \Eqn{eq:r'_map} and the flipping isomorphism $f$, the flipped embedding map $r'_2:\U(5)_\text{2st}\to\Spin(10)$ can be defined as $r'_2=f\circ r'_1$, i.e. $r'_2(T_i)=f(r'_1(T_i))$, which enables us to compare the two embeddings $r'_1$ and $r'_2$ as in \Table{tab:embeddings}.

\begin{table}[!h] %[htp]
\begin{center}
\begin{tabular}{lll}
$T_i$ & $r'_1(T_i)$ (as $r'_1:\mathfrak{u}(5)_\text{1st}\to\mathfrak{spin}(10)$) & $r'_2(T_i)$ (as $r'_2:\mathfrak{u}(5)_\text{2nd}\to\mathfrak{spin}(10)$)\\
\hline
\rowcolor{lightgray}
$T_{1}$ & $\frac{1}{2}(\Gamma_{1\wedge4} - \Gamma_{2\wedge3})$ & $\frac{1}{2}(-\Gamma_{1\wedge4} + \Gamma_{2\wedge3})$ \\ 
$T_{2}$ & $\frac{1}{2}(\Gamma_{1\wedge6} - \Gamma_{2\wedge5})$ & $\frac{1}{2}(\Gamma_{1\wedge6} + \Gamma_{2\wedge5})$ \\ 
$T_{3}$ & $\frac{1}{2}(\Gamma_{1\wedge8} - \Gamma_{2\wedge7})$ & $\frac{1}{2}(\Gamma_{1\wedge8} + \Gamma_{2\wedge7})$ \\ 
$T_{4}$ & $\frac{1}{2}(\Gamma_{1\wedge10} - \Gamma_{2\wedge9})$ & $\frac{1}{2}(\Gamma_{1\wedge10} + \Gamma_{2\wedge9})$ \\ 
$T_{5}$ & $\frac{1}{2}(\Gamma_{3\wedge6} - \Gamma_{4\wedge5})$ & $\frac{1}{2}(\Gamma_{3\wedge6} + \Gamma_{4\wedge5})$ \\ 
$T_{6}$ & $\frac{1}{2}(\Gamma_{3\wedge8} - \Gamma_{4\wedge7})$ & $\frac{1}{2}(\Gamma_{3\wedge8} + \Gamma_{4\wedge7})$ \\ 
$T_{7}$ & $\frac{1}{2}(\Gamma_{3\wedge10} - \Gamma_{4\wedge9})$ & $\frac{1}{2}(\Gamma_{3\wedge10} + \Gamma_{4\wedge9})$ \\ 
\rowcolor{lightgray}
$T_{8}$ & $\frac{1}{2}(\Gamma_{5\wedge8} - \Gamma_{6\wedge7})$ & $\frac{1}{2}(\Gamma_{5\wedge8} - \Gamma_{6\wedge7})$ \\ 
\rowcolor{lightgray}
$T_{9}$ & $\frac{1}{2}(\Gamma_{5\wedge10} - \Gamma_{6\wedge9})$ & $\frac{1}{2}(\Gamma_{5\wedge10} - \Gamma_{6\wedge9})$ \\ 
\rowcolor{lightgray}
$T_{10}$ & $\frac{1}{2}(\Gamma_{7\wedge10} - \Gamma_{8\wedge9})$ & $\frac{1}{2}(\Gamma_{7\wedge10} - \Gamma_{8\wedge9})$ \\ 
\rowcolor{lightgray}
$T_{11}$ & $\frac{1}{2}(\Gamma_{1\wedge3} + \Gamma_{2\wedge4})$ & $\frac{1}{2}(\Gamma_{1\wedge3} + \Gamma_{2\wedge4})$ \\ 
$T_{12}$ & $\frac{1}{2}(\Gamma_{1\wedge5} + \Gamma_{2\wedge6})$ & $\frac{1}{2}(\Gamma_{1\wedge5} - \Gamma_{2\wedge6})$ \\ 
$T_{13}$ & $\frac{1}{2}(\Gamma_{1\wedge7} + \Gamma_{2\wedge8})$ & $\frac{1}{2}(\Gamma_{1\wedge7} - \Gamma_{2\wedge8})$ \\ 
$T_{14}$ & $\frac{1}{2}(\Gamma_{1\wedge9} + \Gamma_{2\wedge10})$ & $\frac{1}{2}(\Gamma_{1\wedge9} - \Gamma_{2\wedge10})$ \\ 
$T_{15}$ & $\frac{1}{2}(\Gamma_{3\wedge5} + \Gamma_{4\wedge6})$ & $\frac{1}{2}(\Gamma_{3\wedge5} - \Gamma_{4\wedge6})$ \\ 
$T_{16}$ & $\frac{1}{2}(\Gamma_{3\wedge7} + \Gamma_{4\wedge8})$ & $\frac{1}{2}(\Gamma_{3\wedge7} - \Gamma_{4\wedge8})$ \\ 
$T_{17}$ & $\frac{1}{2}(\Gamma_{3\wedge9} + \Gamma_{4\wedge10})$ & $\frac{1}{2}(\Gamma_{3\wedge9} - \Gamma_{4\wedge10})$ \\ 
\rowcolor{lightgray}
$T_{18}$ & $\frac{1}{2}(\Gamma_{5\wedge7} + \Gamma_{6\wedge8})$ & $\frac{1}{2}(\Gamma_{5\wedge7} + \Gamma_{6\wedge8})$ \\ 
\rowcolor{lightgray}
$T_{19}$ & $\frac{1}{2}(\Gamma_{5\wedge9} + \Gamma_{6\wedge10})$ & $\frac{1}{2}(\Gamma_{5\wedge9} + \Gamma_{6\wedge10})$ \\ 
\rowcolor{lightgray}
$T_{20}$ & $\frac{1}{2}(\Gamma_{7\wedge9} + \Gamma_{8\wedge10})$ & $\frac{1}{2}(\Gamma_{7\wedge9} + \Gamma_{8\wedge10})$ \\ 
\rowcolor{lightgray}
$T_{21}$ & $\frac{1}{2}(\Gamma_{1\wedge2} - \Gamma_{3\wedge4})$ & $\frac{1}{2}(-\Gamma_{1\wedge2} + \Gamma_{3\wedge4})$ \\ 
\rowcolor{lightgray}
$T_{22}$ & $\frac{1}{2}(2 \Gamma_{5\wedge6} - \Gamma_{7\wedge8} - \Gamma_{9\wedge10})$ & $\frac{1}{2}(2 \Gamma_{5\wedge6} - \Gamma_{7\wedge8} - \Gamma_{9\wedge10})$ \\ 
\rowcolor{lightgray}
$T_{23}$ & $\frac{1}{2}(\Gamma_{7\wedge8} - \Gamma_{9\wedge10})$ & $\frac{1}{2}(\Gamma_{7\wedge8} - \Gamma_{9\wedge10})$ \\ 
\rowcolor{lightgray}
$T_{24}$ & $\frac{1}{2}(-3 \Gamma_{1\wedge2} - 3 \Gamma_{3\wedge4} + 2 \Gamma_{5\wedge6} + 2 \Gamma_{7\wedge8} + 2 \Gamma_{9\wedge10})$ & $\frac{1}{2}(3 \Gamma_{1\wedge2} + 3 \Gamma_{3\wedge4} + 2 \Gamma_{5\wedge6} + 2 \Gamma_{7\wedge8} + 2 \Gamma_{9\wedge10})$ \\ 
\rowcolor{lightgray}
$T_{25}$ & $\Gamma_{1\wedge2} + \Gamma_{3\wedge4} + \Gamma_{5\wedge6} + \Gamma_{7\wedge8} + \Gamma_{9\wedge10}$ & $-\Gamma_{1\wedge2} - \Gamma_{3\wedge4} + \Gamma_{5\wedge6} + \Gamma_{7\wedge8} + \Gamma_{9\wedge10}$
\end{tabular}
\end{center}
\caption{Embeddings of $\mathfrak{u}(5)_\text{1st}$ and $\mathfrak{u}(5)_\text{2nd}$ (both of $q=2$) in $\mathfrak{spin}(10)$.}
\label{tab:embeddings}
\end{table}

\subsection{Intersection and Join}
\label{sec:two-u5-Intersection-Join}

Treat $\mathfrak{u}(5)_\text{1st}$ and $\mathfrak{u}(5)_\text{2nd}$ as Lie subalgebra of $\mathfrak{spin}(10)$. Their intersection is (see the highlighted rows in \Table{tab:embeddings})
\be
\begin{split}
\mathfrak{u}(5)_\text{1st}\cap\mathfrak{u}(5)_\text{2nd}&=\mathrm{span}\{T_{1},T_{8},T_{9},T_{10},T_{11},T_{18},T_{19},T_{20},T_{21},T_{22},T_{23},T_{24},T_{25}\}\\
&=\mathrm{span}\{T_{8},T_{9},T_{10},T_{18},T_{19},T_{20},T_{22},T_{23}\}\oplus 
\mathrm{span}\{T_{1},T_{11},T_{21}\}\\
&\hspace{12pt}\oplus
\mathrm{span}\{T_{24}\}
\oplus
\mathrm{span}\{T_{25}\}\\
&=\mathfrak{su}(3)\oplus\mathfrak{su}(2)\oplus\mathfrak{u}(1)_Y\oplus\mathfrak{u}(1)_X.
\end{split}
\ee
The join of $\mathfrak{u}(5)_\text{1st}$ and $\mathfrak{u}(5)_\text{2nd}$ 
(the  $\mathfrak{u}(5)_\text{1st}$ and $\mathfrak{u}(5)_\text{2nd}$ together generate 
the minimal Lie algebra as their union, which should be understood as the minimal 
vector space generated by both Lie algebras together closed under any of their Lie brackets) is
\be
\begin{split}\label{eq:twoU5-Spin10}
\mathfrak{u}(5)_\text{1st}{\cup}\mathfrak{u}(5)_\text{2nd}
&=(\mathfrak{u}(5)_\text{1st}\cap\mathfrak{u}(5)_\text{2nd}){\cup}\mathrm{span}\{\Gamma_{i\wedge j}|i\in\{1,2,3,4\},j\in\{5,6,7,8,9,10\}\}\\
&=\mathrm{span}\{\Gamma_{i\wedge j}|\{i,j\}\subset\{1,2,\cdots,10\}\}\\
&=\mathfrak{spin}(10).
\end{split}
\ee

On the Lie group level, we have
\be
\begin{tikzcd}
	& {\Spin(10)} \\
	{\U(5)_\text{1st}} && {\U(5)_\text{2nd}} \\
	& {G_\text{SM}\times_{{\Z_5}}\U(1)_X}
	\arrow[hook', from=3-2, to=2-1]
	\arrow[hook, from=3-2, to=2-3]
	\arrow["r'_1", hook, from=2-1, to=1-2]
	\arrow["r'_2"', hook', from=2-3, to=1-2]
	\arrow["f", <->, from=2-1, to=2-3]
\end{tikzcd}
\ee
where $G_\text{SM}=(\SU(3)\times\SU(2))\times_{\Z_6}\U(1)_Y$.
{We have ${G_\text{SM} \times_{{\Z_5}}\U(1)_X}$ because that the 
$\U(1)_Y \times_{{\Z_5}}\U(1)_X$ structure has a shared ${\Z_5} ={\Z_{5,X}}={\Z_{5,Y}}$, see \Table{table:fermionAll}.
}
\subsection{Charge Lattice}

Define the $\U(1)$ charges $Y_1:= r'_1(T_{24})$, $X_1:=r'_1(T_{25})$ in $\U(5)_\text{1st}$, and $Y_2:=r'_2(T_{24})$, $X_2:=r'_2(T_{25})$ in $\U(5)_\text{2nd}$. 
Based on \Table{tab:embeddings}, these charges are related by the flipping isomorphism $f$: %(duality)
\be
\bpm X_1\\Y_1\epm=\frac{1}{5}\bpm1&4\\6&-1\epm\bpm X_2\\Y_2\epm,\quad \bpm X_2\\Y_2\epm=\frac{1}{5}\bpm1&4\\6&-1\epm\bpm X_1\\Y_1\epm.
\ee
See also the discussions in \Sec{sec:SM-GUT-table}.

The two sets of charge lattices intersect at points that matches the charge assignment of fundamental fermions in the SM, as shown in \Fig{fig:charge lattice}.
Moreover, $\U(1)_{X_1}$ and $\U(1)_{X_2}$ shares a $\Z_4$ subgroup, because
\be
\begin{split}
X_1-X_2&=r'_1(T_{25}|_{q=2})-r'_2(T_{25}|_{q=2})\\
&\bumpeq\diag(0, 0, -4, -4, -4, 4, 4, 4, 0, 0, 0, 0, 0, 0, -4, 4)\\
&\overset{\mod 4}{=}0,
\end{split}
\ee
meaning that $\e^{\frac{2\pi\ii}{4}X_1}=\e^{\frac{2\pi\ii}{4}X_2}$, which generates a $\Z_4$ group
\be
\Z_4=\langle \e^{\frac{2\pi\ii}{4}X_1}\rangle = \{\e^{\frac{2\pi\ii}{4}mX_1}|m=0,1,2,3\}.
\ee
This is also the $\Z_4$ center of $\Spin(10)$.

%\newpage
\section{Bibliography}

%\section{Bibliography}
\bibliographystyle{Yang-Mills}
\bibliography{BSM-SU3SU2U1-cobordism-GEQC.bib}

\end{document}